\documentclass[11pt]{article}

\usepackage{amsmath}

\begin{document}

\begin{center}

{\bf
Relativistic aberration effect on the the light  reflection  law \\
and the form of reflecting surface in a moving reference frame \\ [2mm]
V.M. Red'kov, Bernhard Rothenstein,  George J. Spix }
\\[3mm]
Institute of Physics, National  Academy of Sciences
of Belarus\footnote{redkov@dragon.bas-net.by}\\
Politehnica University of Timisoara, Physics Department, Timisoara,
Romania\footnote{bernhard rothenstein@yahoo.com }\\
 BSEE Illinois Institute of Technology, USA \footnote{ gjspix@msn.com}

 \end{center}

\begin{center}
SUMMARY
\end{center}

\begin{quotation}
The influence of the relativistic motion of the reference frame on the light reflection
law is investigated. The method is based on applying the relativistic
aberration affect for three light signals: incident, normal and reflected rays.
 The form of the reflection law in the moving reference frame is substantially modified
 and includes an additional  parameter which is the  velocity vector of the reference frame.
  It is shown that the reflected ray, as measured by a moving observer, would not
  in general be in the same plane as the incident and normal rays.

A general method to transform the form of any rigid surface in 3-dimensional space
with respect to the arbitrary directed relative motion of the reference frame is detailed.
This method is based on the light signals processes and the invariance of the light
velocity under Lorentz transformations. It is shown that a moving observer will measure
a plane surface as a hyperboloid. That observer will also measure a spherical surface
as an ellipsoid. A right line in the plane is seen by a moving observer as a  hyperbola.

The whole analysis  is extended  to the case of a  uniform media.

\end{quotation}


\newpage

\begin{center}
CONTENT
\end{center}

\vspace{5mm}
\noindent
1. Introduction

\vspace{3mm}
\noindent
2. Problem setting

\vspace{3mm}
\noindent
3. Transfer from  the fixed frame $K'$ to  the moving $K$

\vspace{3mm}
\noindent
4. On relativistic form of the light reflection law

\vspace{3mm}
\noindent
5. On describing the form of  a moving  mirror

\vspace{3mm}
\noindent
6. On geometrical form of a  moving mirror

\vspace{3mm}
\noindent
7.  Generalization and simplification, vector form

\vspace{3mm}
\noindent
8. The Lorentz transform with the  arbitrary velocity vector
 ${\bf V}$

\vspace{3mm}
\noindent
9.  Reflection of the light in moving reference frame $K$,
 the  case of an arbitrary velocity vector

\vspace{3mm}
\noindent
10. On  the form of reflection surface in the
moving reference frame

\vspace{3mm}
\noindent
11.  Canonical form of the reflection surface $S$ in the $K$- frame

\vspace{3mm}
\noindent
12.  On the form of spherical mirror
in the  moving reference frame (two-dimensional case)

\vspace{3mm}
\noindent
13.  On the form of spherical mirror in the moving
reference frame, general three-dimensional case

\vspace{3mm}
\noindent
14.  The light reflection law in media

\vspace{3mm}
\noindent
15.  The light and the  tensor formalism of 4-velocities
 $u^{a}$

\vspace{3mm}
\noindent
16. Relativistic velocities and Lobachewski geometry

\vspace{3mm}
\noindent
17. On relativistic transformation of the geometrical form
of  the surface to a moving reference frame in  presence of a media

\vspace{3mm}
\noindent
18.  Modified Lorenz transforms in a uniform media

\vspace{3mm}
\noindent
19.  On transforming the form of rigid surface when going to
a moving reference frame in presence of a uniform media, with modified Lorentz formulas

\vspace{3mm}
\noindent
20. Conclusions

\newpage

\section{ Introduction and overview}

\hspace{5mm}
The main idea of the present work is to follow the role of the relativistic aberration
effect on the light reflection law\footnote{
The same aberration considerations of the reflection law apply to the law of light refraction.}. This will result in answering the question:
 What is the form of the light reflection law to be used by a moving observer?
 The answer is of interest beyond the academic since it has always been assumed
 that the light reflection law is the same irrespective of the notion of the
 reflecting mirror. But this is not necessarily true.

 The problem is old
[1]  and has many aspects  (we plan to review them elsewhere [1-97]).
 In recent years
in the literature have appeared a number of works in this matter.
For instance, the paper by A. Gjurchinovski [27] is started by
words:

{\em  Experiments involving  moving mirrors  are among  the most interesting experiments encountered
in physics. Michelson's  apparatus for measuring the speed of light with a rotating
wheel consisting of mirrored  edges, an array of corner mirrors on the Moon's surface for estimating
the distance between the Earth and the Moon, the Michelson and Morley interferometer  for detecting
the ether , and the rotating Sagnac interferometer for determining the angular velocity of the Earth
are just a few experiments in which  moving mirrors have had prominent roles. In most
textbooks that discuss these experiments, it is implicitly  assumed  that
the ordinary law of reflection
of light is valid, that is, the angles of incidence and reflection are equal.
Our goals in this paper are to show that the ordinary law of reflection does not hold when the mirror
is moving at a constant velocity and to find a correct relation between the incident and the reflected
angle.}

In general points, our approach  coincides with that used by A. Einstein a century ago [1].
Einstein considered  the oblique  incidence of a light ray on a perfectly reflecting mirror whose
velocity was directed perpendicularly to its surface. To derive the equation of the angle of
reflection Einstein transformed with the help of Lorentz formulas the equations describing
the reflection in the reference frame where  the mirror was at rest.

The specificity  of this particular case is that the normal direction is the same for observers at rest and
at motion. Therefore, the problem of  deriving the mathematical form of the reflection law in
the moving reference frame reduces to relativistic aberration effect for two light rays,
incident and
reflected.

In a general case this cannot be so: the surface normal can be oriented arbitrarily
with respect to the direction of the motion of the reference frame. Besides,
the problem must be
stated in a 3-dimensional space and for any surface.

In contrast to  the author  of [27] -- {\em ...
the angle $\phi$ (the inclination
angle of the flat mirror in the rest reference frame) is the real physical quantity,
which, by itself, has nothing to do with relativity ...} -- we  have accepted an opposite view.
Indeed, if we consider only the  rotations of the  rest reference frame
 then we must  rotate correspondingly the vector ${\bf n}$  of the surface normal.
In the same manner, this vector ${\bf n}$ of the surface normal must be changed with respect
to relativistic motion as well.
In the mathematics of relativistic physics there do not exist
any quantities that behave as 3-vectors under Euclidian rotation
that will also be scalar under Lorentz transformations.

For a logical starting point we have chosen to define a light ray
as a physical representative for the normal to the surface in the rest reference frame.
This method from the very beginning contains a logical deficiency:
in the rest reference frame  one can take any of two light signals:
the normal  vector ${\bf n}$  related with the light going normally from the surface,
and the inverse vector ${\bf N} = -{\bf n}$  associated with the light signal
going to the surface.
We have examined both of these variants.

In that formulating,  the problem of the light reflection law in the  moving reference
frame is reduced to examining the details of
the relativistical aberration effect for a triple of light velocity vectors.
All analysis of the present paper
is based only on the kinematics of relativity, that is on Lorentz transformations
and the  exclusive    properties of the the light velocity under them.
We have sought to give a complete treatment of the problem so that
the reader may see many details that may have been regarded
as evident and well-known.

Below we give a short overview of the content by Sections 2-19.

In Section 2, the process of reflecting the light signal on an
inclined mirror in the rest reference frame $K'$, for the simplest plan problem
\begin{eqnarray}
 y' = b+kx'
\label{1.1}
\end{eqnarray}

\noindent     is stated  in terms of a set
of relativistic events. The normal to the inclined  (flat) mirror  is modeled by
a light signal going normally from the mirror.
In Section 3 with the help of Lorentz formulas the coordinates of the above
  set of events in a moving reference frame $K$ are found.

In Section 4, the light reflection law  in this simplest flat arrangement,
  is given in the moving reference frame
\begin{eqnarray}
{ \sin \alpha_{i}  + V\;  (\sin \alpha - \sin \phi_{2})
\over  (1+ V\; \cos \phi_{2} )}
=
{  \sin \alpha_{r}  + V\;  (\sin \alpha - \sin \phi_{3})
\over  (1+ V\; \cos \phi_{3} )}\; \; .
\label{1.2}
\end{eqnarray}

\noindent   It is shown that the modified form of the law, though
 being rather complicated, proves its  invariance under the Lorentz transformations.

In Section 5,  we turn to describing the geometrical form of the inclined mirror
in the moving reference frame $K$.
With  the  inclined flat mirror in the rest reference frame $K'$,
 for the moving  observer $K$ can be associated  a set of space-time
events, each of those  is an arrival of a light signal emitted  from the space-time  point $
(0;0,0)$ toward  the mirror with different angles.
Evidently, all such events will take place on the surface of the  mirror but at different
 times. These time variable can be excluded and a new equation $\varphi(x,y)=0$
  is derived, which  should be  considered as an equation
describing the geometrical form of  the   mirror for the moving observer $K$.
 It should be emphasized that
a  definite and  practically realizable procedure with the use of light signals in the
reference frame $K$  underlies this  equation $\varphi(x,y)=0$.

In Section 6 we show that the given equation $\varphi(x,y)=0$ represents
a second order curve
\begin{eqnarray}
k^{2} \; x^{2} - 2k\cosh \beta \; xy + (1 - k^{2}\sinh^{2} \beta) \; y^{2} +
+ 2bk\; \cosh \beta \; x - 2b\; y + b^{2} = 0 \; .
\label{1.3}
\end{eqnarray}

\noindent  that can be  translated to the  canonical form by the use of  a rotation
and a shift. In this way, we show that the geometrical form of the mirror is a
hyperbola.

In Section 7, we are to extend  the results of the Section 3  to a   vector
form. The  incident, normal, and reflected  light rays  in the rest
reference frame $K'$ are described by respective vectors
\begin{eqnarray}
{\bf a} '= {{\bf W}'_{in} \over c} \; , \;\;
{\bf n} '={  {\bf W}'_{norm} \over c} \; , \qquad {\bf b} '= { {\bf W}'_{out} \over c} \; .
\nonumber
\end{eqnarray}

\noindent all three vectors have  a unit length and their lengths are  invariant
under the Lorentz transformations. The reflection  law in the $K'$ can be written as
\begin{eqnarray}
{\bf a} '\times {\bf n}' =  {\bf b}' \times {\bf n}' \; .
\nonumber
\end{eqnarray}

\noindent After the Lorentz transforming  it will take the  form
\begin{eqnarray}
{{\bf a} \times {\bf n} + {\bf V} \times ({\bf n} - {\bf a}) \over
1 + {\bf a} {\bf V} } = {{\bf b} \times {\bf n} + {\bf V} \times
({\bf n} - {\bf b}) \over 1 + {\bf b} {\bf V} } \; .
\label{1.4}
\end{eqnarray}

In Section 8, some mathematical details on
Lorentz transformations $L({\bf V})$  are given (see much more details of the Lorentz
group theory in the book [17]). With the  notation
$
{\bf V} = {\bf e} \; th\; \beta \;  , \qquad {\bf e} ^{2} = 1 \; $,
arbitrary Lorentz transformation $L({\bf V})$
\begin{eqnarray}
 t = ch\; \beta \; t -   sh\; \beta \; {\bf e}\;   {\bf x} \; ,
\qquad
{\bf x}'= -{\bf e}  \; sh\; \beta  \; t + {\bf x}  + (ch\; \beta
-1)\; {\bf e} \; ({\bf e}  {\bf x})  \;
\nonumber
\end{eqnarray}

\noindent  leads to the  general form of the velocity addition law
\begin{eqnarray}
{\bf w} = { {\bf w}'  - {\bf e}  \; sh\; \beta   + (ch\; \beta
-1)\; {\bf e} ( {\bf e} \; {\bf w}' ) \over ch\; \beta  - sh\;
\beta \; {\bf e}\;  {\bf w}'  } \; .
\label{1.5}
\end{eqnarray}

In Section 9, we derive a general relationship describing  the light
 reflection  in the moving reference frame $K$:
\begin{eqnarray}
{ch\; \beta  \;  ({\bf a} \times {\bf n} )  +
 sh\; \beta  \; ( {\bf a} -{\bf n}  )\times{\bf e}
+  (ch\; \beta -1)\;  {\bf e} \; [ \;{\bf e}  ({\bf n} \times {\bf
a})\; ]
    \over
ch\; \beta  + sh\; \beta \; ({\bf e} {\bf a})  }=
\nonumber
\\
= {ch\; \beta  \;  ({\bf b} \times {\bf n})  +
 sh\; \beta ({\bf b} -{\bf n})  \times {\bf e}
  +   (ch\; \beta -1)\; {\bf e} \; [ \;{\bf e}  ({\bf n} \times {\bf b})\; ]
    \over
ch\; \beta  + sh\; \beta \; ({\bf e} {\bf b})  } \; .
\label{1.6}
\end{eqnarray}

\noindent Take notice of terms including  $[ \;{\bf e}  ({\bf n} \times {\bf a})\; ]$ which
substantially  extend the previous result  (\ref{1.4}) of the plane problem.

In the Section 9, a special attention is given to  one other  aspect of the problem.
In the rest reference frame $K'$ three vectors
${\bf a},{\bf n},{\bf b}$ belong to the same single plane,
but for a moving observer it is not so. To describe this phenomenon,
we have introduced the following quantity (the volumes of the light parallelepiped
$[\; {\bf n} \; ({\bf a} \times   {\bf b} )\; ] $):
\begin{eqnarray}
\Delta'  = {\bf n}' \; ({\bf a}' \times   {\bf b}' )=0 \qquad \mbox{but} \qquad
\Delta = {\bf n} \; ({\bf a} \times   {\bf b} ) =
\nonumber
\\
=  {  - 2 \;sh\;
\beta  \; ({\bf a}' {\bf n}') \; [\; {\bf e} ({\bf n}' \times {\bf a}' ) \;] \;
\over (ch\; \beta  - sh\; \beta \;
{\bf e}{\bf n}' ) \; (ch\; \beta  - sh\; \beta \; {\bf e}{\bf
a}')\;
 (ch\; \beta  - sh\; \beta \; {\bf e}{\bf b}') } \; .
\label{1.8}
\end{eqnarray}

In Section 10 we  turn to examining the form
 of the reflection surface, the plane in the rest reference frame (${\bf d}$ is a normal
 to the plane)
\begin{eqnarray}
S': \qquad {\bf x} ' \; {\bf d}' + D =0 \; ,
\nonumber
\end{eqnarray}

\noindent   in the moving reference frame $K$.
For this  with the reflection plane  is associated  the following set of space-time events:
\begin{eqnarray}
S': \qquad \Longrightarrow \qquad \left \{ \;\; t' = \sqrt{ {\bf x}^{'2}  } \; , \;
 {\bf x} ' \; {\bf d}' + D' =0 \; \;  \right \} \; .
\nonumber
\end{eqnarray}

\noindent This set of events can be readily transformed to the moving reference frame, which
after excluding
the time-variable provides us with a new surface equation
\begin{eqnarray}
 S: \qquad  ({\bf x} {\bf d})   + ( {\bf e} {\bf d} ) \;
 \left [ \;  sh\; \beta  \; (\; \sqrt{ {\bf x}^{2}  } \; ) +
(ch\; \beta -1)\;  ({\bf e}  {\bf x})\;  \right ]    + D =0 \; .
\label{1.9}
\end{eqnarray}

\noindent Take notice than only when ${\bf e} {\bf d}=0$ takes place, then
the plane in the rest frame
will look as the plane in the  moving reference frame too. For all other cases the plane
does not preserve  its geometrical form and is a second order surface.

In Section 11  a canonical form of this second order
 surface has been   established  with the use of a  special rotation and a shift.
 It turns to be a   hyperboloid.  Its symmetry axis is directed  along the vector
\begin{eqnarray}
{\bf f}  = { {\bf d} + (ch\; \beta -1) ({\bf e} {\bf d}) {\bf e} \over
({\bf e} {\bf d}) \; sh \; \beta } \; .
\nonumber
\end{eqnarray}

In Section 12, the geometrical form of the (plane) circle in the rest reference frame has been
transformed to  the moving observer,  it proves  to be the  ellipse
\begin{eqnarray}
{(x- 2R\; sh\; \beta )^{2}\over R^{2}\; ch^{2} \beta}
 +  { y^{2}  \over R^{2}} =1 \; .
\label{1.10}
\end{eqnarray}

\noindent In Section 13, a  3-dimensional problem is solved:
the sphere in the rest reference frame,
being transformed to the  moving reference frame, becomes an  ellipsoid
\begin{eqnarray}
{(X- 2R\; sh\; \beta )^{2}\over R^{2}\; ch^{2} \beta}
 +  { Y^{2}  \over R^{2}} +  { Z^{2}  \over R^{2}} =1 \; ,
\label{1.11}
\end{eqnarray}

\noindent  its symmetry   axis is directed along the vector ${\bf V}$.

In Section 14, we turn to the case when the light ray (incident, reflected, and  normal)
are  propagated in a uniform media with  the refraction index   $n>1$.  In the rest
reference frame $K'$  the reflection  law preserves its form. However,
there arise differences when going to a  moving reference frame
$K$. The fact of the most significance is that when using ordinary  (vacuum-based)
 Lorentz transformations then because  the speed of  light in the rest media is less than c
 ( $kc<c$)
   the modulus of the light velocity  vector  does not preserve
 its value.

 General mathematical  form of the  reflection  law
 in the  moving reference frame
formally stays the same,  however  one  must take into account that the lengths  of the
 vectors involved ${\bf a},{\bf b},{\bf n}$ are different from 1.
The length  of  the light  vector in the  moving reference frame in presence of the uniform
media   is
\begin{eqnarray}
 W   =  \sqrt{1 -
{(1 - k^{2})  \over
(ch\; \beta  -  ({\bf e} {\bf W}')  \; sh\; \beta )^{2} }} \;  .
\label{1.12}
\end{eqnarray}

 It should be especially   noted one other aspect of the problem: the latter equation
  means that the light velocity in the  reference frame  $K$ is a function of
direction of the propagation of the light. This fact is of most significance because it
changes basically
the general structure of special relativity in presence of a media. In such circumstances
there appears an absolute reference frame related to the rest media, the reference frame
 $K'$. In the   reference frame $K'$,
 the light velocity is an isotropic quantity that preserves its
 value in all  space directions. In any  other  reference frame, moving $K$,
 the light velocity is anisotropic --  it is a function of directions.

In Section 15, some aspects the  tensor formalism of 4-velocities
 $u^{a}$ are specified for the  light case.
  In the Section 16.
some geometrical aspects of the relativistic velocity concept  in terms of  the
Lobachewsky 3-geometry are briefly discussed [....].

In Section 17,
the relativistic transformation of the geometrical form
$\varphi({\bf x}')=0$
of any  surface to a moving reference frame in  presence of a media is considered:
\begin{eqnarray}
\varphi({\bf x}')=0 \qquad \Longrightarrow \qquad
 \varphi \left [  {\bf x}  + {\bf e}  \;(\;  sh\; \beta  \;
{  \sqrt{ {\bf x}^{2} }  \over \sqrt{{\bf W}^{2}} }
\;+
(ch\; \beta -1)\;  ({\bf e}  {\bf x})\;  \right  ] = 0 \; ,
\label{1.14}
\end{eqnarray}

In Section 18, we briefly discuss the  scheme of Special Relativity
in a uniform media that can be constructed
on the base of the light velocity in the media $kc$ [....].
Modified Lorentz formulas look as
\begin{eqnarray}
t' =  {   t +    {\bf V} {\bf x} /  k^{2}c^{2}
 \over \sqrt{1 -V^{2} / k^{2} c^{2} } }
 \; , \qquad
{\bf x}'=
  {\bf x}  -  {\bf e} \; ({\bf e}  {\bf x})  +
 {   {\bf e}({\bf e}  {\bf x} )  + {\bf V}  t  \over \sqrt{1 -V^{2} / k^{2} c^{2} } } \; .
\nonumber
\end{eqnarray}

\noindent
The  value of light velocity in the media
is invariant under modified Lorentz formula: $
{\bf W}^{2} = k^{2} c^{2}  \; \Longrightarrow \;  {\bf W}^{'2} = k^{2} c^{2}\; .
$

In Section 19,  the corresponding  scheme of transforming the form of a surface in going to
the  moving reference frame is discussed.
The general method  is the same:
\begin{eqnarray}
S: \qquad \varphi \;[  \;  {\bf x}  + {\bf e}  \;(\;  sh\; \beta  \; \sqrt{ {\bf x}^{2}} +
(ch\; \beta -1)\;  {\bf e}  {\bf x} \; )\;  ] = 0 \; ,
\nonumber
\end{eqnarray}

\noindent
the  media's presence  enters through the hyperbolic functions
\begin{eqnarray}
 ch\;  \beta = {1 \over \sqrt{1 - V^{2}/k^{2}c^{2}}} \; ,
\qquad  sh \;\beta = {V \over kc\; \sqrt{1 - V^{2}/k^{2}c^{2}}} \; .
\nonumber
\end{eqnarray}

\section{ Problem setting   }

\hspace{5mm} Let two observers (two inertial reference frames), $K$
and $K'$, be given. The standard arrangement is shown in the
Fig. 1

\vspace{2mm}

\unitlength=0.6mm
\begin{picture}(100,50)(-55,0)
\special{em:linewidth 0.4pt} \linethickness{0.6pt}

\put(0,0){\vector(+1,0){40}}  \put(+45,-5){$x$}
\put(0,0){\vector(0,+1){40}}  \put(-10,+40){$y$}
\put(0,0){\vector(-1,-1){20}} \put(-30,-20){$z$}

\put(-30,-40){reference frame  $K$ }

\put(80,0){\vector(+1,0){40}}  \put(+125,-5){$x'$}
\put(80,0){\vector(0,+1){40}}  \put(+70,+40){$y'$}
\put(80,0){\vector(-1,-1){20}} \put(+50,-20){$z'$}

\put(+70,-40){reference frame $K'$}

\put(+35,+30){\line(+1,0){20}}  \put(55,+28){>}
\put(45,+35){$\mbox{v}$}

\put(110,0){\line(-1,+1){20}}       \put(100,15){mirror}
\put(110,-0.5){\line(-1,+1){20}}
\put(110,+0.5){\line(-1,+1){20}}

\end{picture}

\vspace{25mm}

\begin{center}
{\bf Fig. 1 Problem setting }
\end{center}

\begin{quotation}

At  the moment of its coincidence, \underline{event $1$} $(t_{0}=0,x_{0} =0)$ and
\underline{event $1'$} $(t_{0}=0,x'_{0} =0)$, observer $K'$ sends  a light ray signal which is reflected by
an inclined flat mirror, unmoving in reference frame $K'$, at the moment
  $t_{2}'$ (\underline{event $2'$}); the third \underline{event $3'$} is  taken
  by symmetry considerations -- see
   in Fig. 2

\end{quotation}

\vspace{32mm}

\unitlength=0.5mm
\begin{picture}(100,50)(-100,0)
\special{em:linewidth 0.4pt} \linethickness{0.6pt}

\put(-70,+100){\underline{$K'$}}

\put(0,0){\vector(+1,0){150}}  \put(+155,-5){$x'$}
\put(0,0){\vector(0,+1){100}}  \put(-10,+100){$y'$}

\put(120,0){\line(-1,+1){110}}
\put(120.5,0){\line(-1,+1){110}}
\put(120.8,0){\line(-1,+1){110}}

\put(0,0){\line(+1,+3){30}}  \put(0,0){\line(-1,+1){60}}
\put(+30,+90){\line(-1,-1){90}}
\put(+30,+90){\line(-3,-1){90}}

\put(-2,-10){$1'$}    \put(+32,+90){$2'$}
\put(-45,+25){$2*'$}  \put(-67,+60){$3'$}

\put(+4,+2){$\phi '_{2}$}  \put(+126,+3){$\gamma '$}
\put(115,0){\oval(+18,+16)[tr]}

\put(+16,+71){$\alpha_{i}'$}  \put(+9,+77){$\alpha_{r}'$}

\put(+60,-10){$L$}  \put(+80,+50){$l$}

\put(100,+70){$\alpha_{i}' = \alpha_{r}'$}

\end{picture}

\vspace{10mm}

\begin{center}
{\bf Fig. 2 Reflection and relativistic events}
\end{center}

\vspace{5mm}

How  will this phenomenon be seen for second  observer $K$?

From geometry considerations in Fig. 2 it follows (take notice in Fig.2 the angles
$\gamma', \phi '_{2}, \phi '_{3}$ are measured from zero degree position at the axis $x$;
  $\gamma'$ and  $\Phi_{3}'$ are  obtuse, whereas  $\phi_{2}'$ is an acute angle)
\begin{eqnarray}
(180^{0} -\gamma' ) + \phi '_{2} + (90^{0} - \alpha'_{i}) = 180^{0}: \qquad
\Longrightarrow
\\
\nonumber
 \alpha_{i}' = \phi '_{2} - \gamma ' + 90^{0} \; . \qquad \qquad
\label{2.1}
\end{eqnarray}

In  $K'$  frame the motion of the light along the line  $1'2'$  is given by parametric formulas
\begin{eqnarray}
x'(t') =  (c  \; \cos \phi ' _{2})\; t '\; , \qquad
y'(t') =  (c \; \sin \phi '_{2} ) \; t '\; ;
\nonumber
\end{eqnarray}

\noindent and the inclined mirror's form  is given by
\begin{eqnarray}
x'  = L' + l' \; \cos \gamma ' \; ,
\qquad y'  =  l' \; \sin \gamma ' \; ,
\label{2.2b}
\end{eqnarray}

\noindent
At the point where the light falls on the mirror \underline{(event $2'$)}
 it must  hold two relations:
\begin{eqnarray}
(c  \; \cos \phi '_{2}) \; t' _{2} = L' + l'_{2} \; \cos \gamma '  \; ,
\nonumber
\\
(c\;  \sin \phi '_{2})  \; t'_{2}  =  l'_{2} \; \sin \gamma '  \; .
\nonumber
\end{eqnarray}

\noindent
This linear system under  $t'_{2},l'_{2}$
\begin{eqnarray}
t' _{2}\; ( c \;\cos \phi '_{2} ) - l'_{2} \; \cos \gamma ' = L' \; ,
\nonumber
\\
t'_{2} ( \;  c\; \sin \phi '_{2})  - l'_{2} \; \sin \gamma ' = 0  \; .
\nonumber
\end{eqnarray}

\noindent  can be easily solved
\begin{eqnarray}
t'_{2} = {L \over c} \; { \sin \gamma ' \over
(\cos \phi' _{2}\; \sin \gamma ' - \sin \phi' _{2}  \;\cos \gamma' ) } =
{L \over c} \; { \sin \gamma ' \over
\sin (\gamma ' - \phi'_{2} ) } \; ,
\nonumber
\\
l'_{2}  =  L  \; {\sin \phi ' _{2}\over
(\cos \phi'_{2} \sin \gamma ' - \sin \phi'_{2} \cos \gamma ')}  =
L  \; {\sin \phi ' _{2} \over
\sin (\gamma ' - \phi'_{2} )   } \; .
\nonumber
\end{eqnarray}

\noindent
Thus, to \underline{the event $2'$ }  there corresponds a set of coordinates:
\begin{eqnarray}
t'_{2} =   {L \over c} \; { \sin \gamma ' \over
\sin (\gamma ' - \phi' _{2})  }  \; ,
\nonumber
\\
\qquad\qquad x_{2}' =
(c  \; \cos \phi '_{2} )\; t '_{2}  =
{ L\; \cos \phi '_{2}\;  \sin \gamma ' \over
\sin (\gamma ' - \phi'_{2} )   } \; ,
\nonumber
\\
\qquad \qquad y'_{2}  =  (c \; \sin \phi '_{2} ) \; t '_{2} =
{ L\; \sin \phi ' _{2} \; \sin \gamma ' \over
\sin (\gamma ' - \phi'_{2} )   } \; ;
\label{2.4}
\end{eqnarray}

\noindent  at an established position of the mirror $(L,\gamma')$ in  $K'$  frame, a given
 angle $\phi'_{2}$  determines  unambiguously coordinates of the event
 $2'$: $\;\;x'_{2},y_{2}',t_{2}'$.

Let us write an equation for light trajectory  $2'3'$ after reflection at the point $2'$.
From geometry considerations in Fig 2. it follows an  expression for
its   directing angle  $\phi'_{3}$:
\begin{eqnarray}
\phi'_{3}  =(\phi'_{2} - \alpha_{i}'  + 180^{0}  - \alpha_{i}' )=\phi'_{2} +180^{0} -2(
\phi'_{2} - \gamma ' + 90^{0}): \qquad \Longrightarrow
\nonumber
\\
\phi'_{3} =  2 \; \gamma' - \phi '_{2}  \hspace{30mm}
\label{2.5}
\end{eqnarray}

\noindent therefore the light at the way $2'3'$ is described by
\begin{eqnarray}
x'(t')- x'_{2} =  (t' - t'_{2}) \;c \; \cos \; \phi '_{3} \; ,
\nonumber
\\
y'(t') - y'_{2} = (t' - t'_{2}) \;c \; \sin \; \phi '_{3} \; .
\label{2.6}
\end{eqnarray}

\noindent

Let us use established convention to   determine event $3'$
 in a symmetrical way with respect to event $1'$:
 $
t_{3}' =  2 t_{2}' \; .
$
As a result \underline{coordinates of the event  $3'$ } are
\begin{eqnarray}
t_{3}'  =  2\; {L \over c} \; {   \sin \gamma ' \over
\sin (\gamma ' - \phi'_{2} ) } \; ,
\nonumber
\\
x'_{3}= x'_{2} + t'_{2} \;c  \;\cos \; \phi '_{3}\; ,
\nonumber
\\
y'_{3}= y'_{2} +  t'_{2} \;c  \;\sin \;  \phi '_{3}\; .
\label{2.8}
\end{eqnarray}

\noindent
It is to be noted  that both  sets $(x'_{2}, y_{2}',t'_{2}, \phi '_{3})$
and   ($x'_{3};, y_{3}',t'_{3}$) are  definite and uniquely  determined functions of the initial  angular
parameter  $\phi '_{2}$.

One other   point: to have possibility to describe a normal to the mirror (with a directing
angle $\alpha ' = \gamma' + 90^{0}$) in terms of  events of space-time  world,
we should introduce one subsidiary event  $2*'$. Let at  the reflection point observer $K'$
send an additional light  signal along normal to the  mirror
$$
\left. \begin{array}{l}
x^{'*}_{2} = x'_{2} + \cos \alpha'  \;(t^{'*}_{2} - t_{2}') \; ,\\[3mm]
y^{'*}_{2} = y'_{2} + \sin \alpha'  \;(t^{'*}_{2} - t_{2}') \; , \\[3mm]
t^{'*}_{2} = 2t'_{2} \; , \qquad \alpha ' = \gamma '+ 90^{0} \; .
\end{array} \right.
\eqno(2.9)
$$

NOTATION

\vspace{2mm}

In the following we are to translate coordinates of four events
$1' - 2' - 2*' - 3'$ to the reference frame  $K$,  with the help
of the Lorentz formulas
\begin{eqnarray}
t= {t' -Vx'/c^{2}  \over \sqrt{1 - V^{2} /c^{2}}} \;, \qquad
x= {x' -Vt' \over \sqrt{1 - V^{2} /c^{2}}} \;, \qquad
y= y' \; .
\label{2.10a}
\end{eqnarray}

\noindent
and then we are to analyze relationships between four events $1-2-2*-3$
in the reference frame  $K$.

 We will now employ the convention of letting  $c = 1 $   to simplify the formulas
(sometimes it is  referred as the use of unit system with  $c=1$):
\begin{eqnarray}
c t \Longrightarrow t , \qquad {V \over c}  \Longrightarrow  V \; \; ,
\nonumber
\end{eqnarray}

\noindent
then Lorentz transforms   (\ref{2.10a}) take the form
\begin{eqnarray}
t= {t' -Vx'  \over \sqrt{1 - V^{2} }} \;, \qquad
x= {x' -Vt' \over \sqrt{1 - V^{2} }} \;, \qquad
y= y' \; .
\label{2.10c}
\end{eqnarray}

With this notation the formulas we need are

\vspace{2mm}
\underline{Events  $2'$ }, $3'$, $2*'$:
\begin{eqnarray}
t'_{2} =   L \; { \sin \gamma ' \over
\sin (\gamma ' - \phi' _{2})  }  \; ,
\nonumber
\\
\qquad\qquad x_{2}' =
 \cos \phi '_{2} \; t '_{2}  =
{ L\; \cos \phi '_{2}\;  \sin \gamma ' \over
\sin (\gamma ' - \phi'_{2} )   } \; ,
\nonumber
\\
\qquad \qquad y'_{2}  =   \sin \phi '_{2}  \; t '_{2} =
{ L\; \sin \phi ' _{2} \; \sin \gamma ' \over
\sin (\gamma ' - \phi'_{2} )   } \; .
\nonumber
\\[2mm]
t_{3}'  =  2\; L  \; {   \sin \gamma ' \over
\sin (\gamma ' - \phi'_{2} ) } \; ,
\nonumber
\\
x'_{3}= x'_{2} + t'_{2} \; \cos \; \phi '_{3}\; ,
\nonumber
\\
y'_{3}= y'_{2} +  t'_{2}  \;\sin \;  \phi '_{3}\; ,
\nonumber
\\[2mm]
t_{2*}' = t_{2}' \; , \qquad
x'_{2*} = x_{2}' - \; t_{2}' \; \cos \alpha_{i}' \; ,
\nonumber
\\
y'_{2*} = y_{2}' - \;t_{2}' \; \sin \alpha_{i}' \; .
\label{2.11c}
\end{eqnarray}

\section{Transition  from  the fixed frame $K'$ to  the moving $K$}

\hspace{5mm}
With the use of (\ref{2.10c}), for coordinates of the event  $2$ one easily gets
\begin{eqnarray}
x_{2} = {x_{2}' -V \;t_{2}' \over \sqrt{1-V^{2}}} = { \cos \phi_{2}'  - V \over
\sqrt{1 - V^{2}}} \; t_{2}'  \; ,
\nonumber
\\
t_{2} = {t'_{2} - V \; x_{2}' \over \sqrt{1 -V^{2}}} =
{1 - V \; \cos \phi_{2}'  \over \sqrt{1 -V^{2}}} \; t_{2}' \; ,
\nonumber
\\
y_{2}= y_{2}'  = \sin \phi_{2} \; t_{2}' \; .
\label{3.1}
\end{eqnarray}

Expressing from second equation  $t_{2}'$   as a function of  $t_{2}$:
\begin{eqnarray}
t'_{2} = {\sqrt{1 - V^{2}} \over  1 - V \; \cos \phi_{2}'   } \; t_{2} \; ,
\nonumber
\end{eqnarray}

\noindent
for coordinates  $(x,y)$ of the event   $2$ we have
\begin{eqnarray}
x_{2} = { \cos \phi_{2}'  - V \over
1 - V \; \cos \phi_{2}'} \; \; t_{2}  \equiv \cos \phi_{2}  \; t_{2} \; ,
\nonumber
\\
y_{2}=  \sin \phi_{2}' \; \; {  \sqrt{1-V^{2}}   \over
1 - V \; \cos \phi_{2}' }  \;\; t_{2} \equiv  \sin \phi_{2} \; t_{2}\; .
\label{3.3}
\end{eqnarray}

It is readily  shown that  the introduction into  (\ref{3.3}) of a   variable angle
$\phi_{2}$ in the reference frame $K$ is correct. Verifying is reduced to
\begin{eqnarray}
(\cos \phi_{2}) ^{2} + (\sin \phi_{2} )^{2} =
({ \cos \phi_{2}'  - V \over
1 - V \; \cos \phi_{2}'} )^{2} +
(\sin \phi_{2}' \; \; {  \sqrt{1-V^{2}}   \over
1 - V \; \cos \phi_{2}' })^{2} =
\nonumber
\\
=
{\cos^{2} \phi_{2}'  - 2 V  \cos \phi_{2}' + V^{2}  + \sin^{2} \phi_{2}' -
\sin^{2} \phi_{2}' V^{2} \over (
1 - V \; \cos \phi_{2}' )^{2} } =
{1   - 2 V  \cos \phi_{2}'   + \cos^{2} \phi_{2}' V^{2} \over (
1 - V \; \cos \phi_{2}' )^{2} } = 1 \; .
\nonumber
\end{eqnarray}

\noindent
In essence, here is tested the famous Einstein's postulate on light velocity constancy:
its invariance  under Lorentz transformations.

Let us write down the formulas for changing the direction of light velocity vector
at transfer from one inertial reference frame  to another
(in other words this phenomenon is called the light aberration):
\begin{eqnarray}
 \cos \phi_{2} = { \cos \phi_{2}'  - V \over
1 - V \; \cos \phi_{2}'} \;  , \qquad
\sin \phi_{2}= \sin \phi_{2}' \; \; {  \sqrt{1-V^{2}}   \over
1 - V \; \cos \phi_{2}' }   \; .
\label{3.4a}
\end{eqnarray}

\noindent
Their inverse is  naturally symmetrical
\begin{eqnarray}
 \cos \phi'_{2} = { \cos \phi_{2}  + V \over
1 + V \; \cos \phi_{2}} \;  , \qquad
\sin \phi'_{2}= \sin \phi_{2} \; \; {  \sqrt{1-V^{2}}   \over
1 + V \; \cos \phi_{2} }   \; .
\nonumber
\end{eqnarray}

\vspace{5mm}

On the same way  we  carry out converting the coordinates of the event
$3'$:
\begin{eqnarray}
\left. \begin{array}{l}
t_{3}'  =  2 t_{2}' \; ,\\[2mm]
x'_{3}= x'_{2} + (t'_{3}-t_{2}') \; \cos \; \phi '_{3}\; , \\[2mm]
y'_{3}= y'_{2} +  (t'_{3} -t'_{2})   \;\sin \;  \phi '_{3}\; .
\end{array} \right \}  \qquad \Longrightarrow
\nonumber
\\
t_{3} = { t'_{3} - V \; x_{3}' \over \sqrt{1 -V^{2}}} =
{ 2t'_{2} - V \; ( \;\cos \phi'_{2}\;  t'_{2} + t_{2}' \; \cos \; \phi '_{3} \; )  \over
 \sqrt{1 -V^{2}}} =
\nonumber
\\
 =
{ 2  - V \; ( \;\cos \phi'_{2}  +  \cos \; \phi '_{3} \; )  \over
 \sqrt{1 -V^{2}}} \; t'_{2} \; ,
\nonumber
\end{eqnarray}

\noindent from where with the use of
\begin{eqnarray}
t_{2} ={1 - V \cos \phi_{2}' \over  \sqrt{1-V^{2}}} \;\; t'_{2}
\nonumber
\end{eqnarray}

\noindent it follows
$$
t_{3}- t_{2} =
{ 1  - V \;  \cos \; \phi '_{3}   \over
 \sqrt{1 -V^{2}}} \;\; t'_{2} \; .
\eqno(3.6c)
$$

\noindent
Analogously, from
\begin{eqnarray}
x_{3} = {x'_{3} - V \; t_{3}' \over \sqrt{1 -V^{2}}} =
{\cos \phi'_{2} \; t'_{2} + t'_{2} \; \cos \; \phi '_{3} - V\; 2t'_{2} \over
\sqrt{1 -V^{2}}} =
\nonumber
\\
=
{ \cos \phi'_{2}  + \cos \; \phi '_{3} - 2V \;   \over
\sqrt{1 -V^{2}}} \; t'_{2} \; ,
\nonumber
\end{eqnarray}

\noindent
with the use of
\begin{eqnarray}
x_{2} = {\cos \phi'_{2} - V \over \sqrt{1 -V^{2}}} \; t_{2}' \; ,
\nonumber
\end{eqnarray}

\noindent one finds
\begin{eqnarray}
x_{3} - x_{2}  =
{ \cos \; \phi '_{3} - V \;   \over
\sqrt{1 -V^{2}}} \; t'_{2} \; .
\label{3.7c}
\end{eqnarray}

\noindent
And finally,
\begin{eqnarray}
y_{3} = y'_{3} = y'_{2} + (t_{3}' - t_{2}') \sin \phi_{3}' = y_{2} +
\; t'_{2} \; \sin \phi_{3}'
\;: \qquad \Longrightarrow
\nonumber
\\
y_{3} - y_{2} =  t'_{2} \; \; \sin \phi_{3}'\;  .
\label{3.8}
\end{eqnarray}

Thus, the trajectory $23$ in the $K$ frame is given by
\begin{eqnarray}
t_{3}- t_{2} =
{ 1  - V \;  \cos \; \phi '_{3} \;   \over
 \sqrt{1 -V^{2}}} \; t'_{2} \; , \qquad \qquad
\nonumber
\\
x_{3} - x_{2}  =
{ \cos \; \phi '_{3} - V \;   \over
1 -V \cos \phi'_{3}} \;  (t_{3}- t_{2}) \equiv
\cos \phi_{3} \;  (t_{3}- t_{2}) \; , \qquad
\nonumber
\\
y_{3} - y_{2} =  \; \sin \phi_{3}'\;  {\sqrt{1 -V^{2}} \over
1 -V \cos \phi'_{3}} \;  (t_{3}- t_{2})
\equiv \sin \phi_{3} \;  (t_{3}- t_{2}) \; .
\label{3.9c}
\end{eqnarray}

\noindent In other words,
the light aberration at the way $2-3$ is described by relations
\begin{eqnarray}
{ \cos \; \phi '_{3} - V \;   \over
1 -V \cos \phi'_{3}} =
\cos \phi_{3} \; , \qquad
\sin \phi_{3}'\;  {\sqrt{1 -V^{2}} \over
1 -V \cos \phi'_{3}} =
\sin \phi_{3}  \; ,
\label{3.10a}
\end{eqnarray}

\noindent
the inverse transform looks as
\begin{eqnarray}
{ \cos \; \phi _{3} + V \;   \over
1 +V \cos \phi_{3}} =
\cos \phi'_{3} \; , \qquad
\sin \phi_{3}\;  {\sqrt{1 -V^{2}} \over
1 +V \cos \phi_{3}} =
\sin \phi'_{3}  \; .
\nonumber
\end{eqnarray}

In the same way, we readily produce the formulas describing
the trajectory $2-2*$ (see. (3.9)):
\begin{eqnarray}
t^{*}_{2}- t_{2} =
{ 1  - V \;  \cos \; \alpha ' \;   \over
 \sqrt{1 -V^{2}} } \; t'_{2} \; , \qquad \qquad
\nonumber
\\
x^{*} _{2} - x_{2}  =
{ \cos \; \alpha ' - V \;   \over
1 -V \cos \alpha ' } \;  (t^{*}_{2}- t_{2}) \equiv
\cos \alpha  \;  (t^{*}_{2}- t_{2}) \; , \qquad
\nonumber
\\
y^{*}_{2} - y_{2} =  \; \sin \alpha '\;  {\sqrt{1 -V^{2}} \over
1 -V \cos \alpha' } \;  (t^{*}_{2}- t_{2})
\equiv \sin \alpha \;  (t^{*}_{2}- t_{2}) \; .
\label{3.11c}
\end{eqnarray}

\noindent
The light aberration on the way  $2-2*$ is given by
\begin{eqnarray}
{ \cos \; \alpha ' - V \;   \over
1 -V \cos \alpha ' } \;  = \cos \alpha  \; , \qquad
\sin \alpha '\;  {\sqrt{1 -V^{2}} \over
1 -V \cos \alpha' } = \sin \alpha \; ,
\label{3.12a}
\end{eqnarray}

\noindent and inverse ones
\begin{eqnarray}
{ \cos \; \alpha  + V \;   \over
1 + V \cos \alpha  } \;  = \cos \alpha'   \; , \qquad
\sin \alpha \;  {\sqrt{1 -V^{2}} \over
1 +V \cos \alpha } = \sin \alpha' \; .
\nonumber
\end{eqnarray}

\section{ On relativistic form of the light reflection law
 }

\hspace{5mm} In the reference frame  $K'$ the light reflection law
is formulated in the equality
\begin{eqnarray}
\alpha'_{i} = \alpha_{r}' \; , \hspace{30mm}
\label{4.1a}
\\
\alpha'_{i} =\phi'_{2} - (\alpha' -180^{0}) \; , \qquad
\alpha_{r}' = \alpha' - \phi'_{3} \; .
\nonumber
\end{eqnarray}

With the use of the Lorentz formulas  one can transform eq.(\ref{4.1a}) to the  moving reference
frame. Thereby it will be obtained a generalized form of the light reflection
on the  moving mirror.

Generally, one might expect  one of the  two following possibilities:

\vspace{5mm}

1) either equation $ \alpha'_{i} = \alpha_{r}'  $  turns out to be invariant under
the Lorentz transformation
\begin{eqnarray}
\alpha'_{i} = \alpha_{r}'    \qquad \Longrightarrow \qquad
\alpha_{i} = \alpha_{r} \; ,
\nonumber
\\
\alpha_{i} = 180^{0} - (\alpha - \phi_{2})  \; , \qquad
\alpha_{r} = (\alpha - \phi_{3} )\; .
\label{4.2b}
\end{eqnarray}

\noindent

2)  or after Lorentz transformation
eqs. (\ref{4.1a}) will take a modified form
\begin{eqnarray}
\alpha'_{i} = \alpha_{r}'    \qquad \Longrightarrow \qquad
\varphi(\alpha_{i},\alpha_{r}; V ) = 0 \; ,
\nonumber
\\
\alpha_{i} = 180^{0} - (\alpha - \phi_{2}) \; , \qquad
\alpha_{r} = ( \alpha - \phi_{3}) \; ,
\label{4.3a}
\end{eqnarray}

\noindent
and  one should expect that this new equation  $\varphi(\alpha_{i},\alpha_{r}; V ) = 0 $
is Lorentz invariant;, in other words  it is covariant under Lorentz transformation.

\vspace{5mm}
Now  we have to carry out the calculation needed.
Firstly, it will be better to re-write eqs.  (\ref{4.1a}) in the form
\begin{eqnarray}
\cos \alpha'_{i}= \cos \alpha_{r}'  \qquad \Longrightarrow \qquad
- \; \cos( \alpha' - \phi_{2}') = \cos (\alpha' - \phi'_{3} )\; ,
\nonumber
\\
\sin \alpha'_{i}= \sin \alpha_{r}'  \qquad
\Longrightarrow \qquad
\;\; \; \sin ( \alpha' - \phi_{2}') = \sin (\alpha' - \phi'_{3} ) \; \;.
\label{4.4b}
\end{eqnarray}

\noindent
Now, these equation are to be transformed to an unprimed quantities  with the use of the formulas
\begin{eqnarray}
\cos \alpha' = {\cos \alpha + V \over 1+ V\; \cos \alpha }\; , \qquad
\sin \alpha' = \sin \alpha\; {  \sqrt{1 -V^{2}} \over 1 + V\; \cos \alpha } \; ,
\nonumber
\\
\cos \phi'_{2} = {\cos \phi_{2} + V \over 1+ V\; \cos \phi_{2} }\; , \qquad
\sin \phi'_{2} = \sin \phi_{2}\; {  \sqrt{1 -V^{2}} \over 1 + V\; \cos \phi_{2} } \; ,
\nonumber
\\
\cos \phi'_{3} = {\cos \phi_{3} + V \over 1+ V\; \cos \phi_{3} }\; , \qquad
\sin \phi'_{3} = \sin \phi_{3}\; {  \sqrt{1 -V^{2}} \over 1 + V\; \cos \phi_{3} } \; .
\label{4.5}
\end{eqnarray}

\noindent The left-hand part of the  first equation in (\ref{4.4b}) becomes
\begin{eqnarray}
- \; \cos( \alpha' - \phi_{2}') =- (\cos \alpha ' \; \cos \phi_{2}' +
\sin \alpha' \; \sin \phi'_{2} ) =
\nonumber
\\
- \left [
\; {\cos \alpha + V \over  (1+ V\; \cos \alpha )}\;
{\cos \phi_{2} + V \over (1+ V\; \cos \phi_{2}) } +
\sin \alpha\; {  \sqrt{1 -V^{2}} \over (1 + V\; \cos \alpha) } \;
\sin \phi_{2}\; {  \sqrt{1 -V^{2}} \over (1 + V\; \cos \phi_{2} ) }  \; \right ] =
\nonumber
\\
= - \; \; {\cos (\alpha - \phi_{2}) +  V\; (\cos \alpha + \cos \phi_{2} ) +
 V^{2}  (1 - \sin \alpha \; \sin \phi_{2}) \over
 (1+ V\; \cos \alpha )\; (1+ V\; \cos \phi_{2}) } \; ,
\label{4.6a}
\end{eqnarray}

\noindent and the right-hand part  will look
\begin{eqnarray}
\cos( \alpha' - \phi_{3}')
= {\cos (\alpha - \phi_{3}) +  V\; (\cos \alpha + \cos \phi_{3} ) +
 V^{2}  (1 - \sin \alpha \; \sin \phi_{3}) \over
 (1+ V\; \cos \alpha )\; (1+ V\; \cos \phi_{3}) } \; .
\label{4.6b}
\end{eqnarray}

\noindent Thus, the first  equation in   (\ref{4.4b}) after Lorentz transformation is
as follows
\begin{eqnarray}
{- \cos (\alpha - \phi_{2}) -  V\; (\cos \alpha + \cos \phi_{2} ) -
 V^{2}  (1 - \sin \alpha \; \sin \phi_{2}) \over
  (1+ V\; \cos \phi_{2}) } =
\nonumber
\\
 =
{\cos (\alpha - \phi_{3}) +  V\; (\cos \alpha + \cos \phi_{3} ) +
 V^{2}  (1 - \sin \alpha \; \sin \phi_{3}) \over
  (1+ V\; \cos \phi_{3}) } \; ,
\label{4.6c}
\end{eqnarray}

\noindent or
\begin{eqnarray}
{ \cos \alpha_{i}  -  V\; (\cos \alpha + \cos \phi_{2} ) -
 V^{2}  (1 - \sin \alpha \; \sin \phi_{2}) \over
  (1+ V\; \cos \phi_{2}) } =
\nonumber
\\
 =
{\cos \alpha_{r}  +  V\; (\cos \alpha + \cos \phi_{3} ) +
 V^{2}  (1 - \sin \alpha \; \sin \phi_{3}) \over
  (1+ V\; \cos \phi_{3}) } \; .
\label{4.6d}
\end{eqnarray}

Analogously, consider other equation in (\ref{4.4b}).
The left-hand and right-hands parts become respectively
\begin{eqnarray}
\sin ( \alpha' - \phi_{2}') =
\sin \alpha' \; \cos \phi'_{2} -  \cos \alpha' \; \sin \phi'_{2} =
\nonumber
\\
=
\sin \alpha\; {  \sqrt{1 -V^{2}} \over  (1 + V\; \cos \alpha ) }\;\;
{\cos \phi_{2} + V \over (1+ V\; \cos \phi_{2} )} -
{\cos \alpha + V \over (1+ V\; \cos \alpha ) }\;\;
\sin \phi_{2}\; {  \sqrt{1 -V^{2}} \over  (1 + V\; \cos \phi_{2} )} =
\nonumber
\\
= {\sqrt{1 -V^{2}} \over (1 + V\; \cos \alpha )\; (1+ V\; \cos \phi_{2} )}\;
\left [ \; \sin (\alpha - \phi_{2}) + V\;  (\sin \alpha - \sin \phi_{2}) \;\right ] \; ,
\label{4.7a}
\end{eqnarray}

\noindent and
\begin{eqnarray}
\sin ( \alpha' - \phi_{3}') =
 {\sqrt{1 -V^{2}} \over (1 + V\; \cos \alpha )\; (1+ V\; \cos \phi_{3} )}\;
\left [ \; \sin (\alpha - \phi_{3}) + V\;  (\sin \alpha - \sin \phi_{3}) \;\right ] \; .
\label{4.7b}
\end{eqnarray}

\noindent So that equation in   (\ref{4.4b}) after Lorentz transformation looks as
\begin{eqnarray}
{ \sin (\alpha - \phi_{2}) + V\;  (\sin \alpha - \sin \phi_{2})
\over  (1+ V\; \cos \phi_{2} )}
=
{  \sin (\alpha - \phi_{3}) + V\;  (\sin \alpha - \sin \phi_{3})
\over  (1+ V\; \cos \phi_{3} )}\; \; ,
\label{4.7c}
\end{eqnarray}

\noindent  or
\begin{eqnarray}
{ \sin \alpha_{i}  + V\;  (\sin \alpha - \sin \phi_{2})
\over  (1+ V\; \cos \phi_{2} )}
=
{  \sin \alpha_{r}  + V\;  (\sin \alpha - \sin \phi_{3})
\over  (1+ V\; \cos \phi_{3} )}\; \; .
\label{4.7d}
\end{eqnarray}

\vspace{5mm}

Thus, the light reflection law in the reference frame $K'$:
\begin{eqnarray}
\cos \alpha'_{i}= \cos \alpha_{r}'  \qquad \Longrightarrow \qquad
- \; \cos( \alpha' - \phi_{2}') = \cos (\alpha' - \phi'_{3} )\; ,
\nonumber
\\
\sin \alpha'_{i}= \sin \alpha_{r}'  \qquad
\Longrightarrow \qquad
\;\; \; \sin ( \alpha' - \phi_{2}') = \sin (\alpha' - \phi'_{3} ) \; \;,
\nonumber
\end{eqnarray}

\noindent after translating to the  moving reference frame $K'$ take on the form
\begin{eqnarray}
{- \cos (\alpha - \phi_{2}) -  V\; (\cos \alpha + \cos \phi_{2} ) -
 V^{2}  (1 - \sin \alpha \; \sin \phi_{2}) \over
  (1+ V\; \cos \phi_{2}) } =
\nonumber
\\
 =
{\cos (\alpha - \phi_{3}) +  V\; (\cos \alpha + \cos \phi_{3} ) +
 V^{2}  (1 - \sin \alpha \; \sin \phi_{3}) \over
  (1+ V\; \cos \phi_{3}) } \; ,
\label{4.8a}
\end{eqnarray}
\begin{eqnarray}
{ \sin (\alpha - \phi_{2}) + V\;  (\sin \alpha - \sin \phi_{2})
\over  (1+ V\; \cos \phi_{2} )}
=
{  \sin (\alpha - \phi_{3}) + V\;  (\sin \alpha - \sin \phi_{3})
\over  (1+ V\; \cos \phi_{3} )}\; \; .
\label{4.8b}
\end{eqnarray}

The relations obtained seem rather cumbersome. Nevertheless they should be taken
seriously because they  display the property of Lorentz-invariance.
Such an additional test  consists in the following:
let the frame  $K$, in turn, be  moving  with the velocity $\tilde{V}$ with respect to  another
reference frame  $\tilde{K}$. We should expect invariance of equations (\ref{4.8a})-(\ref{4.8b}) under
corresponding Lorentz  transformations:
\begin{eqnarray}
\cos \alpha = {\cos \tilde{\alpha} + \tilde{V} \over 1+ \tilde{V}\; \cos \tilde{\alpha} }\; , \qquad
\sin \alpha = \sin \tilde{\alpha} \; {  \sqrt{1 -\tilde{V}^{2}}
\over 1 + \tilde{V}\; \cos \tilde{\alpha} } \; ,
\nonumber
\\
\cos \phi_{2} = {\cos \tilde{\phi}_{2} + \tilde{V} \over
 1+ \tilde{V}\; \cos \tilde{\phi}_{2} }\; , \qquad
\sin \phi_{2} = \sin \tilde{\phi}_{2}\;
{  \sqrt{1 -\tilde{V}^{2}} \over 1 + \tilde{V}\; \cos \tilde{\phi}_{2} } \; ,
\nonumber
\\
\cos \phi_{3} = {\cos \tilde{\phi}_{3} + \tilde{V} \over 1+
\tilde{V}\; \cos \tilde{\phi}_{3} }\; , \qquad \sin \phi_{3} =
\sin \tilde{\phi}_{3}\;
 {  \sqrt{1 -\tilde{V}^{2}} \over 1 + \tilde{V}\; \cos \tilde{\phi}_{3} } \; .
\label{4.9}
\end{eqnarray}

Before proceeding to calculation, one useful change in variables used should be done:
instead of the  light velocity
$\tilde{V}$ (and all other ones) it is better to employ a hyperbolic variable:
\begin{eqnarray}
{1 \over \sqrt{1 -\tilde{V}^{2} }} = \cosh \tilde{\beta}, \qquad
{\tilde{V} \over \sqrt{1 -\tilde{V}^{2} }} = \sinh \tilde{\beta}, \qquad
\tilde{V} = \tanh \tilde{\beta}  \;.
\nonumber
\end{eqnarray}

Eqs.(\ref{4.9})   take the form (the same applies  to  (\ref{4.5}))
\begin{eqnarray}
\cos \alpha = { \cosh \tilde{\beta}\; \cos \tilde{\alpha} + \sinh \tilde{\beta}
 \over \cosh \tilde{\nu} + \sinh \tilde{\beta} \; \cos \tilde{\alpha} }\; , \qquad
\sin \alpha =  {  \sin \tilde{\alpha}
\over
\cosh \tilde{\nu} + \sinh \tilde{\beta} \; \cos \tilde{\alpha} } \; ,
\nonumber
\\
\cos \phi_{2} = { \cosh \tilde{\beta}\; \cos \tilde{\phi}_{2} + \sinh \tilde{\beta}
 \over \cosh \tilde{\nu} + \sinh \tilde{\beta} \; \cos \tilde{\phi}_{2} }\; , \qquad
\sin \phi_{2} =  {  \sin \tilde{\phi}_{2}
\over
\cosh \tilde{\nu} + \sinh \tilde{\beta} \; \cos \tilde{\phi}_{2} } \; ,
\nonumber
\\
\cos \phi_{3} = { \cosh \tilde{\beta}\; \cos \tilde{\phi}_{3} + \sinh \tilde{\beta}
 \over \cosh \tilde{\nu} + \sinh \tilde{\beta} \; \cos \tilde{\phi}_{3} }\; , \qquad
\sin \phi_{2} =  {  \sin \tilde{\phi}_{3} \over \cosh
\tilde{\nu} + \sinh \tilde{\beta} \; \cos \tilde{\phi}_{3} } \; .
\label{4.11}
\end{eqnarray}

Firstly let us consider  behavior of the more simple  formula  (\ref{4.8b}) under
Lorentz transformation  (\ref{4.11}). For the first factor we have
\begin{eqnarray}
{1 \over 1 + V\; \cos \phi_{2} } =
{1 \over 1 + \tanh \beta \;
( \cosh \tilde{\beta}\; \cos \tilde{\phi}_{2} + \sinh \tilde{\beta}) \; /\;
( \cosh \tilde{\nu} + \sinh \tilde{\beta} \; \cos \tilde{\phi}_{2})  } =
\nonumber
\\
{ 1 + \tanh \tilde{\beta} \; \cos \tilde{\phi}_{2} \over  (\tanh \tilde{\beta} +
\tanh \beta ) \cos \tilde{\phi}_{2} + (1+  \tanh \beta \; \tanh \tilde{\beta} )} \; ;
\nonumber
\end{eqnarray}

\noindent from where  with the use of
\begin{eqnarray}
1+ \tanh \beta \; \tanh \tilde{\beta}  = { \tanh \tilde{\beta} +
\tanh \beta  \over \tanh (\beta + \tilde{\beta}) } \; ,
\nonumber
\end{eqnarray}

\noindent get to
\begin{eqnarray}
{1 \over 1 + V\; \cos \phi_{2} } =
{1 + \tanh \tilde{\beta} \; \cos \tilde{\phi}_{2} \over 1 +
\tanh \beta \; \tanh \tilde{\beta} }\;\; {1 \over 1+
\tanh (\beta + \tilde{\beta})\; \cos \tilde{\phi}_{2} \;  }=
\nonumber
\\
=
{\cosh \beta \; ( \cosh \tilde{\beta} + \sinh \tilde{\beta} \; \cos \tilde{\phi}_{2} ) \over
\sinh \beta \; \sinh \tilde{\beta} + \cosh \beta \; \cosh \tilde{\beta} }\;\; {1 \over 1+
 \tanh (\beta + \tilde{\beta})\; \cos \tilde{\phi}_{2} \;  } \; .
\label{4.12a}
\end{eqnarray}

\noindent
Now consider  the term
\begin{eqnarray}
\sin (\alpha - \phi_{2}) = \sin \alpha \; \cos \phi_{2}  -  \cos \alpha\;
\sin \phi_{2} =
\nonumber
\\
=
{  \sin \tilde{\alpha}
\over
(\cosh \tilde{\beta} + \sinh \tilde{\beta} \; \cos \tilde{\alpha} )} \; \;
{ \cosh \tilde{\beta}\; \cos \tilde{\phi}_{2} + \sinh \tilde{\beta}
 \over (\cosh \tilde{\beta} + \sinh \tilde{\beta} \; \cos \tilde{\phi}_{2} )}\;\; -
\nonumber
\\
 - \;\;
{ \cosh \tilde{\beta}\; \cos \tilde{\alpha} + \sinh \tilde{\beta}
 \over (\cosh \tilde{\beta} + \sinh \tilde{\beta} \; \cos \tilde{\alpha}) }\;\;
{  \sin \tilde{\phi}_{2}
\over
(\cosh \tilde{\beta} + \sinh \tilde{\beta} \; \cos \tilde{\phi}_{2}) } =
\nonumber
\\
=
{ \cosh \tilde{\beta} \; \sin (\tilde{\alpha} - \tilde{\phi}_{2}) + \sinh \tilde{\beta} \;\;
(\sin \tilde{\alpha} - \sin \tilde{\phi}_{2} )
 \over
(\cosh \tilde{\beta} + \sinh \tilde{\beta} \; \cos \tilde{\alpha}) \;
(\cosh \tilde{\beta} + \sinh \tilde{\beta} \; \cos \tilde{\phi}_{2})}
\;\;.
\label{4.12b}
\end{eqnarray}

\noindent And finally, the term
\begin{eqnarray}
V\;  (\sin \alpha - \sin \phi_{2})
=
\nonumber
\\
= {\sinh \beta \over  \cosh \beta} \;\; \left [ \;
{  \sin \tilde{\alpha}
\over
\cosh \tilde{\nu} + \sinh \tilde{\beta} \; \cos \tilde{\alpha} }-
{  \sin \tilde{\phi}_{2}
\over
\cosh \tilde{\nu} + \sinh \tilde{\beta} \; \cos \tilde{\phi}_{2} }
 \; \right ] =
\nonumber
\\
=
{\sinh \beta \over  \cosh \beta} \;\;
{ \sinh \tilde{\beta} \; \sin (\tilde{\alpha} -\tilde{\phi}_{2})
+ \cosh \tilde{\beta}\; (\sin \tilde{\alpha} - \sin \tilde{\phi}_{2})
\over
(\cosh \tilde{\beta} + \sinh \tilde{\beta} \; \cos \tilde{\alpha}) \;
(\cosh \tilde{\beta} + \sinh \tilde{\beta} \; \cos \tilde{\phi}_{2})}\; .
\label{4.12c}
\end{eqnarray}

\noindent Combining (\ref{4.12b}) and (\ref{4.12c}),  we have  ( see. (\ref{4.8b}))
\begin{eqnarray}
\sin (\alpha - \phi_{2}) +
V  (\sin \alpha - \sin \phi_{2})= {1 \over
(\cosh \tilde{\beta} + \sinh \tilde{\beta}  \cos \tilde{\alpha})
(\cosh \tilde{\beta} + \sinh \tilde{\beta}  \cos \tilde{\phi}_{2})}\times
\nonumber
\\
\times \left [
\sin (\alpha - \phi_{2})\left (   \cosh \tilde{\beta}
+ {\sinh \beta \over  \cosh \beta}  \sinh \tilde{\beta} \right )  +
(\sin \tilde{\alpha} - \sin \tilde{\phi}_{2} )
\left ( \sinh \tilde{\beta}
+ {\sinh \beta \over  \cosh \beta}  \cosh \tilde{\beta} \right )
 \right ] =
\nonumber
\\
=
{1 \over
(\cosh \tilde{\beta} + \sinh \tilde{\beta}  \cos \tilde{\alpha})
(\cosh \tilde{\beta} + \sinh \tilde{\beta}  \cos \tilde{\phi}_{2})}\times \hspace{25mm}
\nonumber
\\
\times \left [
\sin (\alpha - \phi_{2})  (  { \cosh \beta \cosh \tilde{\beta}
+ \sinh \beta   \sinh \tilde{\beta}  \over  \cosh \beta }   +
  (\sin \tilde{\alpha} - \sin \tilde{\phi}_{2} ) \;\; {
\cosh \beta   \sinh \tilde{\beta} + \sinh \beta  \cosh \tilde{\beta}
\over \cosh \beta  }  \right ] \; .
\nonumber
\end{eqnarray}

\noindent Now  we arrive at
\begin{eqnarray}
{ \sin (\alpha - \phi_{2}) + V\;  (\sin \alpha - \sin \phi_{2})
\over  (1+ V\; \cos \phi_{2} )}
=
{1 \over 1+
 \tanh (\beta + \tilde{\beta})\; \cos \tilde{\phi}_{2} \;  } \times
\nonumber
\\
\times
 \left [  \sin (\tilde{\alpha} - \tilde{\phi}_{2})
+\;
(\sin \tilde{\alpha} - \sin \tilde{\phi}_{2} ) \;\;
{ \cosh \beta  \; \sinh \tilde{\beta} + \sinh \beta \; \cosh \tilde{\beta} \over
\sinh \beta \; \sinh \tilde{\beta} + \cosh \beta \; \cosh \tilde{\beta} }
  \right ]\; .
\nonumber
\end{eqnarray}

\noindent From this, using the hyperbolic function identity
\begin{eqnarray}
 \tanh (\beta + \tilde{\beta}) = { \sinh (\beta + \tilde{\beta})  \over
\cosh (\beta + \tilde{\beta})}  =
 { \cosh \beta  \; \sinh \tilde{\beta} + \sinh \beta \; \cosh \tilde{\beta} \over
\sinh \beta \; \sinh \tilde{\beta} + \cosh \beta \; \cosh \tilde{\beta} }
\; ,
\nonumber
\end{eqnarray}

\noindent for the left-hand part of   (\ref{4.8b}) we arrive at the following
\begin{eqnarray}
{ \sin (\alpha - \phi_{2}) + \tanh \beta \;  (\sin \alpha - \sin \phi_{2})
\over  (1+  \tanh \beta \; \cos \phi_{2} )}
=
\nonumber
\\
=
{ \sin (\tilde{\alpha} - \tilde{\phi}_{2})
+\;
\tanh (\beta + \tilde{\beta})\;  \;(\sin \tilde{\alpha} - \sin \tilde{\phi}_{2} ) \;\;
\over  1 +\tanh (\beta + \tilde{\beta})\; \cos \tilde{\phi}_{2} } \; .
\label{4.13a}
\end{eqnarray}

For the right-hand part of (\ref{4.8b}) , with no additional calculation we  will obtain
\begin{eqnarray}
{ \sin (\alpha - \phi_{3}) + \tanh \beta \;  (\sin \alpha - \sin \phi_{3})
\over  (1+  \tanh \beta \; \cos \phi_{3} )}
=
\nonumber
\\
=
{ \sin (\tilde{\alpha} - \tilde{\phi}_{3})
+\;
\tanh (\beta + \tilde{\beta})\;  \;(\sin \tilde{\alpha} - \sin \tilde{\phi}_{3} ) \;\;
\over  1 +\tanh (\beta + \tilde{\beta})\; \cos \tilde{\phi}_{3} } \; .
\label{4.13b}
\end{eqnarray}

Therefore,
equation  (\ref{4.8b}) after the Lorentz transformation to the reference frame  $\tilde{K}$
displays a  required invariant form
\begin{eqnarray}
{ \sin (\tilde{\alpha} - \tilde{\phi}_{2})
+\;
\tanh (\beta + \tilde{\beta})\;  \;(\sin \tilde{\alpha} - \sin \tilde{\phi}_{2} ) \;\;
\over  1 +\tanh (\beta + \tilde{\beta})\; \cos \tilde{\phi}_{2} }=
\nonumber
\\
=
{ \sin (\tilde{\alpha} - \tilde{\phi}_{3})
+\;
\tanh (\beta + \tilde{\beta})\;  \;(\sin \tilde{\alpha} - \sin \tilde{\phi}_{3} ) \;\;
\over  1 +\tanh (\beta + \tilde{\beta})\; \cos \tilde{\phi}_{3} } \; .
\label{4.14}
\end{eqnarray}

Remember the relativistic velocity addition rule in terms of hyperbolic variables:
\begin{eqnarray}
{V + \tilde{V} \over 1+ V\; \tilde{V} }\; \qquad \Longrightarrow \qquad
\tanh (\beta + \tilde{\beta})  = { \tanh \beta  + \tanh \tilde{\beta} \over
1+ \tanh \beta \; \tanh \tilde{\beta}  } \; .
\nonumber
\end{eqnarray}

\vspace{5mm}

CONCLUSION

\begin{quotation}

The light reflection law on moving mirror (reference frame  $K$)
\begin{eqnarray}
{ \sin (\alpha - \phi_{2}) + V\;  (\sin \alpha - \sin \phi_{2})
\over  (1+ V\; \cos \phi_{2} )}
=
{  \sin (\alpha - \phi_{3}) + V\;  (\sin \alpha - \sin \phi_{3})
\over  (1+ V\; \cos \phi_{3} )}\; \; ,
\nonumber
\end{eqnarray}

\noindent or
\begin{eqnarray}
{ \sin \alpha_{i}  + V\;  (\sin \alpha - \sin \phi_{2})
\over  (1+ V\; \cos \phi_{2} )}
=
{  \sin \alpha_{r}  + V\;  (\sin \alpha - \sin \phi_{3})
\over  (1+ V\; \cos \phi_{3} )}\; \; ,
\nonumber
\end{eqnarray}

\noindent  is invariant under Lorentz transformations.
We will not  verify the  relativistic invariance of the second condition
(\ref{4.8a});  evidently  it is the case.

\end{quotation}

\section{On describing the form of  a moving  mirror}

\hspace{5mm}
Section concerns a
 general problem, described in [31] as follows:

 {\em
As is well known, the notion of a rigid body, which proves so useful in Newtonian
mechanics, is incompatible with the existence of a universal finite upper bound for
all signal velocities [Laue - 1911 ]. As a result, the notion of a perfectly rigid body does not
exist within the framework of SR. However, the notion of a rigid motion does exist.
Intuitively speaking, a body moves rigidly if, locally, the relative spatial distances of
its material constituents are unchanging} .

Let us turn to the geometrical form of the  mirror in unmoving reference frame  $K'$;
 it is described by two  parametric equations
\begin{eqnarray}
x'  = L' + l' \; \cos \gamma ' \; ,
\qquad y'  =  l' \; \sin \gamma ' \; .
\label{5.1a}
\end{eqnarray}

\noindent After evident manipulation we get to
\begin{eqnarray}
{x' -   L' \over  \cos \gamma '} =  l' \;: \qquad \Longrightarrow
\qquad y'  =    {x' -   L' \over  \cos \gamma '} \; \sin \gamma ' \; ,
\nonumber
\\
y' = - L \; \tan \; \gamma ' +  \tan \; \gamma ' \; x' \; ,
\nonumber
\end{eqnarray}

\noindent or
\begin{eqnarray}
y' = b + k \; x' \;
\label{5.1c}
\end{eqnarray}

\noindent where  $b =-L\;\tan \; \gamma '$ and   $k= \tan \; \gamma '$are  fixed parameters
defining the straight line  -- contour  of the  mirror.

For another observer $K$, to the mirror
there corresponds  a set of space-time events of the type $2'$, each of those  is an arrival
of a light signal emitted  from the space-time  point $1': - (0;0,0)$
toward  the mirror with different angles
 $\phi_{2}'$. Evidently, all such events will take place on the surface of the  mirror but at different
 times:
\begin{eqnarray}
t_{2}' = L \; {\sin \gamma ' \over \sin (\gamma' - \phi'_{2}) } \; ,
\nonumber
\\
x_{2}' = \cos \phi_{2}' \;\; t_{2}' \; ,\qquad
y_{2}' = \sin \phi_{2}' \;\; t_{2}' \; .
\label{5.2}
\end{eqnarray}

Space-time coordinates of these events in the reference fame $K$ may be found through the  Lorentz
formulas:
\begin{eqnarray}
t'_{2} = {t_{2} + V x_{2}  \over \sqrt{1 - V^{2} } } =
{x_{2} / \cos \phi _{2} + V x_{2}  \over \sqrt{1 - V^{2} } }  =
{1 / \cos \phi _{2} + V   \over \sqrt{1 - V^{2} } }\; x_{2} \;  ,
\nonumber
\\
x'_{2} = {x_{2} + V t_{2} \over \sqrt{1 - V^{2}}} =
{x_{2} + V x_{2} / \cos \phi_{2}  \over \sqrt{1 - V^{2}}} =
{1 + V  / \cos \phi_{2}  \over \sqrt{1 - V^{2}}}\; x_{2}   \; ,
\nonumber
\\
y_{2}' = y_{2}   \; .
\nonumber
\end{eqnarray}

Taking them  into (\ref{5.1c}) we will arrive  at the equation
\begin{eqnarray}
y_{2} = b + k \;  {1 + V  / \cos \phi_{2}  \over \sqrt{1 - V^{2}}}\; \;x_{2} \; .
\label{5.3b}
\end{eqnarray}

With the  help of
\begin{eqnarray}
x_{2} = \cos \phi_{2} \; t_{2} \; , \qquad y_{2} = \sin \phi_{2} \; t_{2}:
\qquad \tan \phi_{2} = {y_{2} \over x_{2}} \;, \qquad \Longrightarrow
\nonumber
\\
{1 \over \cos \phi_{2} } = \sqrt{1+ \tan^{2} \phi_{2} } = \sqrt{1 + {y_{2}^{2}
\over x_{2}^{2}}}
\nonumber
\end{eqnarray}

\noindent
eq.  (\ref{5.3b}) may be  rewritten as follows:
\begin{eqnarray}
y _{2}= b + k \; \;  {1 + V \sqrt{1 +  y_{2}^{2} /
 x_{2}^{2}}   \over \sqrt{1 - V^{2}}}\; \; x_{2}
\nonumber
\end{eqnarray}

\noindent
or (for simplicity the index 2 at the coordinates will be omitted)
\begin{eqnarray}
y = b + k \;   ( \;
\cosh \beta  + \sinh \beta \; \sqrt{1 +  {y^{2} \over  x^{2}}} \;\;    ) \; x \; .
\label{5.4b}
\end{eqnarray}

\begin{quotation}

Relationship   (\ref{5.4b}) should be  considered as an equation
describing the surface of the  moving mirror, it is not evidently a straight line equation.
It should be emphasized again that
a  definite and  practically realizable procedure with the use of light signals in the
reference frame $K$  underlies this  equation.

\end{quotation}

\section{On geometrical form of a  moving mirror }

\hspace{5mm}
What  curve is given by the equation  (\ref{5.4b}):
\begin{eqnarray}
y = b + k \;  [ \;
\cosh \beta  + \sinh \beta \; \sqrt{1 +  {y^{2} \over x^{2}}} \;\;    ] \; x \; .
\nonumber
\end{eqnarray}

\noindent
After simple rewriting from  (\ref{5.4b}) it follows
\begin{eqnarray}
(y - b - k \cosh \beta \; x)^{2} =  k^{2} \; \sinh^{2}\beta \; (x^{2} + y^{2}) \; ,
\nonumber
\end{eqnarray}

\noindent and further
\begin{eqnarray}k^{2} \; x^{2} - 2k\cosh \beta \; xy + (1 - k^{2}\sinh^{2} \beta) \; y^{2} +
\nonumber
\\
+ 2bk\; \cosh \beta \; x - 2b\; y + b^{2} = 0 \; .
\label{6.1c}
\end{eqnarray}

This is of second order curve. In order to establish its explicit geometrical form
eq. (\ref{6.1c})  should be translated to the  canonical form.
We will proceed  in accordance with standard procedures.
As a first step let us find certain  $(x,y)$-rotation that
will eliminate a coefficient at the cross term  $xy$ from the curve  equation.
To this end, instead of $(x,y)$  one should introduce new  (rotated) variables
 $(X',Y')$
\begin{eqnarray}
\left | \begin{array}{c}
x \\ y
\end{array} \right | =
\left | \begin{array}{cc}
 \cos \phi & - \sin \phi \\
 \sin \phi & \cos \phi
 \end{array} \right |
\left | \begin{array}{c}
X' \\ Y'
\end{array} \right | \; .
\label{6.2}
\end{eqnarray}

\noindent Then eq.  (\ref{6.1c}) will take the form
\begin{eqnarray}
k^{2} \; (\cos \phi\; X' - \sin \phi \;Y') ^{2}
 +  (1 - k^{2}\sinh^{2} \beta) (\sin \phi\; X' +\cos \phi \; Y')^{2}
-
\nonumber
\\
 - 2k\cosh \beta \; (\cos \phi\; X' - \sin \phi\;  Y') (\sin \phi\; X' + \cos \phi \; Y')+
\nonumber
\\
+2bk\; \cosh \beta \; (\cos \phi\; X' - \sin \phi \;Y') -
\nonumber
\\
-  2b\;
(\sin \phi\; X' + \cos \phi\;  Y') + b^{2} =0 \;
\nonumber
\end{eqnarray}

\noindent
or
\begin{eqnarray}
X^{'2}  [ \; k^{2} \cos^{2} \phi  + ( 1 - k^{2}  \sinh^{2} \beta ) \sin^{2} \phi -
2k\cosh \beta  \cos \phi  \sin \phi \; ]  +
\nonumber
\\
+   Y^{'2}  [ \;k^{2} \sin^{2} \phi  +
( 1 - k^{2}  \sinh^{2} \beta )  \cos^{2} \phi +
2k\cosh \beta  \cos \phi  \sin \phi \; ] \; +
\nonumber
\\
+  X'Y'
[ -2 \sin \phi  \cos \phi   k^{2} + (1 - k^{2}  \sinh^{2} \beta )
2 \sin \phi \cos \phi -
2k  \cosh \beta  (\cos^{2} \phi - \sin^{2} \phi )  ]  +
\nonumber
\\
+  X'  [  2bk   \cosh \beta  \cos \phi - 2b  \sin \phi  ] + Y'
[
2bk  \cosh \beta  \sin \phi -2b  \cos \phi  ]
+ b^{2} =0 \; .
\nonumber
\end{eqnarray}
\begin{eqnarray}
\label{6.3b}
\end{eqnarray}

\noindent
The rotation angle  $\phi = \phi_{0}$ should be determined from special requirement
that the coefficient at  $X'Y'$ be zero:
\begin{eqnarray}
\sin 2\phi_{0} \; [\;  (1 - k^{2} \sinh^{2} \beta ) \; - \; k^{2}\; ] -
2k\; \cosh \beta \; \cos 2\phi_{0} = 0 \; ,
\nonumber
\end{eqnarray}

\noindent from where it follows
\begin{eqnarray}
\tan 2\phi_{0} = { 2k\; \cosh \beta \over 1 - k^{2} \; \cosh^{2} \beta } \; .
\label{6.4b}
\end{eqnarray}

\noindent
With the aid of the known relation
\begin{eqnarray}
\tan 2\phi_{0} = { 2 \; \tan \phi_{0} \over 1 - \tan^{2} \phi_{0} }
\nonumber
\end{eqnarray}

\noindent  we arrive at a simple representation for  $\tan \phi_{0}$:
\begin{eqnarray}
\tan \phi_{0} = k \; \cosh \beta\; , \qquad \mbox{or }\;\;
\tan \phi_{0} =  \tan \gamma' \;  \cosh \beta\;   \; ,
\label{6.4c}
\end{eqnarray}

\noindent and also expressions for
\begin{eqnarray}
\sin \phi_{0} = {\tan \phi_{0} \over \sqrt{1 + \tan^{2} \phi_{0}}} =
{
k\cosh \beta \over \sqrt{1 + k^{2} \cosh^{2} \beta }} \; ,
\nonumber
\\
\cos \phi_{0} = {1 \over \sqrt{1 + \tan^{2} \phi_{0}}} =
{ 1 \over \sqrt{1 + k^{2} \cosh^{2} \beta }} \; .
\nonumber
\end{eqnarray}

\noindent
Now, eq.  (\ref{6.3b}) becomes
\begin{eqnarray}
X^{'2} \; { k^{2} + (1 - k^{2} \sinh^{2} \beta ) \;
k^{2} \cosh^{2} \beta - 2k \cosh  \beta \; k \cosh \beta \over
1 + k^{2} \cosh^{2} \beta } +
\nonumber
\\
+ Y^{'2}\; { k^{2} \; k^{2} \cosh^{2} \beta + (1 - k^{2} \sinh^{2} \beta ) +
2k\cosh \beta \; k \cosh \beta \over 1 + k^{2} \cosh^{2} \beta } +
\nonumber
\\
+ {X' \over \sqrt{1 + k^{2}\cosh^{2} \beta }}
 \; ( 2bk \; \cosh \beta - 2bk\; \cosh \beta )  +
\nonumber
\\
+
{Y' \over \sqrt{1 + k^{2}\cosh^{2} \beta }}
 \; ( -2bk \; \cosh \beta  \; k \cosh \beta - 2b ) \; + b^{2} = 0 \; .
\label{6.5a}
\end{eqnarray}

\noindent From this, after simple computation
we get to the equation of second order:
\begin{eqnarray}
- X^{'2} \; k^{2} \sinh^{2} \beta + Y^{'2} \; (1 + k^{2} ) -
\nonumber
\\
-  2b\; \sqrt{1 + k^{2} \cosh^{2} \beta } \; Y'  + b^{2} = 0 \; ;
\label{6.5b}
\end{eqnarray}

\noindent
which is an equation of a hyperbola.  Its canonical form will   be achieved
by a definite displacement along the axis $Y'$:
\begin{eqnarray}
- X^{'2}  k^{2} \sinh^{2} \beta + (1 + k^{2})
 [   Y' - b  { \sqrt{1 + k^{2}  \cosh^{2} \beta}  \over 1 + k^{2} }\;   ]^{2} -
b^{2} { 1 + k^{2} \cosh^{2} \beta  \over 1 + k^{2}  } +  b^{2} =0 \; ;
\nonumber
\end{eqnarray}

\noindent that is
\begin{eqnarray}
- X^{'2}  k^{2} \sinh^{2} \beta + (1 + k^{2})
[   Y' - b  { \sqrt{1 + k^{2}  \cosh^{2} \beta}  \over 1 + k^{2} } \;   ]^{2}
=  { b^{2} k^{2} \sinh^{2} \beta  \over 1 + k^{2} } \; .
\label{6.5c}
\end{eqnarray}

\noindent
So, the  canonical equation for a geometrical form of  the moving mirror in the frame $K$
looks
\begin{eqnarray}
- {X^{'2} \over  A^{2}  } +
{ ( Y' - C ) ^{2}   \over B^{2}}
    = +1 \; , \hspace{30mm}
\nonumber
\\
A^{2} = {b^{2} \over  1+k^{2} } \; , \qquad
B^{2} = { b^{2} k^{2} \sinh^{2} \beta \over  (1+k^{2})^{2} } \; ,
\qquad C = { b\;\sqrt{1 + k^{2}  \cosh^{2} \beta}  \over   (1 + k^{2}) }\; .
\label{6.6b}
\end{eqnarray}

Let us trace again  establishing  canonical equation: it consists of two
steps: rotation and  displacement:
\begin{eqnarray}
x = \cos \phi_{0} \; X' - \sin \phi_{0} \; Y' \; ,\qquad
y = \sin \phi_{0} \; X' + \cos \phi_{0} \; Y' \; ;
\nonumber
\\
X' = X'' \; , \qquad Y''= Y' - C \; .
\nonumber
\end{eqnarray}

\noindent After performing these two transforms the equation of the moving  mirror
will represent the hyperbola:
\begin{eqnarray}
- {X^{''2} \over  A^{2}  } +
{  Y^{''2}   \over B^{2}}
    = +1 \;;
\nonumber
\end{eqnarray}

The above rotation  (6.7a) may be  clarified by  Figure 3:

\unitlength=0.8mm
\begin{picture}(100,50)(-55,0)
\special{em:linewidth 0.4pt} \linethickness{0.6pt}

\put(0,0){\vector(+1,0){45}}  \put(+50,-5){$x$}
\put(0,0){\vector(0,+1){45}}  \put(+5,+42){$y$}

\put(0,0){\vector(+3,+1){50}}  \put(+52,+12){$X'$}
\put(0,0){\vector(-1,+3){15}}  \put(-23,+40){$Y'$}

\put(+15,+1){$\phi_{0}$}

\end{picture}

\vspace{5mm}

\begin{center}
{\bf Fig. 3 Additional rotation}
\end{center}

\noindent
To the total transform    there corresponds to Figure 4

\vspace{15mm}

\unitlength=0.9mm
\begin{picture}(100,50)(-55,0)
\special{em:linewidth 0.4pt} \linethickness{0.6pt}

\put(0,0){\vector(+1,0){45}}  \put(+50,-5){$x$}
\put(0,0){\vector(0,+1){45}}  \put(+5,+42){$y$}

\put(-5,+15){\circle*{2}}
\put(-5,+15){\vector(+3,+1){50}}  \put(+48,+25){$X''$}
\put(-5,+15){\vector(-1,+3){15}}  \put(-28,+54){$Y''$}

\put(+10,+16){$\phi_{0}$}
\put(-5,+13){$- - - - - - - - - - - $}
\put(-5,+15){\line(+1,-3){2}}
\put(+5,-15){\line(-1,+3){3}}
\put(0,0){\line(-1,+3){2}}

\put(-3,+9){\circle*{1}}
\put(-1,+8.7){\circle*{1}}
\put(+1,+8.3){\circle*{1}}
\put(+3,+7.5){\circle*{1}}
\put(+5,+7){\circle*{1}}
\put(+7,+6.2){\circle*{1}}
\put(+9,+5){\circle*{1}}
\put(+11,+4.2){\circle*{1}}
\put(+13,+3.3){\circle*{1}}
\put(+15,+2){\circle*{1}}  \put(0,0){\circle*{2}}
\put(+17,+1){\circle*{1}}
\put(+19,0){\circle*{1}}

\put(+30,+5){$
- {X^{''2} \over  A^{2}  } +
{  Y^{''2}   \over B^{2}}
    = +1 $}

\end{picture}

\vspace{15mm}

\begin{center}
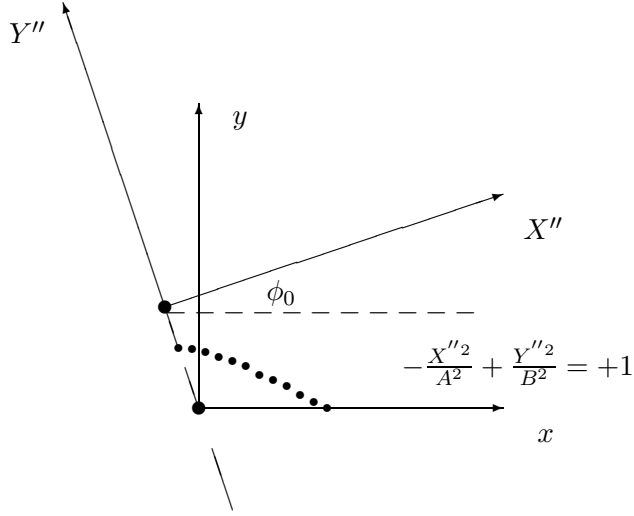
{\bf Fig. 4 The  mirror's form is a hyperbola in  ($X'',Y'')$ frame}
\end{center}

\vspace{5mm}
OVERALL RESULT:
\vspace{5mm}

{\em
1) the moving mirror for the observer  $K$ looks as
\begin{eqnarray}
k^{2} \; x^{2} - 2k\cosh \beta \; xy + (1 - k^{2}\sinh^{2} \beta) \; y^{2} +
\nonumber
\\
+ 2bk\; \cosh \beta \; x - 2b\; y + b^{2} = 0 \; ;
\nonumber
\end{eqnarray}

2)
in the reference frame  $K$  one needs to take the new coordinates
$X'',Y''$
\begin{eqnarray}
x =  \cos \phi_{0} \; X'' - \sin \phi_{0} \; (Y'' + C) =
\nonumber
\\
=
- \sin \phi_{0} \; C  + (\cos \phi_{0} \; X'' - \sin \phi_{0} \; Y'' )\;,
\nonumber
\\
y =  \sin \phi_{0} \; X'' + \cos \phi_{0} \; (Y'' + Y'_{0}) =
\nonumber
\\
=
+ \cos  \phi_{0} \; C  + (\sin \phi_{0} \; X'' + \cos  \phi_{0} \; Y'') \;,
\nonumber
\\
\tan \phi_{0} = k \; \cosh \beta\; , \qquad
C = { b\;\sqrt{1 + k^{2}  \cosh^{2} \beta}  \over   (1 + k^{2}) }\; ;
\nonumber
\end{eqnarray}

3)  then eq. (6.8a)  will become a canonical form of a hyperbola
\begin{eqnarray}
- {X^{'2} \over  A^{2}  } +
{  Y^{''2}   \over B^{2}}
    = +1 \;;
\qquad
A^{2} = {b^{2} \over  1+k^{2} } \; , \qquad
B^{2} = { b^{2} k^{2} \sinh^{2} \beta \over  (1+k^{2})^{2} } \; .
\nonumber
\end{eqnarray}

}

\section{ Generalization and simplification, vector form }

\hspace{5mm}
Now we are to extend the results above to a  more general vector
form. Let us start with the conventional designation.
The  incident and reflected  light rays  in the rest
reference frame $K'$ are described by respective vectors of unit length
 ${\bf a}'$ and  ${\bf b}'$:
\begin{eqnarray}
{\bf a} '= {{\bf W}'_{in} \over c} \; , \;\;  {\bf a}^{'2} = 1 \;
, \qquad {\bf b} '= { {\bf W}'_{out} \over c} \; , \; \; {\bf
b}^{'2} = 1 \; ;
\label{7.1a}
\end{eqnarray}

\noindent  With the  surface  perpendicular  there can be associated the unit
vector ${\bf n}'$  (normal  light ray -- see Fig. 5)
\begin{eqnarray}
{\bf n} '={  {\bf W}'_{norm} \over c} \; , \qquad {\bf n}^{'2} = 1
\; .
\label{7.1b}
\end{eqnarray}

\noindent
The designation introduced  can be clarified by  Fig. 5

\vspace{10mm}

\unitlength=0.6mm
\begin{picture}(100,50)(-115,0)
\special{em:linewidth 0.4pt} \linethickness{0.6pt}

\put(0,0){\circle*{3}}

\put(-30,+40){\vector(+3,-4){29}}  \put(-35,+45){${\bf a}'$}
\put(0.5,0.5){\vector(+3,+4){30}}  \put(-5,+50){${\bf n}'$}
\put(0,0.5){\vector(0,+1){50}}     \put(+35,+45){${\bf b}'$}

\put(-40,0){\line(+1,0){80}} \put(-40,-0.3){\line(+1,0){80}}
\put(-40,-0.6){\line(+1,0){80}}

\end{picture}

\vspace{10mm}

\begin{center}
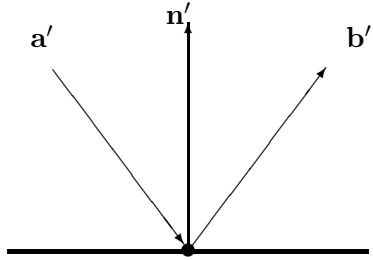
{\bf Fig 5.  Reflection  in the rest reference frame $K'$}
\end{center}

The reflection  law in the $K'$ can be mathematically
described by means of the following vector formula
(all three vectors have  a unit length)
\begin{eqnarray}
{\bf a} '\times {\bf n}' =  {\bf b}' \times {\bf n}' \; .
\label{7.2}
\end{eqnarray}

This  relationship includes both

\vspace{5mm}

 1)
  $\alpha_{i}' = \alpha_{r}' = \alpha' $ and

  2) all three vectors belong to the same plane.

\vspace{5mm}

In addition the following relation is in effect:
\begin{eqnarray}
\cos \alpha_{i}' = \cos \alpha_{r}' \qquad \Longrightarrow \qquad
{\bf a} ' \;  {\bf n}' +  {\bf b}' \; {\bf n}'  = 0 \; .
\label{7.3}
\end{eqnarray}

To describe the same reflection  process in the  moving reference frame $K$ means
that the relation (\ref{7.2})  is to   be transformed to this new frame through the Lorentz  formula.
The Lorentz  formulas we need are
\begin{eqnarray}
a_{1} ' = { a_{1} + V \over 1 + a_{1} V } \; , \qquad a_{2}' =
{\sqrt{1-V^{2}} \over 1 + a_{1} V } \; a_{2} \; , \qquad a_{3}' =
{\sqrt{1-V^{2}} \over 1 + a_{1} V } \; a_{3} \; ,
\nonumber
\\
b_{1} ' = { b_{1} + V \over 1 + b_{1} V } \; , \qquad b_{2}' =
{\sqrt{1-V^{2}} \over 1 + b_{1} V } \; b_{2} \; , \qquad b_{3}' =
{\sqrt{1-V^{2}} \over 1 + b_{1} V } \; b_{3} \; ,
\nonumber
\\
n_{1} ' = { n_{1} + V \over 1 + n_{1} V } \; , \qquad n_{2}' =
{\sqrt{1-V^{2}} \over 1 + n_{1} V } \; n_{2} \; , \qquad n_{3}' =
{\sqrt{1-V^{2}} \over 1 + n_{1} V } \; n_{3} \; .
\label{7.4}
\end{eqnarray}

The most significant property of all the three light velocity vectors consists in the following:
their length is invariant under the Lorentz transformation\footnote{This statement is true only if
all light rays are propagated in the vacuum.}.
Indeed, for example let it be
$(n_{1}^{'2} +  n_{2}^{'2} +  n_{3}^{'2})=1$, then
\begin{eqnarray}
n_{1}^{2} + n_{2}^{2} + n_{3}^{2} = {( n'_{1} - V )^{2}\over (1 -
n'_{1} V )^{2} })  + {1-V^{2} \over (1 - nэ_{1} V )^{2} } \;
n_{2}^{'2} + {1-V^{2} \over (1 - nэ_{1} V )^{2} } \; n_{3}^{'2}
\nonumber
\\
={1 \over (1 - nэ_{1} V )^{2} } \;
 \; \left [
 (n_{1}^{'2} +  n_{2}^{'2} +  n_{3}^{'2})  - 2 n_{1}' V + V^{2} \;  (1 -  n_{2}^{'2} -
  n_{2}^{'2} \; \right ] = 1 \; .
\nonumber
\end{eqnarray}

The vector relation (\ref{7.2}) in coordinate form looks as three ones
\begin{eqnarray}
1: \qquad \qquad \qquad  a'_{2} n'_{3} -  a'_{3} n'_{2} =  b'_{2} n'_{3} -
b'_{3} n'_{2} \; ,
\nonumber
\\
2: \qquad a'_{3} n'_{1} -  a'_{1} n'_{3} =  b'_{3} n'_{1} -
b'_{1} n'_{3} \; ,
\nonumber
\\
3: \qquad a'_{1} n'_{2} -  a'_{2} n'_{1} =  b'_{1} n'_{2} -
b'_{2} n'_{1} \; .
\nonumber
\end{eqnarray}

\noindent
From these, substituting eqs.  (\ref{7.4}), we get to
\begin{eqnarray}
1: \qquad {a_{2} n_{3} -a_{3} n_{2} \over  1 + a_{1} V } = {b_{2}
n_{3} -b_{3} n_{2} \over  1 + b_{1} V } \; ,
\nonumber
\\
2: \qquad { a_{3} (n_{1}+V)  -  (a_{1} +V)  n_{3}\over 1 + a_{1} V
} =
 { b_{3} (n_{1}+V)  -  (b_{1} +V)  n_{3}\over 1 + b_{1} V }\; ,
\nonumber
\\
3: \qquad  { (a_{1} +V) n_{2} -  a_{2} (n_{1}+V) \over 1 + a_{1} V
} =
 { (b_{1} +V) n_{2} -  b_{2} (n_{1}+V) \over 1 + b_{1} V } \; ,
\nonumber
\end{eqnarray}

\noindent or differently
\begin{eqnarray}
1: \qquad {a_{2} n_{3} -a_{3} n_{2} \over  1 + a_{1} V } = {b_{2}
n_{3} -b_{3} n_{2} \over  1 + b_{1} V } \; ,
\nonumber
\\
2: \qquad { (a_{3} n_{1} -a_{1} n_{3})  -V  (   n_{3} - a_{3}  )
\over 1 + a_{1} V } = { (b_{3} n_{1} -b_{1} n_{3})  -V  (   n_{3}
- b_{3}  ) \over 1 + b_{1} V } \; ,
\nonumber
\\
3: \qquad  { (a_{1} n_{2} -a_{2} n_{1}) +V ( n_{2} -  a_{2}) \over
1 + a_{1} V } =
 { (b_{1} n_{2} -b_{2} n_{1}) +V ( n_{2} -  b_{2}) \over 1 + b_{1} V } \; .
\label{7.5}
\end{eqnarray}

\noindent
Now, having in mind
\begin{eqnarray}
{\bf V} = (V,0,0) \; ,
\nonumber
\\
{\bf V} \times (\; {\bf n} - {\bf a}) = \left (0, -V(   n_{3} -
a_{3}  ), V (   n_{2} - a_{2}  ) \; \right ) \; ,
\nonumber
\\
{\bf V} \times ({\bf n} - {\bf b}) = \left ( \; 0, -V(   n_{3} -
b_{3}  ), V (   n_{2} - b_{2}  ) \; \right ) \; ,
\nonumber
\end{eqnarray}

\noindent the three equations  (\ref{7.5}) can be written as a vector one
\begin{eqnarray}
{{\bf a} \times {\bf n} + {\bf V} \times ({\bf n} - {\bf a}) \over
1 + {\bf a} {\bf V} } = {{\bf b} \times {\bf n} + {\bf V} \times
({\bf n} - {\bf b}) \over 1 + {\bf b} {\bf V} } \; .
\label{7.6}
\end{eqnarray}

\begin{quotation}

Because
\begin{eqnarray}
\mid {\bf a} \times {\bf n} \mid = \sin \alpha_{i} , \qquad
\mid {\bf b} \times {\bf n} \mid = \sin \alpha_{r} \; ,
\nonumber
\end{eqnarray}

\noindent the formula (\ref{7.6}) represents the  reflection
law in the  moving reference frame $K$.

\end{quotation}

Let us compare the generalized   formula (\ref{7.6}) with previously examined  a particular
case when
\begin{eqnarray}
a_{1} = \cos \phi_{2} \; , \qquad a_{2} = \sin \phi_{2} \; ,
\qquad a_{3} = 0 \; ,
\nonumber
\\
b_{1} = \cos \phi_{3} \; , \qquad b_{2} = \sin \phi_{3} \;  ,
\qquad b_{3} =0 \; ,
\nonumber
\\
n_{1} = \cos \alpha\;  , \qquad n_{2} = \sin \alpha \;, \qquad
n_{3} = 0 \; ,
\label{7.7a}
\end{eqnarray}

\noindent
 and  eqs. (\ref{7.5})   will take the form
\begin{eqnarray}
 0=0 \; , \qquad \qquad
\qquad 0=0
 \; ,
\nonumber
\\
  { ( \cos \phi_{2}  \sin \alpha  - \sin \phi_{2} \cos \alpha ) +
V ( \sin \alpha  -  \sin \phi_{2} ) \over 1 + \cos \phi_{2} V } =
\nonumber
\\
={ ( \cos \phi_{3}  \sin \alpha  - \sin \phi_{3} \cos \alpha ) + V
( \sin \alpha  -  \sin \phi_{3} ) \over 1 + \cos \phi_{3} V } \; ,
\nonumber
\end{eqnarray}

\noindent or
\begin{eqnarray}
 {  \sin (\alpha - \phi_{2} )  +
V ( \sin \alpha  -  \sin \phi_{2} ) \over 1 + \cos \phi_{2} V } =
{  \sin ( \alpha - \phi_{3})  + V ( \sin \alpha  -  \sin \phi_{3}
) \over 1 + \cos \phi_{3} V }
 \; ,
\label{7.7b}
\end{eqnarray}

\noindent which coincides with  (\ref{4.7c}).

Now let us transform to the reference frame $K$ the second relevant relationship
(\ref{7.3}):
\begin{eqnarray}
 {\bf a} ' \;  {\bf n}' +  {\bf b}' \; {\bf n}' =0 \; , \qquad \Longrightarrow \qquad \qquad
\nonumber
\\
{a_{1} + V \over 1 + a_{1} V} \; {n_{1}+V \over 1 +n_{1} V} +
{(1-V^{2})(a_{2}n_{2} + a_{3}n_{3})  \over (1 + a_{1} V )(1 +n_{1}
V) }  \; +
\nonumber
\\
+ \; {b_{1} + V \over 1 + b_{1} V} \; {n_{1}+V \over 1 +n_{1} V} +
{(1-V^{2})(b_{2}n_{2} + b_{3}n_{3})  \over (1 + b_{1} V )(1 +n_{1}
V) }=0 \; ,
\nonumber
\end{eqnarray}

\noindent from this it follows
\begin{eqnarray}
{(a_{1}n_{1} +a_{2}n_{2} +a_{3} n_{3}) + (a_{1} +n_{1}) V + V^{2}(
1 - a_{2}n_{2} -a_{3}n_{3}) \over 1+a_{1} V} +
\nonumber
\\
+ \; {(b_{1}n_{1} +b_{2}n_{2} +b_{3} n_{3}) + (b_{1} +n_{1}) V +
V^{2}( 1 - b_{2}n_{2} -b_{3}n_{3}) \over 1+b_{1} V} =0\; .
\label{7.8}
\end{eqnarray}

\vspace{5mm}

For the case (\ref{7.7a}) it becomes
\begin{eqnarray}
{( \cos \phi_{2} \cos \alpha  + \sin \phi_{2} \sin \alpha  ) +
 (\cos \phi_{2}  + \cos \alpha ) V + V^{2}( 1 - \sin \phi_{2} \sin \alpha  ) \over
1+ \cos \phi_{2}  V} +
\nonumber
\\
+ \; {(\cos \phi_{3} \cos \alpha  + \sin \phi_{3} \sin \alpha  ) +
(\cos \phi_{3}  + \cos \alpha) V +
 V^{2}( 1 - \sin \phi_{3} \sin \alpha) \over
1+ \cos \phi_{3} V}=0
\nonumber
\end{eqnarray}

\noindent or
\begin{eqnarray}
{ \cos  (\alpha - \phi_{2}) +
 (\cos \phi_{2}  + \cos \alpha ) V + V^{2}( 1 - \sin \phi_{2} \sin \alpha  ) \over
1+ \cos \phi_{2}  V} +
\nonumber
\\
+ \; {\cos (\alpha - \phi_{3} ) + (\cos \phi_{3}  + \cos \alpha) V
+
 V^{2}( 1 - \sin \phi_{3} \sin \alpha) \over
1+ \cos \phi_{3} V} =0
\label{7.9}
\end{eqnarray}

\noindent which coincides with eq.  (\ref{4.8a}):
\begin{eqnarray}
{- \cos (\alpha - \phi_{2}) -  V\; (\cos \alpha + \cos \phi_{2} )
-
 V^{2}  (1 - \sin \alpha \; \sin \phi_{2}) \over
  (1+ V\; \cos \phi_{2}) } =
\nonumber
\\
 =
{\cos (\alpha - \phi_{3}) +  V\; (\cos \alpha + \cos \phi_{3} ) +
 V^{2}  (1 - \sin \alpha \; \sin \phi_{3}) \over
  (1+ V\; \cos \phi_{3}) } =0 \; .
\nonumber
\end{eqnarray}

\section{The Lorentz transform with the  arbitrary velocity vector
 ${\bf V}$ }

We shall now investigate the case of an arbitrary velocity vector ${\bf V}$.
The velocity equations of (\ref{7.5}) and rewritten as one equation in (\ref{7.6})
are true only when
${\bf V} = (V,0,0)$.

Firstly, in the following we will need an explicit form of the Lorentz transform
with an arbitrary ${\bf V}$. With the notation
\begin{eqnarray}
{\bf V} = {\bf e} \; th\; \beta \;  , \qquad {\bf e} ^{2} = 1 \; , \qquad \qquad
\nonumber
\\
{1 \over \sqrt{1 - V^{2} }} = ch\; \beta, \qquad {V \over \sqrt{1
- V^{2} }} = sh\; \beta, \qquad
\label{8.1}
\end{eqnarray}

\noindent the Lorentz matrix is [86]
\begin{eqnarray}
(L_{a}^{\;\;b} )=  \hspace{40mm}
\nonumber
\\
\left | \begin{array}{cccc}
ch\; \beta       &   e_{1} sh\; \beta   &  e_{2} sh \;\beta  e_{2}  &  e_{3}sh\; \beta  \\
e_{1} sh\; \beta  & ch\; \beta -(ch\; \beta -1)(e_{2}^{2}
+e_{3}^{2} ) &
(ch\; \beta -1) e_{1} e_{2} &  (ch \; \beta -1) e_{2}e_{3}\\
 e_{2} sh\; \beta   & (ch\; \beta -1) e_{1}e_{2} &  ch\; \beta -(ch\; \beta -1)
 (e_{1}^{2} +e_{3}^{2} )&
(ch\; \beta -1) e_{2}e_{3} \\
 e_{3} sh\; \beta   & (ch\; \beta -1) e_{1}e_{3} & (ch\; \beta -1) e_{2}e_{3} &
ch\; \beta -(ch\; \beta -1)(e_{1}^{2} +e_{2}^{2} )
\end{array} \right |  .
\nonumber
\end{eqnarray}

\noindent In three  cases the matrix $L$ looks especially simple:
\begin{eqnarray}
{\bf e} =(1,0,0), \qquad L = \left | \begin{array}{cccc}
ch\; \beta       &   sh\; \beta   &  0    &  0  \\
sh\; \beta       & ch\; \beta     &  0    &  0  \\
0                & 0              &  1    &  0  \\
0                & 0              &  0    &  1
\end{array} \right | \; ,
\nonumber
\\
{\bf e} =(0,1,0), \qquad L = \left | \begin{array}{cccc}
ch\; \beta       &  0    &   sh\; \beta    &  0  \\
0       &  1    &  0    &  0  \\
 sh\; \beta                 & 0              &  ch\; \beta     &  0  \\
0                & 0              &  0    &  1
\end{array} \right | \; ,
\nonumber
\\
{\bf e} =(0,0,1), \qquad L = \left | \begin{array}{cccc}
ch\; \beta       &  0    &   0    &  sh\; \beta   \\
0       &  1    &  0    &  0  \\
0             & 0              & 1    &  0  \\
sh\; \beta                 & 0              &   0       &   ch\;
\beta
\end{array} \right | \; .
\nonumber
\end{eqnarray}

\noindent
The matrix  $L$  can be rewritten as follows
\begin{eqnarray}
(L_{a}^{\;\;b} )= \left | \begin{array}{cccc}
ch\; \beta       &   e_{1} sh\; \beta   &  e_{2} sh \;\beta  e_{2}  &  e_{3}sh\; \beta  \\
e_{1} sh\; \beta  &  1 + (ch\; \beta -1) e_{1}^{2} &
(ch\; \beta -1) e_{1} e_{2} &  (ch \; \beta -1) e_{2}e_{3}\\
 e_{2} sh\; \beta   & (ch\; \beta -1) e_{1}e_{2} &1 + (ch\; \beta -1) e_{2}^{2}  &
(ch\; \beta -1) e_{2}e_{3} \\
 e_{3} sh\; \beta   & (ch\; \beta -1) e_{1}e_{3} & (ch\; \beta -1) e_{2}e_{3} &
1 + (ch\; \beta -1) e_{3}^{2}
\end{array} \right |
\nonumber
\end{eqnarray}

\noindent  or in a  more formal symbolical manner
\begin{eqnarray}
L = \left | \begin{array}{cc}
ch \; \beta & {\bf e}\;  sh\; \beta \\
{\bf e}  \; sh\; \beta & [\delta_{ij} + (ch\; \beta -1) e_{i}
e_{j}]\;
\end{array} \right | \; .
\label{8.4b}
\end{eqnarray}

The Lorentz transform  (\ref{8.4b}) acts on space-time coordinates  $(t,{\bf x})$  in
accordance with
\begin{eqnarray}
L: \qquad t' = ch\; \beta \; t +   sh\; \beta \; {\bf e}\;   {\bf x} \; ,
\nonumber
\\
{\bf x}'= {\bf e}  \; sh\; \beta  \; t + {\bf x}  + (ch\; \beta
-1)\; {\bf e} \; ({\bf e}  {\bf x})  \; .
\label{8.5a}
\end{eqnarray}

\noindent
For the inverse transform we have
\begin{eqnarray}
L^{-1} : \qquad t = ch\; \beta \; t' -   sh\; \beta \; {\bf e}\;   {\bf x}'\; ,
\nonumber
\\
{\bf x} = -{\bf e}  \; sh\; \beta  \; t + {\bf x}'  + (ch\; \beta
-1)\; {\bf e} \; ({\bf e}  {\bf x}') \; .
\nonumber
\end{eqnarray}

\noindent
Indeed, let us consider the time variable  $t$:
\begin{eqnarray}
t = ch\; \beta \; t' -   sh\; \beta \; {\bf e}\;   {\bf x}' =
\nonumber
\\
= ch\; \beta \; (ch\; \beta \; t +   sh\; \beta \; {\bf e}\;
{\bf x}) -
  sh\; \beta \; {\bf e}\;  ({\bf e}  \; sh\; \beta  \; t + {\bf x}  +
(ch\; \beta -1)\; {\bf e} \; ({\bf e}  {\bf x})) =
\nonumber
\\
= t  + ch\; \beta \;   sh\; \beta \; ({\bf e}\;   {\bf x}) - sh\;
\beta \; {\bf e}\;  [\;  {\bf x}  + (ch\; \beta -1)\; {\bf e} \;
({\bf e}  {\bf x}) \; ]=
\nonumber
\\
= t  + ch\; \beta \;   sh\; \beta \; ({\bf e}\;   {\bf x}) - sh\;
\beta \; ( {\bf e}\;   {\bf x})  -  sh\; \beta \; \; ch\; \beta
({\bf e}  {\bf x}) +  sh\; \beta \;  \; ({\bf e}  {\bf x}) = t \;
.
\nonumber
\end{eqnarray}

\noindent
In the  same manner, for the space variable  ${\bf x}$ we have
\begin{eqnarray}
{\bf x}= -{\bf e}  \; sh\; \beta  \; t' + {\bf x}'  + (ch\; \beta
-1)\; {\bf e} \; ({\bf e}  {\bf x}') =
\nonumber
\\
= -{\bf e}  \; sh\; \beta  \;  [\; ch\; \beta \; t +   sh\; \beta
\; ({\bf e}\;   {\bf x}) \; ] +
\nonumber
\\
+{\bf e}  \; sh\; \beta  \; t + {\bf x}  + (ch\; \beta -1)\; {\bf
e} \; ({\bf e}  {\bf x})    +
\nonumber
\\
+ (ch\; \beta -1)\; {\bf e} \; [\; {\bf e} \; [\; ({\bf e}  \;
sh\; \beta  \; t + {\bf x}  + (ch\; \beta -1)\; {\bf e} \;  ({\bf
e} {\bf x}) \;] \; ] =
\nonumber
\\
= - {\bf e}  \; sh\; \beta  \;  ch\; \beta \; t -    \; sh^{2}
\beta{\bf e}  \;({\bf e}\;{\bf x}) + {\bf e}  \; sh\; \beta  \; t
+ {\bf x}  + (ch\; \beta -1) \; {\bf e} \; ({\bf e}  {\bf x})
+
\nonumber
\\
+ (ch\; \beta -1)\; {\bf e} \; [\;   (  \; sh\; \beta  \; t  +
ch\; \beta   ({\bf e}{\bf x} )] =
\nonumber
\\
= {\bf x}   +  [ \; - sh^{2} \beta
  + ch\; \beta -1       +
ch^{2} \; \beta - ch\; \beta  \;] \; {\bf e}  \;({\bf e}\;{\bf
x})= {\bf x} \; .
\nonumber
\end{eqnarray}

It can be shown by straightforward calculation that
the  general Lorentz transformation  (\ref{8.5a}) leaves invariant the so called
relativistic length of 4-vector  $(t,{\bf x})$:
\begin{eqnarray}
t^{'2} -{\bf x}^{'2} = [ \; ch\; \beta \; t +   sh\; \beta \;
({\bf e}\;   {\bf x}) \; ]^{2} -
\nonumber
\\
- \left [ \; {\bf e}  \; sh\; \beta  \; t + {\bf x}  + (ch\; \beta
-1)\; {\bf e} \; ({\bf e}  {\bf x})\; \right  ]^{2} =
\nonumber
\\
= ch^{2}  \beta \;  t^{2} + 2 ch\; \beta \; sh\; \beta \; t ({\bf
e}\;   {\bf x}) + sh^{2}  \beta \; ({\bf e}\;   {\bf x})^{2} -
\nonumber
\\
-  sh^{2} \beta \; t^{2} - 2 sh\; \beta t \; ({\bf e} {\bf x})  -
{\bf x}^{2}\; -
\nonumber
\\
- \; 2 sh \; \beta (ch\; \beta -1) t\;  ( {\bf e} {\bf x} ) - 2
(ch\; \beta -1) \;  ({\bf e} {\bf x} )^{2} -
\nonumber
\\
- (ch\; \beta -1)^{2} \; ({\bf e}  {\bf x})^{2} =
\nonumber
\\
= (t^{2} -{\bf x}^{2}) + [\;  2 ch\; \beta \; sh\; \beta \; t
({\bf e}\;   {\bf x}) + sh^{2}  \beta \; ({\bf e}\;   {\bf x})^{2}
- 2 sh\; \beta t \; ({\bf e} {\bf x}) -
\nonumber
\\
- \; 2 sh \; \beta ch\; \beta  t'\;  ( {\bf e} {\bf x}' )   + 2sh
\; \beta \;  t'\;  ( {\bf e} {\bf x}' ) - 2 (ch\; \beta -1) \;
({\bf e} {\bf x}' )^{2} - (ch\; \beta -1)^{2} \; ({\bf e}  {\bf
x}')^{2} =
\nonumber
\\
= (t^{'2} -{\bf x}^{'2}) + [\; sh^{2}  \beta \;  - 2 (ch\; \beta
-1)  - (ch\; \beta -1)^{2}  ] \; ({\bf e}\;   {\bf x}')^{2}=
\nonumber
\\
= (t^{'2} -{\bf x}^{'2}) + [\; sh^{2}  \beta \;  - 2 ch\; \beta +
2   - ch\; ^{2} \beta + 2 ch \; \beta -1
 ] \; ({\bf e}\;   {\bf x}')^{2}=
\nonumber
\\
= (t^{'2} -{\bf x}^{'2}) + 0 \; ,
\nonumber
\end{eqnarray}

\noindent so that
\begin{eqnarray}
(t^{2} -{\bf x}^{2})= (t^{'2} -{\bf x}^{'2})  \; .
\nonumber
\end{eqnarray}

\vspace{5mm} The formulas  (\ref{8.5a}) can  be rewritten differently with special notation for
longitudinal and perpendicular constituents:
\begin{eqnarray}
{\bf e} \; ({\bf e}  {\bf x}' ) = {\bf x}'_{\|} \; , \qquad ({\bf
e}  {\bf x}' ) = x'_{\|}, \qquad {\bf x}' - {\bf e} \; ({\bf e}
{\bf x}' ) = {\bf x}'_{\bot} \;
\label{8.8}
\end{eqnarray}

\noindent then
\begin{eqnarray}
t = ch\; \beta \; t' +   sh\; \beta \;  x'_{\|} \; ,
\nonumber
\\
{\bf x}=  {\bf e}   (
 \; sh\; \beta  \; t' + ch \; \beta \;     x'  _{\|} ) \;  + \; {\bf x}'_{\bot} \; ,
\nonumber
\end{eqnarray}

\noindent or
\begin{eqnarray}
t = ch\; \beta \; t' +   sh\; \beta \;  x'_{\|} \; , \qquad {\bf
x}_{\bot} = {\bf x}'_{\bot}  \; ,
\nonumber
\\
{\bf x}_{\|}=   {\bf e} \; ({\bf e}  {\bf x} )= {\bf e}(
 sh\; \beta  \; t' + ch \; \beta \;     x'  _{\|} ) \;.
\label{8.10}
\end{eqnarray}

\vspace{5mm}

Now, with general expression for Lorentz transform in hands,
we may quite easily obtain a  general form of velocity addition law.
Let us take two events on the trajectory  of a moving particle
\begin{eqnarray}
(t_{1}', {\bf x}'_{1}) \qquad   \mbox{and}  \qquad   (t_{2}',  {\bf
x}'_{2} ) \; ;
\nonumber
\end{eqnarray}

\noindent the  velocity vector in the reference frame  $K'$
 is defined by the  relation
\begin{eqnarray}
{\bf W}' = { {\bf x}'_{2} - {\bf x}'_{1} \over t_{2}' - t_{1}' } \
; .
\nonumber
\end{eqnarray}

\noindent
In the same manner, the velocity vector in the reference frame  $K$ is given by
\begin{eqnarray}
{\bf W} = { {\bf x}_{2} - {\bf x}_{1} \over t_{2} - t_{1} } = \hspace{30mm}
\nonumber
\\
= {- {\bf e}  \; sh\; \beta  \; t'_{2} + {\bf x}'_{2}   + (ch\;
\beta -1)\; {\bf e} \; ({\bf e}  {\bf x}'_{2}) + {\bf e}  \; sh\;
\beta  \; t'_{1} -{\bf x}'_{1}   - (ch\; \beta -1)\; {\bf e} \;
({\bf e}  {\bf x}'_{1} ) \over ch\; \beta \; t'_{2} -   sh\; \beta
\; {\bf e}\;   {\bf x}'_{2} - ch\; \beta \; t'_{1} +   sh\; \beta
\; {\bf e}\;   {\bf x}'_{1} }=
\nonumber
\\
={ ({\bf x}'_{2} - {\bf x}'_{1}) - {\bf e}  \; sh\; \beta  \;(
t'_{2}-t_{1}') + (ch\; \beta -1)\; {\bf e} \; [{\bf e}  ( {\bf
x}'_{2} -{\bf x}'_{1}) ] \over ch\; \beta \; (t'_{2} -t_{1}') -
sh\; \beta \; {\bf e}\;  ( {\bf x}'_{2} -{\bf x}'_{1}) }\; ;
\nonumber
\end{eqnarray}

\noindent  from this it follows the  velocity addition rule in the most general form
\begin{eqnarray}
{\bf W} = { {\bf W}'  - {\bf e}  \; sh\; \beta   + (ch\; \beta
-1)\; {\bf e} ( {\bf e} \; {\bf W}' ) \over ch\; \beta  - sh\;
\beta \; {\bf e}\;  {\bf W}'  } \; .
\label{8.11}
\end{eqnarray}

\noindent
Eq. (\ref{8.11}) may be rewritten as
\begin{eqnarray}
{\bf W} = {  {\bf W}' -{\bf e} ( {\bf e} \; {\bf W}' )  \over ch\;
\beta  - sh\; \beta \; {\bf e}\;  {\bf W}'  }  \; + \; { ch\;
\beta \; {\bf e} ( {\bf e} \; {\bf W}' ) - {\bf e}  \; sh\; \beta
\over ch\; \beta  -sh\; \beta \; {\bf e}\;  {\bf W}'  } \; ,
\nonumber
\end{eqnarray}

\noindent which, with the  notation
\begin{eqnarray}
{\bf e} ( {\bf e} \; {\bf W}' ) = {\bf W}'_{\|} , \qquad
  {\bf W}' -{\bf e} ( {\bf e} \; {\bf W}' )  = {\bf W}'_{\bot} \; ,
\nonumber
\end{eqnarray}

\noindent  will give
\begin{eqnarray}
{\bf W} = { \sqrt{1-V^{2}} \over 1 -   {\bf W}'{\bf V}  }  \; {\bf
W}'_{\bot} \;\;  + \; \; {  {\bf W}' _{\|} -{\bf V}      \over 1
-   {\bf W}' {\bf V}  } \; .
\label{8.13}
\end{eqnarray}

\noindent
Evidently, eqs. (\ref{8.13}) and   (\ref{8.11}) are equivalent to each other: eq.  (\ref{8.13})
seems more understandable from physical standpoint,  though eq. (\ref{8.11})
is more convenient at
general calculating.

\section{ Reflection of the light in moving reference frame $K$,\\
 the  case of an arbitrary velocity vector ${\bf V}$
}

\hspace{5mm} Three vectors in the  reflection law (see (\ref{7.2}))
\begin{eqnarray}
{\bf a} '\times {\bf n}' =  {\bf b}' \times {\bf n}' \;
\label{9.1}
\end{eqnarray}

\noindent change at Lorentz transform $L({\bf V})$  as follows
\begin{eqnarray}
{\bf a}' = { {\bf a}  + {\bf e}  \; [ \; sh\; \beta   + (ch\;
\beta -1)\;  ( {\bf e} \; {\bf a} ) \; ]  \over ch\; \beta  + sh\;
\beta \; ({\bf e} {\bf a} ) } \; ,
\nonumber
\\
{\bf b}' = { {\bf b}  + {\bf e}  \;[\;  sh\; \beta   + (ch\; \beta
-1)\;  ( {\bf e} \; {\bf b} )\; ] \over ch\; \beta  + sh\; \beta
\; ({\bf e}{\bf b})  } \; ,
\nonumber
\\
{\bf n}' = { {\bf n}  + {\bf e}  \;[\;  sh\; \beta   + (ch\; \beta
-1)\;  ( {\bf e} \; {\bf n} ) \; ] \over ch\; \beta  + sh\; \beta
\; ({\bf e}{\bf n})  } \; .
\label{9.2}
\end{eqnarray}

\noindent Substituting eqs. (\ref{9.2}) into  (\ref{9.1}) we  get to
\begin{eqnarray}
{ {\bf a}  + {\bf e}  \; [\; sh\; \beta   + (ch\; \beta -1)\; (
{\bf e} \; {\bf a} )\; ]  \over ch\; \beta  - sh\; \beta \; {\bf
e}\;  {\bf a}  }  \; \times \; {  {\bf n}  +  {\bf e}  \;[\;  sh\;
\beta   + (ch\; \beta -1)\; ( {\bf e} \; {\bf n} )\; ] \over  ch\;
\beta  + sh\; (\beta  {\bf e}  {\bf n})  }=
\nonumber
\\
= { {\bf b}  + {\bf e}  \;[\;  sh\; \beta   + (ch\; \beta -1)\; (
{\bf e} \; {\bf b} )\; ] \over ch\; \beta  + sh\; \beta \; {\bf
e}\;  {\bf b}  } \;  \times \; { {\bf n}  + {\bf e}  \;[\;  sh\;
\beta   + (ch\; \beta -1)\; ( {\bf e} \; {\bf n} ) \; ]\over ch\;
\beta  + sh\; ( \beta {\bf e} {\bf n})  }
\nonumber
\end{eqnarray}

\noindent and further
\begin{eqnarray}
{ {\bf a} \times  {\bf n}  +  ({\bf a} \times{\bf e}) \; [\;
sh\; \beta   + (ch\; \beta -1) \;( {\bf e}  {\bf n} ) \;] + ({\bf
e}  \times {\bf n}) \; [\; sh\; \beta   + (ch\; \beta -1) \; (
{\bf e}  {\bf a} ) \; ] \over ch\; \beta  + sh\; \beta  ({\bf e}
{\bf a})  }=
\nonumber
\\
= { {\bf b} \times  {\bf n}  +  ({\bf b} \times{\bf e}) \; [\;
sh\; \beta   + (ch\; \beta -1) \; ( {\bf e}  {\bf n} ) \;] + ({\bf
e}  \times {\bf n}) \; [\; sh\; \beta   + (ch\; \beta -1) \;  (
{\bf e}  {\bf b} ) \; ] \over ch\; \beta  + sh\; \beta  ({\bf e}
{\bf b} ) } \; .
\nonumber
\nonumber
\end{eqnarray}

\noindent After simple manipulation it gives
\begin{eqnarray}
{ {\bf a} \times  {\bf n}  +  ({\bf a} \times{\bf e} +  {\bf e}
\times  {\bf n} )\;
 sh\; \beta   +
[ ({\bf a} \times{\bf e}) \;( {\bf e}  {\bf n} ) +({\bf e}
\times{\bf n}) \;( {\bf e}  {\bf a} )\;] (ch\; \beta -1)
 \over
ch\; \beta  + sh\; \beta  ({\bf e}  {\bf a})  }=
\nonumber
\\
{ {\bf b} \times  {\bf n}  +  ({\bf b} \times{\bf e} +  {\bf e}
\times  {\bf n} )\;
 sh\; \beta   +
[ ({\bf b} \times{\bf e}) \;( {\bf e}  {\bf n} ) +({\bf e}
\times{\bf n}) \;( {\bf e}  {\bf b} )\;] (ch\; \beta -1)
 \over
ch\; \beta  + sh\; \beta  ({\bf e}  {\bf b})  }\; .
\label{9.3b}
\end{eqnarray}

Taking the  known relations for double vector product:
\begin{eqnarray}
{\bf e} \times ({\bf n} \times {\bf a}) = {\bf  n} ({\bf  e}{\bf
a}) -  {\bf  a} ({\bf  e}{\bf n}) \; ,
\nonumber
\\
{\bf e} \times ({\bf n} \times {\bf b}) = {\bf  n} ({\bf  e}{\bf
b}) -  {\bf  b} ({\bf  e}{\bf n}) \; ,
\nonumber
\end{eqnarray}

\noindent eq. (\ref{9.3b}) may be taken to the form
\begin{eqnarray}
{\;  {\bf a} \times  {\bf n}    + ({\bf a} \times{\bf e} +  {\bf
e}  \times  {\bf n} ) \; sh\; \beta +   {\bf e}\times ( {\bf e}
\times ({\bf n} \times {\bf a})  )  (ch\; \beta -1)  \over ch\;
\beta  + sh\; \beta \; ({\bf e} {\bf a})  }=
\nonumber
\\
={\;  {\bf b} \times  {\bf n}    + ({\bf b} \times{\bf e} +  {\bf
e}  \times  {\bf n} )\; sh\; \beta +   {\bf e}\times ( {\bf e}
\times ({\bf n} \times {\bf b})  ) (ch\; \beta -1)  \over ch\;
\beta  + sh\; \beta \; ({\bf e}  {\bf b})  }\; .
\label{9.3c}
\end{eqnarray}

\noindent
Again using the known identities
\begin{eqnarray}
{\bf e}\times ( {\bf e} \times ({\bf n} \times {\bf a})  ) = {\bf
e} \; [ \;{\bf e}  ({\bf n} \times {\bf a})\;] - ({\bf n} \times
{\bf a}) \; ,
\nonumber
\\
{\bf e}\times ( {\bf e} \times ({\bf n} \times {\bf b})  ) = {\bf
e} \; [ \;{\bf e}  ({\bf n} \times {\bf b})\; ] - ({\bf n} \times
{\bf a}) \; ,
\nonumber
\end{eqnarray}

\noindent eq. (\ref{9.3c}) will read
\begin{eqnarray}
{ch\; \beta  \;  ({\bf a} \times {\bf n} )  +
 sh\; \beta  \; ( {\bf a} -{\bf n}  )\times{\bf e}
+  (ch\; \beta -1)\;  {\bf e} \; [ \;{\bf e}  ({\bf n} \times {\bf
a})\; ]
    \over
ch\; \beta  + sh\; \beta \; ({\bf e} {\bf a})  }=
\nonumber
\\
= {ch\; \beta  \;  ({\bf b} \times {\bf n})  +
 sh\; \beta ({\bf b} -{\bf n})  \times {\bf e}
  +   (ch\; \beta -1)\; {\bf e} \; [ \;{\bf e}  ({\bf n} \times {\bf b})\; ]
    \over
ch\; \beta  + sh\; \beta \; ({\bf e} {\bf b})  } \; ,
\nonumber
\end{eqnarray}

\noindent  or
\begin{eqnarray}
{ {\bf a} \times {\bf n}   +
 th\; \beta  \; ( {\bf a} -{\bf n}  )\times{\bf e}
+  (1  - ch^{-1} \beta)\;  {\bf e} \; [ \;{\bf e}  ({\bf n} \times
{\bf a})\; ]
    \over
1  + th\; \beta \; ({\bf e} {\bf a})  }=
\nonumber
\\
= { {\bf b} \times {\bf n}  +
 th\; \beta ({\bf b} -{\bf n})  \times {\bf e}
  +   (1  - ch^{-1} \beta)\; {\bf e} \; [ \;{\bf e}  ({\bf n} \times {\bf b})\; ]
    \over
1  + th\; \beta \; ({\bf e} {\bf b})  } \; .
\label{9.5}
\end{eqnarray}

\begin{quotation}

{\em These equations  provides us with mathematical form

of the  reflection law in the moving reference frame $K$.}

\end{quotation}

One may note that if four vector  ${\bf a}, {\bf b},
{\bf n}, {\bf e}$ belong the same plane (previously obtained eq.  (\ref{7.6})
relates to just that situation) then two identities hold
\begin{eqnarray}
[ \;{\bf e}  ({\bf n} \times {\bf a})\; ] = 0 , \qquad [ \;{\bf e}
({\bf n} \times {\bf b})\; ] = 0 \;  ,
\label{9.6}
\end{eqnarray}

\noindent at this eq. (\ref{9.5}) will  become  much simpler;
\begin{eqnarray}
{{\bf a} \times {\bf n}  +
 th\; \beta  ({\bf a}  - {\bf n}) \times{\bf e}     \over
1  + th\; \beta \; ({\bf e} {\bf a})  }= { {\bf b} \times {\bf n}
+
 th\; \beta  \; ( {\bf b}  -{\bf n}) \times {\bf e}
  \over
1  + th\; \beta \; ({\bf e} {\bf b})  } \; ,
\label{9.7a}
\end{eqnarray}

\noindent what coincides with eq. (\ref{7.6}):
\begin{eqnarray}
{{\bf a} \times {\bf n}  +
 ({\bf a}  - {\bf n}) \times{\bf V}     \over
1  +  {\bf a} {\bf V}   }= { {\bf b} \times {\bf n}  +
  ( {\bf b}  -{\bf n}) \times {\bf V}
  \over
1  +   {\bf b} {\bf V}   } \; .
\nonumber
\end{eqnarray}

\vspace{5mm}

Now let us perform some additional  calculation that  permit us to decompose
the vector form of the reflection law (\ref{9.5}) into two more simple equations: one
along the ${\bf e}$  vector and other transversely to ${\bf e}$.
Indeed, forming the dot product of eq (\ref{9.5}) and the
 vector   ${\bf e}$ we obtain
\begin{eqnarray}
 {  {\bf e}\;({\bf a} \times {\bf n} )  +
 th\; \beta  \;  {\bf e}\;[\; ( {\bf a} -{\bf n}  )\times{\bf e}\; ]
+  (1  - ch^{-1} \beta)\;   [ \;{\bf e}  ({\bf n} \times {\bf
a})\; ]
    \over
1  + th\; \beta \; ({\bf e} {\bf a})  }=
\nonumber
\\
= { {\bf e}\; ({\bf b} \times {\bf n})  +
 th\; \beta {\bf e}\; [\; ({\bf b} -{\bf n})  \times {\bf e} \; ]
  +   (1  - ch^{-1} \beta)\;  [ \;{\bf e}  ({\bf n} \times {\bf b})\; ]
    \over
1  + th\; \beta \; ({\bf e} {\bf b})  } \; .
\nonumber
\end{eqnarray}

\noindent  from this it follows
\begin{eqnarray}
(1  - ch^{-1} \beta)\; {   [  {\bf e}  ({\bf n} \times {\bf a}) ]
    \over
1  + th\; \beta \; ({\bf e} {\bf a})  }= (1  - ch^{-1} \beta)\;
 { [ {\bf e}  ({\bf n} \times {\bf b}) ]
    \over
1  + th\; \beta \; ({\bf e} {\bf b})  } \; ,
\nonumber
\end{eqnarray}

\noindent and further
\begin{eqnarray}
{    {\bf V}  ({\bf n} \times {\bf a})
    \over
1  +  ({\bf V} {\bf a})  }=
 { {\bf V}  ({\bf n} \times {\bf b})
    \over
1  +  ({\bf V} {\bf b})  } \; .
\label{9.8b}
\end{eqnarray}

\noindent Taking this in mind , we see that in eq.  (\ref{9.5}) some terms on the left and on the right
 canceled each other so  we  will obtain
 (\ref{9.7a}).
More clarity  may be reached with the use of symbolical  vector notation for eq. (\ref{9.5}):
\begin{eqnarray}
{\bf A} ={\bf B} \; .
\nonumber
\end{eqnarray}

The above calculation consists of two step:

\begin{eqnarray}
\underline{first} \qquad
{\bf e} {\bf A} = {\bf e} {\bf B} \; , \qquad
\underline{ second} \qquad
{\bf A}  - {\bf e}({\bf e} {\bf A})={\bf B} -{\bf e}({\bf e} {\bf B}) \ ;.
\nonumber
\end{eqnarray}

{\em So,
\noindent  the light  reflection law   (\ref{9.5}) in the  moving reference

 frame $K$  may be presented with the help of two equations
 }
\begin{eqnarray}
{    [{\bf V}  ({\bf n} \times {\bf a})]
    \over
1  +  ({\bf V} {\bf a})  }=
 {[ {\bf V}  ({\bf n} \times {\bf b})]
    \over
1  +  ({\bf V} {\bf b})  } \; , \qquad
\nonumber
\\
{{\bf a} \times {\bf n}  +
 ({\bf a}  - {\bf n}) \times{\bf V}     \over
1  +  {\bf a} {\bf V}   }= { {\bf b} \times {\bf n}  +
  ( {\bf b}  -{\bf n}) \times {\bf V}
  \over
1  +   {\bf b} {\bf V}   } \; .
\label{9.9b}
\end{eqnarray}

\vspace{5mm}

Now let us consider second (scalar) equation (\ref{7.3})
\begin{eqnarray}
{\bf a}' {\bf n} ' + {\bf b}' {\bf n}' = 0 \; .
\label{9.10}
\end{eqnarray}

\noindent From  (\ref{9.10}), with the use of (\ref{9.2}), one gets
\begin{eqnarray}
{ {\bf a}  + {\bf e}  \; [ \; sh\; \beta   + (ch\; \beta -1)\;  (
{\bf e} \; {\bf a} ) \; ]  \over ch\; \beta  + sh\; \beta \; ({\bf
e} {\bf a} ) } \;\; { {\bf n}  + {\bf e}  \;[\;  sh\; \beta
+ (ch\; \beta -1)\;  ( {\bf e} \; {\bf n} ) \; ] \over ch\; \beta
+ sh\; \beta \; ({\bf e}{\bf n})  } \;+
\nonumber
\\
+ { {\bf b}  + {\bf e}  \;[\;  sh\; \beta   + (ch\; \beta -1)\;  (
{\bf e} \; {\bf b} )\; ] \over ch\; \beta  + sh\; \beta \; ({\bf
e}{\bf b})  } \;  \; { {\bf n}  + {\bf e}  \;[\;  sh\;
\beta   + (ch\; \beta -1)\;  ( {\bf e} \; {\bf n} ) \; ] \over
ch\; \beta  + sh\; \beta \; ({\bf e}{\bf n})  } =0\; ;
\nonumber
\end{eqnarray}

\noindent from where after evident calculation it follows
\begin{eqnarray}
{ {\bf a}  {\bf n}  +  ({\bf a} +  {\bf n} ){\bf e}
  sh \beta   + 2({\bf a} {\bf e}) ({\bf n} {\bf e})
(ch \beta -1)     + [  sh \beta   + (ch \beta -1)   {\bf e}  {\bf
n}  ] [  sh \beta   + (ch \beta -1)  {\bf e}  {\bf a}  ]  \over ch
\beta  + sh \beta \; {\bf e} {\bf a}  } +
\nonumber
\\
+ { {\bf b}  {\bf n}  +  ({\bf b}  +  {\bf n} ){\bf e})
  sh \beta   +  2({\bf b} {\bf e}) ({\bf n} {\bf e})
(ch \beta -1)     + [  sh \beta   + (ch \beta -1)   {\bf e}  {\bf
n}  ] [  sh \beta   + (ch \beta -1)  {\bf e}  {\bf b}  ]  \over ch
\beta  + sh \beta  \;{\bf e} {\bf b}  }
=0
\nonumber
\end{eqnarray}

\noindent and  further
\begin{eqnarray}
{ {\bf a}  {\bf n}  +  ({\bf a} +  {\bf n} ){\bf e}\;
  sh \beta  ch \beta    + [\; 1 + ({\bf e} {\bf n}) ({\bf e} {\bf a})  \; ] \; sh^{2} \beta
 \over
ch \beta  + sh \beta  {\bf e} {\bf a}  }\; +
\nonumber
\\
+\; { {\bf b}  {\bf n}  +  ({\bf b} +  {\bf n} ){\bf e}\;
sh \beta  ch \beta    + [\; 1 + ({\bf e} {\bf n}) ({\bf e} {\bf b})  \; ] \; sh^{2} \beta
\over
ch \beta  + sh \beta  {\bf e} {\bf b}  } = 0 \; .
\label{9.11}
\end{eqnarray}

Let us verify that this general formula  (\ref{9.11}) is reduced to the  established result
for a particular simpler case. Indeed. let it be
\begin{eqnarray}
{\bf e} = (1,0,0) , \qquad {\bf a} = (a_{1},a_{2}, a_{3}) ,
\qquad {\bf b} = (b_{1}, b_{2}, b_{3})
\nonumber
\end{eqnarray}

\noindent  then  eq. (\ref{9.11}) reads
\begin{eqnarray}
{ ( a_{1} n_{1} + a_{2}n_{2}+ a_{3}n_{3}) (ch^{2}\beta - sh^{2}
\beta)
  +  ( a_{1} +  n_{1} )
  sh \beta  ch \beta    + ( 1 + a_{1} n_{1}) \; sh^{2} \beta
 \over
ch \beta  + sh \beta   a_{1}  } +
\nonumber
\\
+ { ( b_{1} n_{1} + b_{2}n_{2}+ b_{3}n_{3}) (ch^{2}\beta - sh^{2}
\beta)
  +  ( b_{1} +  n_{1} )
  sh \beta  ch \beta    + ( 1 + b_{1} n_{1}) \; sh^{2} \beta
 \over ch \beta  + sh \beta   b_{1}  } = 0 \;
\nonumber
\end{eqnarray}

\noindent from  where it follows
\begin{eqnarray}
{ ( a_{1} n_{1} + a_{2}n_{2}+ a_{3}n_{3}) ch^{2}\beta
+  ( a_{1} +  n_{1} )
sh \beta  ch \beta    + ( 1 - a_{2} n_{2}- a_{3} n_{3} ) \; sh^{2} \beta
\over
ch \beta  + sh \beta   a_{1}  } +
\nonumber
\\
+ { ( b_{1} n_{1} + b_{2}n_{2}+ b_{3}n_{3}) ch^{2}\beta   +  (
b_{1} +  n_{1} )
  sh \beta  ch \beta    + ( 1 - b_{2} n_{2}- b_{3} n_{3}) \; sh^{2} \beta
 \over ch \beta  + sh \beta   b_{1}  } = 0 \;
\nonumber
\end{eqnarray}

\noindent which after dividing by  $ch^{2} \beta $   gives
the yet  known result (\ref{7.8}):
\begin{eqnarray}
{(a_{1}n_{1} +a_{2}n_{2} +a_{3} n_{3}) + (a_{1} +n_{1}) V + V^{2}(
1 - a_{2}n_{2} -a_{3}n_{3}) \over 1+a_{1} V} +
\nonumber
\\
+ \; {(b_{1}n_{1} +b_{2}n_{2} +b_{3} n_{3}) + (b_{1} +n_{1}) V +
V^{2}( 1 - b_{2}n_{2} -b_{3}n_{3}) \over 1+b_{1} V} =0\; .
\label{9.12b}
\end{eqnarray}

Bearing in mind the previous analysis, eq.  (\ref{9.11}) can readily be changed to the form
\begin{eqnarray}
{ {\bf a}  {\bf n}  (1-V^{2})  +  ({\bf a} +  {\bf n} ){\bf V}
      + [\; V^{2} + ({\bf a} {\bf V}) ({\bf n} {\bf V})  \; ]
 \over
1  +  {\bf a} {\bf V}   }\; +
\nonumber
\\
+\; { {\bf b}  {\bf n}  +  ({\bf b} +  {\bf n} ){\bf V}
     + [\; V^{2} + ({\bf b} {\bf V}) ({\bf n} {\bf V})  \; ]
 \over
1 +  {\bf b} {\bf V}   } = 0 \; ;
\label{9.13}
\end{eqnarray}

\noindent eq.  (\ref{9.13}) is the result of converting eq. (\ref{9.10})
to moving reference frame.

Now let us give special attention to  one other  aspect of the problem under consideration.
In the starting (rest) reference frame $K'$ three vectors
${\bf a},{\bf n},{\bf b}$ belong to the same single plane, as a result one can derive a
decomposition of  ${\bf b}$ into a linear combination of two remaining vectors:
see Fig. 6.

\vspace{10mm}

\unitlength=0.6mm
\begin{picture}(100,50)(-95,0)
\special{em:linewidth 0.4pt} \linethickness{0.6pt}

\put(0,0){\circle*{3}}

\put(-30,+40){\vector(+3,-4){29}}  \put(+33,-42){${\bf a}'$}
\put(0.5,0.5){\vector(+3,+4){30}}  \put(-5,+50){${\bf n}'$}
\put(0,0.5){\vector(0,+1){50}}     \put(+35,+45){${\bf b}'$}

\put(-40,0){\line(+1,0){80}} \put(-40,-0.3){\line(+1,0){80}}
\put(-40,-0.6){\line(+1,0){80}} \put(0,0){\vector(+3,-4){29}}

\end{picture}

\vspace{25mm}

\begin{center}
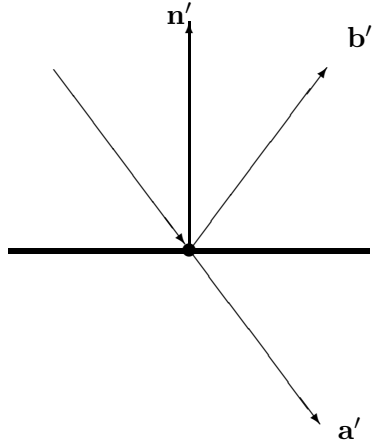
{\bf Fig. 6  The plane of incident, reflected, and normal  rays  }
\end{center}

Let it be
\begin{eqnarray}
{\bf b}' = \alpha \;  {\bf a}' + \nu  \; {\bf n}' .
\nonumber
\end{eqnarray}

\noindent
With the use of
\begin{eqnarray}
{\bf a}' \times {\bf n}'  = (\alpha \;  {\bf a}' + \nu  \; {\bf
n}' ) \times  {\bf n}' , \qquad \Longrightarrow \qquad \alpha = +1
\; ,
 \; {\bf b}' =  {\bf a}' + \nu  \; {\bf n}' .
\nonumber
\end{eqnarray}

\noindent and
\begin{eqnarray}
 ({\bf a}' + {\bf b}) '{\bf n} '= 0 \; ,
\qquad \Longrightarrow \qquad
(2 {\bf a}' +  \nu  {\bf n}' )\; {\bf n}' = 0 \; ,
\nonumber
\end{eqnarray}

\noindent we get to
\begin{eqnarray}
\nu = - 2 ({\bf a}' {\bf n}') = - 2 \cos \phi \; .
\nonumber
\end{eqnarray}

\begin{quotation}
{\em  Therefore, in the rest reference frame $K'$

the  reflection
law
can be formulated as follows

}
\end{quotation}
\begin{eqnarray}
{\bf b}' =  {\bf a}' -  2 ({\bf a}' {\bf n}')  \; {\bf n}' \; .
\label{9.14}
\end{eqnarray}

\vspace{5mm}
We are to convert this equation to the  moving reference frame $K$:
\begin{eqnarray}
{ {\bf b}  + {\bf e}  \;[\;  sh\; \beta   + (ch\; \beta -1)\;  (
{\bf e} \; {\bf b} )\; ] \over ch\; \beta  + sh\; \beta \; ({\bf
e}{\bf b})  }  =
 { {\bf a}  + {\bf e}  \; [ \; sh\; \beta   +
(ch\; \beta -1)\;  ( {\bf e} \; {\bf a} ) \; ]  \over ch\; \beta
+ sh\; \beta \; ({\bf e} {\bf a} ) }  -
\nonumber
\\
-  2 \left (  { {\bf a}  + {\bf e}  \; [ \; sh\; \beta   + (ch\;
\beta -1)\;  ( {\bf e} \; {\bf a} ) \; ]  \over ch\; \beta  + sh\;
\beta \; ({\bf e} {\bf a} ) }\;\; { {\bf n}  + {\bf e}  \;[\;
sh\; \beta   + (ch\; \beta -1)\;  ( {\bf e} \; {\bf n} ) \; ]
\over ch\; \beta  + sh\; \beta \; ({\bf e}{\bf n})  } \; \right )
\;
\nonumber
\\
(\times ) \; { {\bf n}  + {\bf e}  \;[\;  sh\; \beta   + (ch\;
\beta -1)\;  ( {\bf e} \; {\bf n} ) \; ] \over ch\; \beta  + sh\;
\beta \; ({\bf e}{\bf n})  } \; ;
\nonumber
\end{eqnarray}

\noindent or in a more short form
\begin{eqnarray}
{ {\bf b}  + {\bf e}  \;[\;  sh\; \beta   + (ch\; \beta -1)\;  (
{\bf e} \; {\bf b} ) \; ] \over ch\; \beta  + sh\; \beta \; ({\bf
e}{\bf b})  }  =
\nonumber
\\
=
 { {\bf a}  + {\bf e}  \; [ \; sh\; \beta   +
(ch\; \beta -1)\;  ( {\bf e} \; {\bf a} ) \; ]  \over ch\; \beta
+ sh\; \beta \; ({\bf e} {\bf a} ) }  + \nu
 \; { {\bf n}  + {\bf e}  \;[\;  sh\; \beta   +
(ch\; \beta -1)\;  ( {\bf e} \; {\bf n} ) \; ] \over ch\; \beta  +
sh\; \beta \; ({\bf e}{\bf n})  } \; .
\label{9.15b}
\end{eqnarray}

\vspace{5mm} It should be especially noted that the  general structure of the relation
obtained is
\begin{eqnarray}
{\bf b} = A \; {\bf a} +  B \; {\bf n} + C \;{\bf e} \; ;
\nonumber
\end{eqnarray}

\noindent we see that there appears  a constituent along ${\bf e}$ direction which means that
the vector ${\bf b}$ (reflected light ray) is not situated in the plane
of vectors ${\bf a}$ and ${\bf n}$.

In general, eq. (\ref{9.15b}) can be resolved with respect to
${\bf b}$. To this end,
let us form the dot product of eq. (\ref{9.15b}) and vector ${\bf e}$:
\begin{eqnarray}
{  sh\; \beta   + ch\; \beta  ( {\bf e} \; {\bf b} ) \over ch\;
\beta  + sh\; \beta \; ({\bf e}{\bf b})  }  =
 {  sh\; \beta   +
ch\; \beta  ( {\bf e} \; {\bf a} )   \over ch\; \beta  + sh\;
\beta \; ({\bf e} {\bf a} ) } \; + \;\nu \;
 {   sh\; \beta   +
ch\; \beta  ( {\bf e}  {\bf n} ) \; ] \over ch\; \beta  + sh\;
\beta  ({\bf e}{\bf n})  }  \equiv B \; ;
\label{9.16}
\end{eqnarray}

\noindent the right hand side of eq.  (\ref{9.16}) is designated by $B$.
From eq. (\ref{9.16})  it follows
\begin{eqnarray}
 sh\; \beta   +
ch\; \beta  ( {\bf e} \; {\bf b} )= B \; ch\; \beta  +B\;  sh\;
\beta \; ({\bf e}{\bf b}) \; , \qquad \Longrightarrow \qquad {\bf
e} {\bf b} = { B - th \; \beta \over  1 - B \; th\; \beta } \; .
\label{9.17}
\end{eqnarray}

\noindent The relation (\ref{9.16})  is in accordance with the identity
 $ B \equiv {\bf b}' {\bf e}$.
 Besides, taking into account eq. (\ref{9.16}), the previous relation (\ref{9.15b})  can be
 much simplified:
\begin{eqnarray}
{ {\bf b}  - {\bf e} \;  ( {\bf e} \; {\bf b} ) \over ch\; \beta
+ sh\; \beta \; ({\bf e}{\bf b})  }  =
 { {\bf a}  - {\bf e}  \;   ( {\bf e} \; {\bf a} )   \over
ch\; \beta  + sh\; \beta \; ({\bf e} {\bf a} ) }  + \nu  \; { {\bf
n}  - {\bf e}  \;  ( {\bf e} \; {\bf n} ) \; ] \over ch\; \beta  +
sh\; \beta \; ({\bf e}{\bf n})  }
\nonumber
\end{eqnarray}

\noindent or
\begin{eqnarray}
{ {\bf b}_{\bot}    \over 1 +  {\bf b}{\bf V}  }  =
 { {\bf a}_{\bot}     \over
1  +  {\bf a} {\bf V}  }  + \nu  \; { {\bf n}_{\bot}     \over 1
+  {\bf n}{\bf V}  } \; .
\label{9.18b}
\end{eqnarray}

It is quite understandable that
 (\ref{9.16}) and  (\ref{9.18b})  are the projections of eq.  (\ref{9.15b})
onto direction of  ${\bf e}$ and a plane  orthogonal to it.

In the end let us calculate an additional characteristic of the  reflection law,
as sketched in Fig. 7:
\begin{eqnarray}
\Delta' = {\bf n}' \; ({\bf a}' \times   {\bf b}' ) = 0 \; , \qquad \Longrightarrow \qquad
\Delta = {\bf n} \; ({\bf a} \times   {\bf b} ) \; .
\nonumber
\end{eqnarray}

\vspace{12mm}

\unitlength=0.2mm
\begin{picture}(100,50)(-190,0)
\special{em:linewidth 0.4pt} \linethickness{0.6pt}

\put(-20,+80){${\bf n}$}  \put(+60,-60){${\bf a}$}  \put(+85,+5){${\bf b}$}

\put(0,0){\vector(0,+1){90}}
\put(0,0){\vector(+3,-2){60}} \put(0,90){\line(+3,-2){60}}
\put(0,0){\vector(+3,+1){90}} \put(0,90){\line(+3,+1){90}}
\put(60,-40){\line(+3,+1){90}}
\put(60,+50){\line(+3,+1){90}}   \put(60,+50){\line(0,-1){90}}

\put(150,+80){\line(0,-1){90}}

\put(90,+30){\line(+3,-2){60}}  \put(90,+120){\line(+3,-2){60}} \put(90,+120){\line(0,-1){90}}

\end{picture}

\vspace{7mm}

\begin{center}
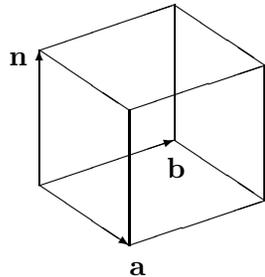
{\bf Fig 7.  The parallelepiped  of incident, normal, reflected rays}
\end{center}

With the Lorentz formulas
\begin{eqnarray}
{\bf a} = { {\bf a}'  + {\bf e}  \; [ \; -sh\; \beta   + (ch\;
\beta -1)\;  ( {\bf e} \; {\bf a}' ) \; ]  \over ch\; \beta  -
sh\; \beta \; {\bf e} {\bf a}'  } \; ,
\nonumber
\\
{\bf b} = { {\bf b}'  + {\bf e}  \;[\; - sh\; \beta   + (ch\;
\beta -1)\;  ( {\bf e} \; {\bf b}' )\; ] \over ch\; \beta  - sh\;
\beta \; {\bf e}{\bf b}'  } \; ,
\nonumber
\\
{\bf n} = { {\bf n}  + {\bf e}  \;[\; - sh\; \beta   + (ch\; \beta
-1)\;  ( {\bf e} \; {\bf n}' ) \; ] \over ch\; \beta  - sh\; \beta
\; {\bf e}{\bf n}'  } \;
\nonumber
\end{eqnarray}

\noindent we get
\begin{eqnarray}
\Delta = { {\bf n}'  + {\bf e}  \;[\; - sh\; \beta   + (ch\; \beta
-1)\;  ( {\bf e} \; {\bf n}' ) \; ] \over ch\; \beta  - sh\; \beta
\; {\bf e}{\bf n}'  }  \bullet
\nonumber
\\
\bullet \; \left ( { {\bf a}'  + {\bf e}  \; [ \; -sh\; \beta   +
(ch\; \beta -1)\;  ( {\bf e} \; {\bf a}' ) \; ]  \over ch\; \beta
- sh\; \beta \; {\bf e} {\bf a}'  } \times { {\bf b}'  + {\bf e}
\;[\; - sh\; \beta   + (ch\; \beta -1)\;  ( {\bf e} \; {\bf b}'
)\; ] \over ch\; \beta  - sh\; \beta \; {\bf e}{\bf b}'  } \right
) =
\nonumber
\\
= { {\bf n}'  + {\bf e}  \;[\; - sh\; \beta   + (ch\; \beta -1)\;
( {\bf e} \; {\bf n}' ) \; ] \over ch\; \beta  - sh\; \beta \;
{\bf e}{\bf n}'  }  \bullet
\nonumber
\\
\bullet {{\bf a} ' \times  {\bf b}' + {\bf a}' \times {\bf e}
\;[\; - sh\; \beta   + (ch\; \beta -1)\;  ( {\bf e} \; {\bf b}'
)\; ] + {\bf e} \times {\bf b}'  \; [ \; -sh\; \beta   + (ch\;
\beta -1)\;  ( {\bf e} \; {\bf a}' ) \; ] \over (ch\; \beta  -
sh\; \beta \; {\bf e} {\bf a}'  ) (ch\; \beta  - sh\; \beta \;
{\bf e}{\bf b}') } \; ,
\nonumber
\end{eqnarray}

\noindent and further
\begin{eqnarray}
\Delta = {1 \over (ch\; \beta  - sh\; \beta \; {\bf e}{\bf n}' )
\; (ch\; \beta  - sh\; \beta \; {\bf e}{\bf a}')
 (ch\; \beta  - sh\; \beta \; {\bf e}{\bf b}') }
\nonumber
\\
\{ \;\; [\; {\bf n}' ({\bf a}' \times {\bf e} ) \;] \; [\; - sh
\beta   + (ch \beta -1)  \; {\bf e}  {\bf b}'  \;]  \; +
\nonumber
\\
+\; [\; {\bf n}' ( {\bf e} \times {\bf b}' ) \;]  \;  [ \; -sh
\beta   + (ch \beta -1) \;  {\bf e}  {\bf a}'\;  ] +
\nonumber
\\
+ \; [\; {\bf e} ({\bf a}' \times {\bf b}') \; ]  \; [ - sh \beta
+ (ch \beta -1) \; {\bf e}  {\bf n}'   \; ] \; \} \; .
\label{9.20}
\end{eqnarray}

\noindent
Bearing in mind eq. (\ref{9.14})
\begin{eqnarray}
{\bf b}' =  {\bf a}' -  2 ({\bf a}' {\bf n}')  \; {\bf n}' = {\bf
a}' + \nu \; {\bf n}'\; ,
\nonumber
\end{eqnarray}

\noindent eq. (\ref{9.20}) reads
\begin{eqnarray}
\Delta = {1 \over (ch\; \beta  - sh\; \beta \; {\bf e}{\bf n}' )
\; (ch\; \beta  - sh\; \beta \; {\bf e}{\bf a}')\;
 (ch\; \beta  - sh\; \beta \; {\bf e}{\bf b}') }\; (\times)
\nonumber
\\
\{ \;\; [\; {\bf n}' ({\bf a}' \times {\bf e} ) \;] \; [\; - sh
\beta   + (ch \beta -1)  ( {\bf e}  {\bf a}' + \nu \;  {\bf e}
{\bf n}' ) \;]  \; +
\nonumber
\\
+\; [\; {\bf n}' ( {\bf e} \times {\bf a}' ) \;]  \;  [ \; -sh
\beta   + (ch \beta -1)  ( {\bf e}  {\bf a}' )  ] +
\nonumber
\\
+ \; \nu \; [\; {\bf e} ({\bf a}' \times {\bf n}') \; ]  \; [ - sh
\beta   + (ch \beta -1)  ( {\bf e}  {\bf n}' )  \; ] \; \} \; .
\label{9.21}
\end{eqnarray}

From this,  with the help of the known  symmetry properties of the mixed vector
product,  we produce
\begin{eqnarray}
\Delta = { [\; {\bf e} ({\bf n}' \times {\bf a}' ) \;] \over (ch\;
\beta  - sh\; \beta \; {\bf e}{\bf n}' ) \; (ch\; \beta  - sh\;
\beta \; {\bf e}{\bf a}')\;
 (ch\; \beta  - sh\; \beta \; {\bf e}{\bf b}') } \; (\times)
\nonumber
\\
\;[ \;   - sh \beta   + (ch \beta -1)  ( {\bf e}  {\bf a}' + \nu \;
{\bf e} {\bf n}' ) + sh \beta   - (ch \beta -1) \;  {\bf e}  {\bf
a}' +  \nu  sh \beta   - \nu (ch \beta -1) \;  {\bf e}  {\bf n}'
\; ] =
\nonumber
\\
= {  \nu \; sh\; \beta  \;\; [\; {\bf e} ({\bf n}' \times {\bf a}'
) \;] \over (ch\; \beta  - sh\; \beta \; {\bf e}{\bf n}' ) \;
(ch\; \beta  - sh\; \beta \; {\bf e}{\bf a}')\;
 (ch\; \beta  - sh\; \beta \; {\bf e}{\bf b}') }.
\nonumber
\end{eqnarray}

So, finally we have obtained  the result
\begin{eqnarray}
\Delta = {\bf n} \; ({\bf a} \times   {\bf b} ) =
\nonumber
\\
=  {  - 2 \;sh\;
\beta  \; ({\bf a}' {\bf n}') \; [\; {\bf e} ({\bf n}' \times {\bf a}' ) \;] \;
\over (ch\; \beta  - sh\; \beta \;
{\bf e}{\bf n}' ) \; (ch\; \beta  - sh\; \beta \; {\bf e}{\bf
a}')\;
 (ch\; \beta  - sh\; \beta \; {\bf e}{\bf b}') } \; .
\label{9.22}
\end{eqnarray}

\begin{quotation}

{\em This relation gives the volumes of the light parallelepiped

$$[\; {\bf n} \; ({\bf a} \times   {\bf b} )\; ] $$

\noindent in the moving reference frame
as a function of the $ (\beta, {\bf e} ;\; {\bf n}' , \;{\bf a}' ) \;$.
We must note that in contrast to the ordinary reflection law  (when
the  expression $ A {\bf a}' + B  {\bf n} '+C {\bf b}'$ in fact presents
a  2-dimensional object -- plane) now
in the moving reference  frame $K$ we  will   have in general   a 3-space object
$ A {\bf a} + B  {\bf n} +C {\bf b}$ .
}

\end{quotation}

Now, one other aspect of the problem under consideration should be discussed.
The  obtained form of the light reflection law in the  moving reference frame
depends, strictly  speaking, on  the occasional choice of the direction of the normal
light  motion:  namely, from the surface. However, one might chose opposite direction, that is the light
ray going to the surface.

The matter is that the simple relation between two corresponding vectors
\begin{eqnarray}
{\bf N} '= - \; {\bf n}' \; ,
\label{9.23}
\end{eqnarray}

\noindent will not preserve its form with respect to Lorentz transformation.
Indeed,  in the moving reference frame we will have
\begin{eqnarray}
{\bf n} =
{ {\bf n}'  + {\bf e}  \;[\;-  sh\; \beta   +
(ch\; \beta -1)\;  ( {\bf e} \; {\bf n}' ) \; ] \over
ch\; \beta  - sh\; \beta \; ({\bf e}{\bf n}')  } \; ,
\nonumber
\\
{\bf N} =
{ -{\bf n} ' + {\bf e}  \;[\;-  sh\; \beta   -
(ch\; \beta -1)\;  ( {\bf e} \; {\bf n}' ) \; ] \over
ch\; \beta  + sh\; \beta \; ({\bf e}{\bf n}')  } \; .
\label{9.24b}
\end{eqnarray}

\noindent Two derivative characteristics  can be calculated:
\begin{eqnarray}
{\bf n} \times {\bf N} = { 2 sh\; \beta  \over
ch^{2}\; \beta  - sh^{2}\; \beta \; ({\bf e}{\bf n}')^{2}  } \; {\bf e} \times {\bf n} \; ,
\qquad
{\bf n} \bullet  {\bf N} = 1 - {2 \over ch^{2}\;\beta - sh^{2}\beta ({\bf e} {\bf d})^{2} } \; .
\nonumber
\end{eqnarray}

\noindent As it must be expected
when  $\beta=0$  the equations  give
\begin{eqnarray}
\beta=0 \qquad \Longrightarrow \qquad
{\bf n} \times {\bf N} =0\; , \;\; {\bf n} \bullet {\bf N} = -1 \; .
\nonumber
\end{eqnarray}

\noindent
It is readily verified the identity:
\begin{eqnarray}
({\bf n} \times {\bf N} ) \bullet( {\bf n} \times {\bf N}) = 1 - ({\bf n}
\bullet  {\bf N})^{2}\;.
\nonumber
\end{eqnarray}

\hspace{5mm}
If, instead of the vector ${\bf n}$,  one uses in formulating the light reflection law
in the rest reference frame the opposite vector
 ${\bf N}$
\begin{eqnarray}
 {\bf a} '\times {\bf N}' =   {\bf b}' \times {\bf N}' \; , \qquad
  {\bf a}  \bullet {\bf N}  + {\bf b} \bullet  {\bf N} =0 \; ,
\label{9.27a}
\end{eqnarray}

\noindent then further in calculation no serious  change will not appear:
everywhere instead of ${\bf n}$  one will write another symbol ${\bf N}$.
So that instead of (\ref{9.5})  and (\ref{9.13}) one has
\begin{eqnarray}
{ch\; \beta  \;  ({\bf a} \times {\bf N} )  +
 sh\; \beta  \; ( {\bf a} -{\bf N}  )\times{\bf e}
+  (ch\; \beta -1)\;  {\bf e} \; [ \;{\bf e}  ({\bf N} \times {\bf a})\; ]
    \over
ch\; \beta  + sh\; \beta \; ({\bf e} {\bf a})  }=
\nonumber
\\
=
{ch\; \beta  \;  ({\bf b} \times {\bf N})  +
 sh\; \beta ({\bf b} -{\bf N})  \times {\bf e}
  +   (ch\; \beta -1)\; {\bf e} \; [ \;{\bf e}  ({\bf N} \times {\bf b})\; ]
    \over
ch\; \beta  + sh\; \beta \; ({\bf e} {\bf b})  } \; ,
\label{9.27b}
\\
{ {\bf a}  {\bf N}  (1-V^{2})  +  ({\bf a} +  {\bf N} ){\bf V}
      + [\; V^{2} + ({\bf a} {\bf V}) ({\bf N} {\bf V})  \; ]
 \over
1  +  {\bf a} {\bf V}   }\;
+
\nonumber
\\
+\;
{ {\bf b}  {\bf N}  +  ({\bf b} +  {\bf N} ){\bf V}
     + [\; V^{2} + ({\bf b} {\bf V}) ({\bf N} {\bf V})  \; ]
 \over
1 +  {\bf b} {\bf V}   } = 0 \; .
\label{9.27с}
\end{eqnarray}

\noindent
In turn, instead of (\ref{9.22}) one  derives
\begin{eqnarray}
[{\bf N} \; ({\bf a} \times   {\bf b} ) ]= \hspace{40mm}
\nonumber
\\
= [\; {\bf e} ({\bf N}' \times {\bf a}' ) \;] \;
{  - 2 \;sh\; \beta  \; ({\bf a}' {\bf N}')
\over (ch\; \beta  - sh\; \beta \; {\bf e}{\bf N}' ) \;
(ch\; \beta  - sh\; \beta \; {\bf e}{\bf a}')\;
 (ch\; \beta  - sh\; \beta \; {\bf e}{\bf b}') }=
\nonumber
\\
=
[\; {\bf e} ({\bf n}' \times {\bf a}' ) \;] \;
{  - 2 \;sh\; \beta  \; ({\bf a}' {\bf n}')
\over (ch\; \beta  + sh\; \beta \; {\bf e}{\bf n}' ) \;
(ch\; \beta  - sh\; \beta \; {\bf e}{\bf a}')\;
 (ch\; \beta  - sh\; \beta \; {\bf e}{\bf b}') }\; .
\label{9.28b}
\end{eqnarray}

\begin{quotation}

Existence of the two substantially different normal vectors in the  moving reference frame
makes us to  consider the formulas obtained for the light reflection in the
 moving reference frame as ones  describing the relativistic aberration effect
 for  two different    triples of light velocity vectors:
${\bf a},{\bf b},{\bf n}$ and ${\bf a},{\bf b},{\bf N}$.
One may expect to reach more clarity  in looking at the geometrical properties
of the reflecting surfaces with respect to relativistic motion of the reference frame.

\end{quotation}

\section{On  the form of reflection surface in the
moving reference frame
}

\hspace{5mm}
In the  rest reference frame $K'$, the reflection surface is a plane.
Let  ${\bf d} $ be its (unit) normal vector, the the known form of the  plane
is given by (here  $D$ is the distance from the origin $O$ to the plane)
\begin{eqnarray}
S': \qquad {\bf x} ' \; {\bf d}' + D =0 \; ,
\label{10.1}
\end{eqnarray}

\vspace{-1mm}

\unitlength=0.6mm
\begin{picture}(100,50)(-50,0)
\special{em:linewidth 0.4pt} \linethickness{0.6pt}

\put(0,0){\circle*{3}}

\put(-4,+2){$O$}

\put(0,-30){\line(+1,+2){40}}
\put(0,-31){\line(+1,+2){40}}
\put(0,-30.5){\line(+1,+2){40}}

\put(0,0){\vector(+1,1){30}}
\put(30,30){\vector(-2,+1){7}}  \put(22,35){${\bf d}'$}
 \put(0,0){\line(+2,-1){12}}    \put(20,25){${\bf x}'$}
                                \put(3,-9){$D$}
\put(+60,+10){$ {\bf x} ' \; {\bf d}' = -\mid x ' \mid \sin \alpha_{i}' = - D $}

\end{picture}

\vspace{14mm}

\begin{center}
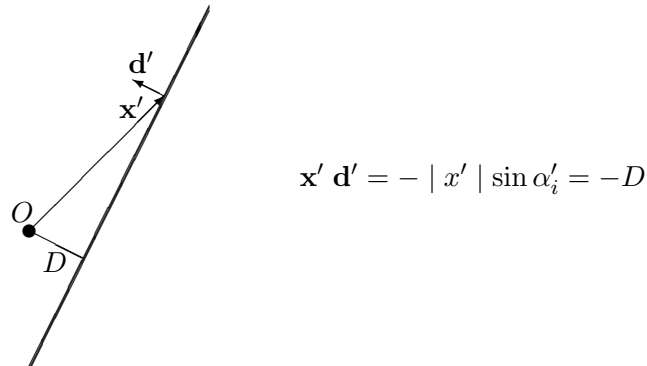
{\bf Fig. 8:  The reflection plane in the rest reference frame $K'$}
\end{center}

Now we  should associate with this  plane $S'$, purely geometrical object,  a
set of events in space-time:
\begin{eqnarray}
S': \qquad \mbox{the plane}\;({\bf d},D)  \qquad \Longrightarrow \qquad
\{ \; (t', {\bf x}' = (x'_{1},x_{2}',x_{3}')  \; \} \; .
\nonumber
\end{eqnarray}

\noindent To this end, at the moment $t'=0$ let light signals  be sent at all directions
to the plane  -- the arriving of any particular  signal to the plane gives rise to an event:
\begin{eqnarray}
 {\bf x}' = {\bf W}' \; t', \qquad \Longrightarrow \qquad
 {\bf x}^{'2} = {\bf W}^{'2}  \; t^{'2} , \qquad
t' = \sqrt{ {\bf x}^{'2}  } \; .
\label{10.2}
\end{eqnarray}

\noindent
Therefore, to the reflection plane  corresponds the following set of space-time events:
\begin{eqnarray}
S': \qquad \Longrightarrow \qquad \left \{ \;\; t' = \sqrt{ {\bf x}^{'2}  } \; , \;
 {\bf x} ' \; {\bf d}' + D' =0 \; \;  \right \} \; .
\label{10.3}
\end{eqnarray}

\noindent
The same  set of event can be observed in the moving
reference frame:
\begin{eqnarray}
{\bf x} = {\bf W} \; t, \qquad \Longrightarrow \qquad
 {\bf x}^{2} = {\bf W}^{2}  \; t^{2} , \qquad
t = \sqrt{ {\bf x}^{2}  } \; .
\label{10.4}
\end{eqnarray}

As a result, the above equation of the plane  (\ref{10.3}) will take a new form
(take notice that here the Lorentz transform acts only onto ${\bf x}'$ whereas
the two quantities   ${\bf d}$  and  $D'$ are considered as  given parameters -- below we will
omit the {\em prime}  symbol  $'$ at them)
\begin{eqnarray}
\left \{ \;\; t = \sqrt{ {\bf x}^{2}  } \; , \;
 [\;  {\bf x}  + {\bf e}  \;(\;  sh\; \beta  \; t +
(ch\; \beta -1)\;  ({\bf e}  {\bf x})\; ) \;
] \; {\bf d} + D =0 \; \;  \right \} \; ,
\label{10.5}
\end{eqnarray}

To obtain an equation for geometrical surface in the reference  frame $K$
one should exclude the variable $t$ with the help of the first relation in (\ref{10.5}):
\begin{eqnarray}
 S: \qquad  ({\bf x} {\bf d})   + ( {\bf e} {\bf d} ) \;
 \left [ \;  sh\; \beta  \; (\; \sqrt{ {\bf x}^{2}  } \; ) +
(ch\; \beta -1)\;  ({\bf e}  {\bf x})\;  \right ]    + D =0 \; .
\label{10.6}
\end{eqnarray}

\begin{quotation}
{\em
This equation describes the geometry of the reflection surface $S$ in the moving reference frame $K$.
In the rest reference frame $K'$ (when  $\beta =0$) eq. (\ref{10.6}) becomes an
equation of the  plane.

To avoid misunderstanding it should be noted one special case: when vector ${\bf e}$ is orthogonal
to the normal vector ${\bf d}$ we have the identity  $( {\bf e} {\bf d} ) =0 $  which means that
in this case the plane in the rest reference frame  does not change its geometrical shape
in the  moving reference frame:
\begin{eqnarray}
 S: \qquad  ({\bf x} {\bf d})      + D =0 \; .
\label{10.7}
\nonumber
\end{eqnarray}
 }
\end{quotation}

Let us check our calculation by seing form of this general result (\ref{10.6})  for a
 particular case
\begin{eqnarray}
{\bf x} = (x,y,0) \;, \qquad {\bf d} = ( \cos \alpha  , \sin \alpha ,0 )\; , \qquad
{\bf e} = (1,0,0) \; ,
\label{10.8a}
\end{eqnarray}

\noindent  at this eq.  (\ref{10.6}) gives
\begin{eqnarray}
 x \cos \alpha + y \sin \alpha + \cos \alpha \; [\;  sh \; \beta  \sqrt{x^{2} + y^{2}} +
(ch\; \beta -1) x\; ]  + D =0 \; ,
\nonumber
\end{eqnarray}

\noindent and  further
\begin{eqnarray}
 y \sin \alpha + \cos \alpha ( sh \; \beta  \sqrt{x^{2} + y^{2}} +
ch\; \beta \;  x\; )   + D =0 \; .
\nonumber
\end{eqnarray}

\noindent
Dividing it by  $\sin \alpha$, after simple regrouping all the terms we will
obtain
\begin{eqnarray}
 y  = -  {D \over  \sin \alpha }  - {1 \over \tan  \alpha }  \;
 \left [\;   ch\; \beta  + sh \; \beta  \sqrt{1 + y^{2} /  x^{2} } +
\;   \right ] \; x     \; ,
\label{10.9a}
\end{eqnarray}

\noindent
what coincides with the previously found equation  in Section 5:
\begin{eqnarray}
y = b + k \;  \left [ \;
ch  \; \beta  + sh\;  \beta \; \sqrt{1 +  y^{2} / x^{2}}\;\; \right   ] \; x \; .
\nonumber
\end{eqnarray}

\vspace{5mm}
It may be easily verified that eq. (\ref{10.6}) corresponds to a second order surface.
Indeed,
\begin{eqnarray}
-( {\bf e} {\bf d} ) \; sh\; \beta  \; (\; \sqrt{ {\bf x}^{2}  } \; ) =
  ({\bf x} {\bf d})   + ( {\bf e} {\bf d} )  ({\bf e}  {\bf x})\;
(ch\; \beta -1)  + D \; ,
\nonumber
\end{eqnarray}

 \noindent  and further we get to
\begin{eqnarray}
( {\bf e} {\bf d} )^{2}   sh^{2}  \beta  \;  {\bf x}^{2}  =
[  ({\bf x} {\bf d})   + ( {\bf e} {\bf d} )  ({\bf e}  {\bf x})\;
(ch\; \beta -1)  + D  ]^{2} \; .
\label{10.10}
\end{eqnarray}

\section{Canonical form of the reflection surface $S$ in the $K$-frame}

\hspace{5mm}
Now we are to perform analysis extending the Section 5:
with the use of a special rotation in 3-space and then a special
shift in 3-space the equation of the reflection surface $S$
\begin{eqnarray}
( {\bf e} {\bf d} )^{2}   sh^{2}  \beta  \;  {\bf x}^{2}  =
[  ({\bf x} {\bf d})   + ( {\bf e} {\bf d} )  ({\bf e}  {\bf x})\;
(ch\; \beta -1)  + D  ]^{2} \;
\label{11.1}
\end{eqnarray}

\noindent is to be taken to a canonical form (supposedly it is a hyperboloid).

We need some fact on rotations in 3-space. An arbitrary rotation matrix can be given as follows
(more details  see in [86]):
\begin{eqnarray}
\left. \begin{array}{lll}
O_{1}^{\;\;1} = 1 -2 (c_{2}^{2} + c_{3}^{2}) \; ,  & \qquad
O_{1}^{\;\;2} = -2c_{0}c_{3} + 2c_{1}c_{2}  \; ,       & \qquad
O_{1}^{\;\;3} = +2c_{0}c_{2} + 2c_{1}c_{3}  \; ,       \\[2mm]
O_{2}^{\;\;1} = +2c_{0}c_{3} + 2c_{1}c_{2}  \; ,       & \qquad
O_{2}^{\;\;2} = 1 -2 (c_{3}^{2} + c_{1}^{2}) \; ,  & \qquad
O_{2}^{\;\;3} = -2c_{0}c_{1} + 2c_{2}c_{3}  \; ,       \\[2mm]
O_{3}^{\;\;1} = -2c_{0}c_{2} + 2c_{1}c_{3}  \; ,       &\qquad
O_{3}^{\;\;2} = +2c_{0}c_{1} + 2c_{2}c_{3}  \; ,       &\qquad
O_{3}^{\;\;3} = 1 -2 (c_{1}^{2} + c_{2}^{2}) \; ,
\end{array} \right.
\label{11.2a}
\end{eqnarray}

\noindent
where parameters $b_{a}$ obey the condition
\begin{eqnarray}
c_{0}^{2}  + c_{1}^{2} + c_{2}^{2} + c_{3}^{2} = +1  \; .
\nonumber
\end{eqnarray}

\noindent
The rotation matrix $O$ may be presented in the form more convenient at further calculation:
\begin{eqnarray}
O = \left| \begin{array}{ccc}
 1   &  0  &  0       \\[2mm]
 0 & 1  & 0   \\[2mm]
0      &  0      & 1 \end{array} \right | +
2c_{0} \; \left | \begin{array}{lll}
 0    &   - c_{3}   & c_{2}        \\[2mm]
 c_{3}        & 0  &  -c_{1}     \\[2mm]
 -c_{2}      &  c_{1}         &\qquad 0
\end{array} \right | +
 \left | \begin{array}{ccc}
  -2 (c_{2}^{2} + c_{3}^{2})    &   2c_{1}c_{2}    &   2c_{1}c_{3}         \\[2mm]
  2c_{1}c_{2}        & -2 (c_{3}^{2} + c_{1}^{2}) &   2c_{2}c_{3}     \\[2mm]
  2c_{1}c_{3}       &  2c_{2}c_{3}         &   -2 (c_{1}^{2} + c_{2}^{2}) \; .
\end{array} \right |
\nonumber
\end{eqnarray}

\noindent
With the notation
\begin{eqnarray}
{\bf c}^{\times}  =
\left | \begin{array}{lll}
 0    & \qquad  - c_{3}   & \qquad  c_{2}        \\[2mm]
 c_{3}        & \qquad 0  & \qquad  -c_{1}     \\[2mm]
 -c_{2}      &\qquad  c_{1}         &\qquad 0
\end{array} \right | \; ,
\nonumber
\end{eqnarray}

\noindent the matrix $O$  reads as
\begin{eqnarray}
O = I  + 2 c_{0} \; {\bf c}^{\times} +  2({\bf c}^{\times})^{2} \; .
\label{11.3b}
\end{eqnarray}

\noindent
The variable $c_{0}$ can be excluded:
\begin{eqnarray}
{\bf C} = {{\bf c} \over c_{0} } , \qquad
O = I  + 2 \; {  {\bf C}^{\times} +  ({\bf C}^{\times})^{2} \over 1 + {\bf C}^{2}} \; ,
\label{11.3c}
\end{eqnarray}

\noindent
eq. (\ref{11.3c}) provides us with the known formula by Gibbs that is  very helpful
in practical calculation\footnote{More details see in [.....].}; in particular,
the composition rule has the form
\begin{eqnarray}
O({\bf C}') \;O({\bf C}) = O({\bf C}'') , \qquad
{\bf C}'' = { {\bf C}' + {\bf C} + {\bf C}' \times {\bf C} \over 1 -
{\bf C}'{\bf C}  } \; .
\label{11.3d}
\end{eqnarray}

It is readily verified the identity
\begin{eqnarray}
{\bf c}^{\times} \; {\bf A} = {\bf c} \times {\bf A} \; ,
\nonumber
\end{eqnarray}

\noindent so that the matrix $O$ acts on a vector in accordance with the rule
\begin{eqnarray}
O ({\bf c}) {\bf A} =   {\bf A} + 2 c_{0} \; {\bf c} \times {\bf A}  +2
{\bf c} \times ({\bf c} \times {\bf A} )\; .
\label{11.4}
\end{eqnarray}

\noindent
Instead of four parameters  $c_{a}$  one can use an angular variable
 $\phi/2 $  and a unit vector ${\bf o})$:
\begin{eqnarray}
c_{0} = \cos {\phi\over 2} \; , \qquad {\bf c} = \sin {\phi \over 2} \; \;{\bf o} ,
\qquad {\bf o}^{2} = 1 \; ,
\label{11.5a}
\end{eqnarray}

\noindent then eq. (\ref{11.4}) looks
\begin{eqnarray}
O \; {\bf A} =   {\bf A} + \sin \phi \; ( {\bf o} \times {\bf A} )  +(1 - \cos \phi ) \;
{\bf o} \times ({\bf o} \times {\bf A} )\; .
\label{11.5b}
\end{eqnarray}

With the known relation
\begin{eqnarray}
{\bf o} \times ({\bf o} \times {\bf A} ) =
 {\bf o} \; ({\bf o} {\bf A}) - {\b A} \; ,
\nonumber
\end{eqnarray}

\noindent
eq. (\ref{11.5b}) becomes
\begin{eqnarray}
O  (\phi,{\bf o}) \; {\bf A} =   {\bf A} +  \sin \phi \;( {\bf o} \times {\bf A})  +
(1 - \cos \phi) \; [  {\bf o} \; ({\bf o} {\bf A}) - {\bf A} ]\; .
\nonumber
\end{eqnarray}

\noindent
Thus,
\vspace{5mm}

{\em  the final formula for an arbitrary 3-rotation
is as follows}:
\begin{eqnarray}
O (\phi,{\bf o}) \;  {\bf A} =   \cos \phi \; {\bf A} +  \sin \phi \; ({\bf o} \times {\bf A} ) +
(1 - \cos \phi) \; ({\bf o} {\bf A})\; {\bf o} \; \;  .
\label{11.6}
\end{eqnarray}

{\em
in the right side we have a linear decomposition of the vector in terms of  those}
\begin{eqnarray}
 {\bf o} \; , \qquad  {\bf A}, \qquad   ( {\bf A} \times {\bf o}) \; .
\nonumber
\end{eqnarray}

There are three simple cases:
\begin{eqnarray}
\underline{ \phi, \; {\bf o}  =(1,0,0) \;}   \qquad \qquad
O {\bf A} =
\nonumber
\\
= \cos \phi
\left | \begin{array}{c}
A_{1} \\A_{2}\\  A_{3}
\end{array} \right |
+ \sin \phi
\left | \begin{array}{c}
0 \\ - A_{3}\\ + A_{2}
\end{array} \right | + (1 - \cos \phi ) \left | \begin{array}{c}
A_{1} \\ 0\\  0
\end{array} \right | =
\left | \begin{array}{ccc}
1 & 0 & 0 \\
0 & \cos \phi &  -\sin \phi \\
0 & \sin \phi & \cos \phi
\end{array} \right |  \left | \begin{array}{c}
A_{1} \\A_{2}\\  A_{3}
\end{array} \right | \; ,
\nonumber
\\
\underline{\phi, \; {\bf o}  =(0,1,0) \;}   \qquad \qquad
O {\bf A} =
\nonumber
\\
= \cos \phi
\left | \begin{array}{c}
A_{1} \\A_{2}\\  A_{3}
\end{array} \right |
+ \sin \phi
\left | \begin{array}{c}
 A_{3}  \\ 0 \\ - A_{1}
\end{array} \right | + (1 - \cos \phi ) \left | \begin{array}{c}
0 \\ A_{2} \\  0
\end{array} \right | =
\left | \begin{array}{ccc}
\cos \phi  & 0 & \sin \phi  \\
0 & 1 &   0  \\
-\sin \phi  & 0 & \cos \phi
\end{array} \right |  \left | \begin{array}{c}
A_{1} \\A_{2}\\  A_{3}
\end{array} \right | \; ,
\nonumber
\\
\underline{\phi, \; {\bf o}  =(0,0,1) \;}   \qquad \qquad
O {\bf A} =
\nonumber
\\
= \cos \phi
\left | \begin{array}{c}
A_{1} \\A_{2}\\  A_{3}
\end{array} \right |
+ \sin \phi
\left | \begin{array}{c}
 - A_{2} \\ + A_{1}\\ 0
\end{array} \right | + (1 - \cos \phi ) \left | \begin{array}{c}
0  \\ 0\\  A_{3}
\end{array} \right | =
\left | \begin{array}{ccc}
\cos \phi  & - \sin \phi  & 0 \\
\sin \phi & \cos \phi  &  0 \\
0 &  0 & 1 \end{array} \right |  \left | \begin{array}{c}
A_{1} \\A_{2}\\  A_{3}
\end{array} \right | \; .
\nonumber
\end{eqnarray}

We  return to the equation  (\ref{11.1}) now  written in the form
\begin{eqnarray}
{\bf x}^{2} = ( {\bf x} \bullet  {\bf  f} + F)^{2} ,
\label{11.8a}
\end{eqnarray}

\noindent  where
\begin{eqnarray}
{\bf f} = { {\bf d} + (ch\; \beta -1) ({\bf e} {\bf d}) {\bf e} \over
({\bf e} {\bf d}) \; sh \; \beta } \; , \qquad F = {D \over
({\bf e} {\bf d}) \; sh \; \beta } \; .
\nonumber
\end{eqnarray}

Let us introduce  new (rotated) variables  ${\bf X}'$:
\begin{eqnarray}
{\bf X} = O\; {\bf x} \; , \; \qquad  {\bf x} = O^{-1}\; {\bf X} \; ,
\label{11.9a}
\end{eqnarray}

\noindent  parameters  $c_{a}$   will be determined below. Take notice that any
rotation matrix obeys the so called   orthogonality condition:
$ O^{-1} = \tilde{O} \; .
$
In the  new variables,  eq.  (\ref{11.8a}) looks as
\begin{eqnarray}
(O^{-1} {\bf X})^{2} = ( O^{-1} {\bf X}  \bullet {\bf  f} + F)^{2} ,
\nonumber
\end{eqnarray}

\noindent  or
\begin{eqnarray}
({\bf X})^{2} = ( {\bf X}  \bullet O {\bf  f} + F)^{2} .
\label{11.9c}
\end{eqnarray}

It is the point to determine a rotation  needed: let the identity
hold
\begin{eqnarray}
O(c) \;  {\bf f} = (0,0,f) \;  , \qquad f>0 \; ;
\label{11.10}
\end{eqnarray}

\noindent  such a rotation   $O(c)$  provides us with a new Cartesian coordinate system
$(X,Y,Z)$ for which the axis $Z$ is located  along the vector ${\bf f}$. An explicit
 form of that rotation can be found quite easily\footnote{Some details see below.}
 As a result, eq. (\ref{11.9c})  reads
\begin{eqnarray}
X^{2} + Y^{2} +Z^{2} = Z^{2} f^{2} + 2ZFf + F^{2} \; ;
\label{11.11a}
\end{eqnarray}

\noindent or after simple calculation
\begin{eqnarray}
X^{2} + Y^{2} + (1 - f^{2}) \; \left  [
Z^{2}  - 2 Z {Ff \over  1 - f^{2} }
+  {F^{2} f ^{2} \over  (1 - f^{2})^{2}  }  \; \right ]
 =  F^{2}  + {F^{2} f ^{2} \over  1 - f^{2} } \; ,
\nonumber
\end{eqnarray}

\noindent and further
\begin{eqnarray}
X^{2} + Y^{2} + (1 - f^{2}) \; \left  [
Z  - {Ff \over  1 - f^{2} }
\; \right ] ^{2}
 =   {F^{2}  \over  1 - f^{2} } \; .
\nonumber
\end{eqnarray}

So we have arrived at the equation for reflection surface in the  moving
reference frame:
\begin{eqnarray}
{1 - f^{2} \over F^{2} } \; X^{2} + {1 - f^{2} \over F^{2} } \; Y^{2} +
{(1 - f^{2})^{2} \over F^{2}}  \; \left  [
Z  - {Ff \over  1 - f^{2} }
\; \right ] ^{2}
 =  1  \; ;
\label{11.11b}
\end{eqnarray}

\noindent It remains to add some details for  $1-f^{2}$. With the help of:
\begin{eqnarray}
{\bf f} = { {\bf d} + (ch\; \beta -1) ({\bf e} {\bf d}) {\bf e} \over
({\bf e} {\bf d}) \; sh \; \beta } \; , \qquad
1-f^{2} = 1-
{ 1    +2 (ch\; \beta -1) ({\bf e} {\bf d})^{2}    + (ch\; \beta -1)^{2} ({\bf e} {\bf d})^{2}\over
({\bf e} {\bf d}) ^{2} \; sh ^{2} \; \beta } = \nonumber
\\
= {1 \over  ({\bf e} {\bf d}) ^{2} \; sh ^{2} \; \beta  } \;
\left [ \; ({\bf e} {\bf d}) ^{2} \; sh ^{2} \; \beta - 1  - ({\bf e} {\bf d}) ^{2} \; [\;
2 \; ch \; \beta - 2 + ch^{2} \beta -2 \;ch\; \beta  +1 \;  ] \; \right  ] \; ,
\qquad \Longrightarrow
\nonumber
\\
1-f^{2} =   {- 1 \over
({\bf e} {\bf d}) ^{2} \; sh ^{2} \; \beta  } \hspace{60mm}
\label{11.12b}
\end{eqnarray}

\noindent and also remembering
\begin{eqnarray}
F^{2} = {D^{2} \over ({\bf e} {\bf d}) ^{2} \; sh ^{2} \; \beta  } \; ,
\nonumber
\end{eqnarray}

\noindent  to eq.  (\ref{11.11b}) can be given  the form
\begin{eqnarray}
-X^{2}  - Y^{2}  + {(Z-Z_{0})^{2}  \over ({\bf e} {\bf d}) ^{2} \; sh ^{2}\beta}
 = D^{2} \; ,
\label{11.13}
\end{eqnarray}

\noindent which  is  a canonical equation for hyperboloid. Its symmetry
axis is located along the
vector
\begin{eqnarray}
{\bf f}  = { {\bf d} + (ch\; \beta -1) ({\bf e} {\bf d}) {\bf e} \over
({\bf e} {\bf d}) \; sh \; \beta } \; , \qquad F = {D \over
({\bf e} {\bf d}) \; sh \; \beta } \; .
\label{11.14}
\end{eqnarray}

\noindent
These formulas are correct only if ${\bf e} {\bf d} \neq 0$ .
\vspace{5mm}

Now let us briefly clarify how one can handle with a rotation
relating two given vectors (of the same
length)
\begin{eqnarray}
{\bf K} = K \; {\bf k},  \qquad
{\bf K}' = K \; {\bf k}', \qquad {\bf k} ^{2} = {\bf k} ^{'2} = 1,
\nonumber
\\
O (\phi, {\bf e})\; {\bf K} =  {\bf K} ' \; .
\label{11.15}
\end{eqnarray}

\noindent
One very simple solution of the  problem follows immediately from the geometrical scheme
(see Fig. 9)

\vspace{2mm}

\unitlength=0.6mm
\begin{picture}(100,50)(-55,0)
\special{em:linewidth 0.4pt} \linethickness{0.6pt}

\put(0,0){\vector(+3,1){40}}  \put(+40,+5){${\bf k}$}
\put(0,0){\vector(3,+2){30}}  \put(+33,+25){${\bf k}'$}
\put(0,0){\vector(-1,+3){10}} \put(-15,+25){${\bf e}$}
\put(+15,+7){$\phi$}

\put(80,+15){$ {\bf e} = {\bf k} \times {\bf k}' $}

\put(80,+5){$ \cos \phi = {\bf k} \bullet {\bf k}'  $}

\end{picture}

\vspace{5mm}

\begin{center}
{\bf Fig. 9 Additional rotation}
\end{center}

\noindent Evidently, such a vector  $ {\bf C} = \tan {\phi \over 2} \;{\bf e}$ might be
constructed by the formula
\begin{eqnarray}
O (\phi, {\bf e})\; {\bf K} =  {\bf K} ' \; , \qquad
{\bf C} = \tan {\phi \over 2} \;{\bf e}  = {{\bf K} \times {\bf K}' \over
 {\bf K} \bullet ({\bf K} +  {\bf K}') } =
 {{\bf k} \times {\bf k}' \over
 1 +{\bf k} \bullet {\bf k}' }  \; .
\label{11.16}
\end{eqnarray}

\noindent
It is  quite understandable that having done this simplest rotation
after that one can rotate additionally   over the  axis
 ${\bf k}'$ -- at this the vector  ${\bf K}'$ itself leaves unchanged:
\begin{eqnarray}
 O( \varphi , {\bf k}' ) {\bf K}' = {\bf K}' \; .
\nonumber
\end{eqnarray}

\section{ On the form of spherical mirror
in the  moving reference frame (2-dimensional case)}

\hspace{5mm}
Firstly, let us us  consider 2-dimension spherical surface (circle)
to which in the rest reference frame   $K'$ corresponds an equation
\begin{eqnarray}
x^{'2} + y^{'2}  = R^{2} \; .
\nonumber
\end{eqnarray}

\noindent  Through light signals, with this geometrical circle can be associated
the following set of events  $S'$  in space-time:
\begin{eqnarray}
S' =  \left \{ \; t'=\sqrt{x^{'2} + y^{'2}}  \; ,
x^{'2} + y^{'2}  = R^{2} \; \right  \} \; .
\label{12.2}
\end{eqnarray}

\noindent
The Lorentz transform
\begin{eqnarray}
t'=  ch\; \beta \;t + sh\; \beta \;x \;, \qquad
x'= ch \; \beta\; x + sh\; \beta \;t \;, \qquad y'=y
\nonumber
\end{eqnarray}

\noindent
will take this events set $S'$  to the form
\begin{eqnarray}
S: \qquad  \left \{ \; t=\sqrt{x^{2} + y^{2}} \;,
(ch \; \beta\; x + sh\; \beta \;t)^{2} + y^{2} =R^{2} \;  \right \} \; .
\label{12.4}
\end{eqnarray}

\noindent
Excluding the time variable $t$, one gets for geometrical surface $S$  in the  moving reference frame $K$
\begin{eqnarray}
(\;ch \; \beta\; x + sh\; \beta \; \sqrt{x^{2} + y^{2}} \;)^{2} + y^{2} =R^{2}
\; .
\label{12.5a}
\end{eqnarray}

\noindent
From where it follows
\begin{eqnarray}
ch^{2} \beta\; x^{2} +  2 ch \; \beta\; sh\; \beta \; x \sqrt{x^{2} +y^{2}} +
sh^{2}\beta \;  x^{2}  + sh^{2}\beta \;  y^{2}  + y^{2} = R^{2} \; ,
\nonumber
\end{eqnarray}

\noindent or
\begin{eqnarray}
sh\; 2\beta \; x \; \sqrt{x^{2} +y^{2}}=
R^{2} - ch\;2\beta \; x^{2} - ch^{2} \beta \; y^{2}  \; .
\nonumber
\end{eqnarray}

\noindent
Squaring this equation
\begin{eqnarray}
sh^{2}  2\beta \; x^{4}  + sh^{2}  2\beta \; x^{2}   y^{2} =
\nonumber
\\
= R^{4}   +
ch^{2}2\beta \; x^{4} + 2 \;ch\; 2\beta \; ch^{2} \beta \;  x^{2}y^{2}
 + ch^{4}\beta \; y^{4} - 2R^{2} ch\;2\beta \; x^{2} -2R^{2} ch^{2} \beta \; y^{2}\; ,
\nonumber
\end{eqnarray}

\noindent  one  gets  to
\begin{eqnarray}
( x^{4} + 2\; ch^{2} \beta \;  x^{2} y^{2}+
ch^{4} \beta \;y^{4} )  +R^{4} - 2R^{2} \; (2 \;sh^{2} \beta +1)\; x^{2} - 2R^{2} \; ch^{2} \beta \; y^ {2}
 = 0 \; .
\nonumber
\end{eqnarray}

\noindent
The equation obtained can be rewritten differently
\begin{eqnarray}
( x^{2} +  ch^{2} \beta \; y^{2} )^{2}   - 2R^{2} ( x^{2} + ch^{2} \beta y^{2}) +R^{4}
 =  \; 4 R^{2} \; sh^{2} \beta \; x^ {2}     \; ,
\nonumber
\end{eqnarray}

\noindent  that is
\begin{eqnarray}
[\; x^{2} +  ch^{2} \beta \; y^{2} - R^{2} ]^{2} = [\; 2 R \; sh\; \beta \; x\; ] ^{2}\; .
\nonumber
\end{eqnarray}

So, we arrive at a second order equation
\begin{eqnarray}
 x^{2} +  ch^{2} \beta \; y^{2} - R^{2}  =  2 R \; sh\; \beta \; x\; \; ;
\nonumber
\end{eqnarray}

\noindent
it easily can be  simplified to the form
\begin{eqnarray}
(x- 2R\; sh\; \beta )^{2} + ch^{2} \beta \; y^{2} = R^{2}\; ch^{2} \beta \; .
\nonumber
\end{eqnarray}

\noindent
what is a canonical equation for ellipse
\begin{eqnarray}
{(x- 2R\; sh\; \beta )^{2}\over R^{2}\; ch^{2} \beta}
 +  { y^{2}  \over R^{2}} =1 \; .
\label{12.7}
\end{eqnarray}

\section{On the form of spherical mirror in the moving
reference frame, general 3-dimensional case}

\hspace{5mm}
A spherical mirror in the rest reference frame $K'$
is described by the equation
\begin{eqnarray}
{\bf x}' {\bf x}' = R^{2} \; ;
\nonumber
\end{eqnarray}

\noindent  with it can be related  the special set of events   $S'$ in space-time:
\begin{eqnarray}
S' = \left \{ \; t'=\sqrt{{\bf x}' {\bf x}'}=R, \; {\bf x}' {\bf x}' = R^{2} \;  \right \}\; .
\label{13.2}
\end{eqnarray}

The Lorentz transform change it to other form
\begin{eqnarray}
S: \qquad  \left \{ \; t=\sqrt{{\bf x} {\bf x}}=R, \;
[\;  {\bf x}  + {\bf e}  \;(\;  sh\; \beta  \; t +
(ch\; \beta -1)\;  ({\bf e}  {\bf x})\; ) \;
] ^{2}   = R^{2} \; \;  \right \} \; ,
\nonumber
\end{eqnarray}

Excluding the time variable $t$, one gets  an equation corresponding to a geometrical
form of the same surface in  the  moving reference frame
\begin{eqnarray}
[\;  {\bf x}  + {\bf e}  \;(\;  sh\; \beta  \;  \sqrt{{\bf x} {\bf x}} +
(ch\; \beta -1)\;  ({\bf e}  {\bf x})\; ) \;
] ^{2}   = R^{2} \; .
\label{13.4}
\nonumber
\end{eqnarray}

\noindent
From this it follows
\begin{eqnarray}
{\bf x}^{2} +2 \; ( {\bf e} {\bf x})
[\;  sh\; \beta  \;  \sqrt{{\bf x} {\bf x}} +
(ch\; \beta -1)\;  ({\bf e}  {\bf x})\; ] +
(\;  sh\; \beta  \;  \sqrt{{\bf x} {\bf x}} +
(ch\; \beta -1)\;  ({\bf e}  {\bf x})\; )^{2} = R^{2} \; ,
\nonumber
\end{eqnarray}

\noindent or
\begin{eqnarray}
{\bf x}^{2} +2 \; sh\; \beta  \; ( {\bf e} {\bf x})
    \sqrt{{\bf x} {\bf x}} +
2 (ch\; \beta -1)\;  ({\bf e}  {\bf x})^{2}  +
\nonumber
\\
+
 sh^{2} \beta  \;  {\bf x} ^{2}  + 2 sh\; \beta (ch\; \beta -1)\;
({\bf e}  {\bf x})\;  \sqrt{{\bf x} {\bf x}}    +
(ch\; \beta -1)^{2} \;  ({\bf e}  {\bf x})^{2} = R^{2} \; ,
\nonumber
\end{eqnarray}

\noindent and further
\begin{eqnarray}
ch^{2} \beta \;{\bf x}^{2}  +
sh^{2}\beta \;  ({\bf e}  {\bf x})^{2}  +
 2 sh\; \beta \; ch\; \beta \;
 \sqrt{{\bf x} {\bf x}}  \;   ({\bf e}  {\bf x})  = R^{2} \; .
\nonumber
\end{eqnarray}

\noindent
Thus, we have arrived at the equation
\begin{eqnarray}
 2 sh\; \beta \; ch\; \beta \;
 \sqrt{{\bf x} {\bf x}}  \;   ({\bf e}  {\bf x})  = R^{2} -
 ch^{2} \beta \;{\bf x}^{2}  -
sh^{2}\beta \;  ({\bf e}  {\bf x})^{2}  \; .
\label{13.5}
\end{eqnarray}

\noindent
It make sense  to employ special coordinate system --
such that ${\bf e}$  become oriented along  the first axis:
\begin{eqnarray}
O{\bf e} = (1,0,0), \qquad
{\bf e} {\bf x} =  {\bf e} \tilde{O} O{\bf x} = (O{\bf e}) \bullet {\bf X}= X
\label{13.6}
\end{eqnarray}

\noindent as a result eq.  (\ref{13.5}) will take a  simpler form
\begin{eqnarray}
 sh\; 2\beta \; X\; \sqrt{X^{2} + Y^{2} + Z^{2} }    =  R^{2} -
 ch^{2} \beta \;(X^{2} +Y^{2} +Z^{2})  -
sh^{2}\beta \; X\;  \; ;
\nonumber
\end{eqnarray}

\noindent
which can be presented as
\begin{eqnarray}
 sh\; 2\beta \sqrt{X^{2} + (Y^{2} + Z^{2}) } \; \;  X   =  R^{2} -
 ch \; 2 \beta \;X^{2} -   ch^{2} \beta \; (Y^{2} +Z^{2})   \; .
\label{13.7a}
\end{eqnarray}

\noindent
The relation obtained is readily compared with previously established eq.  (\ref{12.5a})
\begin{eqnarray}
sh\; 2\beta \; x \; \sqrt{x^{2} +y^{2}}=
R^{2} - ch\;2\beta \; x^{2} - ch^{2} \beta \; y^{2}  \;
\nonumber
\end{eqnarray}

\noindent  through the  formal changes
\begin{eqnarray}
x \;\;\;
\Longrightarrow \;\;\;X\; , \qquad y^{2}  \;\;\; \Longrightarrow  \;\;\;  (Y^{2} + Z^{2}) \; .
\nonumber
\end{eqnarray}

\noindent
One  may use the previously determined  solution  -- see. (\ref{12.7}):
\begin{eqnarray}
{(x- 2R\; sh\; \beta )^{2}\over R^{2}\; ch^{2} \beta}
 +  { y^{2}  \over R^{2}} =1 \; ;
\nonumber
\end{eqnarray}

\noindent  from where  one arrives at
\begin{eqnarray}
{(X- 2R\; sh\; \beta )^{2}\over R^{2}\; ch^{2} \beta}
 +  { Y^{2}  \over R^{2}} +  { Z^{2}  \over R^{2}} =1 \; .
\label{13.9}
\end{eqnarray}

\begin{quotation}
\noindent
{\em
Thus,  for the  moving observer,
 the spherical surface becomes an ellipsoid $K$.
}
\end{quotation}

\section{ The light reflection law in media}

\hspace{5mm} Let a light ray (incident and reflected) be propagated in
a uniform media with refraction index   $n>1$.  In the rest
reference frame $K'$  the reflection  law preserves
its form. However,
there arise differences when going to a  moving reference frame
$K$.

The fact of the most significance is that when using ordinary  (vacuum-based)
 Lorentz transformations
then because
 the speed of  light in the rest media is less than c ( $kc<c$)
   the modulus of the light velocity  does not preserve
 its value, as a result  meaning of  the formulas becomes very different.

 Instead of eq. (\ref{7.1a}) now we have
\begin{eqnarray}
{\bf a} '= {{\bf W}'_{in} \over c} \; , \;\;  {\bf a}^{'2} <  1 \;
, \qquad {\bf b} '= { {\bf W}'_{out} \over c} \; , \; \; {\bf
b}^{'2} < 1 \; ;
\nonumber
\\
{\bf n} '={  {\bf W}'_{norm} \over c} \; , \qquad {\bf n}^{'2} < 1
\; .
\label{14.1}
\end{eqnarray}

\noindent
General mathematical  form of the  reflection  law in the  moving reference frame
(\ref{9.5}) formally stays the same
\begin{eqnarray}
{ {\bf a} \times {\bf n}   +
 th\; \beta  \; ( {\bf a} -{\bf n}  )\times{\bf e}
+  (1  - ch^{-1} \beta)\;  {\bf e} \; [ \;{\bf e}  ({\bf n} \times
{\bf a})\; ]
    \over
1  + th\; \beta \; ({\bf e} {\bf a})  }=
\nonumber
\\
= { {\bf b} \times {\bf n}  +
 th\; \beta ({\bf b} -{\bf n})  \times {\bf e}
  +   (1  - ch^{-1} \beta)\; {\bf e} \; [ \;{\bf e}  ({\bf n} \times {\bf b})\; ]
    \over
1  + th\; \beta \; ({\bf e} {\bf b})  } \; ,
\label{14.2}
\\
{ {\bf a}  {\bf n}  (1-V^{2})  +  ({\bf a} +  {\bf n} ){\bf V}
      + [\; V^{2} + ({\bf a} {\bf V}) ({\bf n} {\bf V})  \; ]
 \over
1  +  {\bf a} {\bf V}   }\; +
\nonumber
\\
+\; { {\bf b}  {\bf n}  +  ({\bf b} +  {\bf n} ){\bf V}
     + [\; V^{2} + ({\bf b} {\bf V}) ({\bf n} {\bf V})  \; ]
 \over
1 +  {\bf b} {\bf V}   } = 0 \; ,
\label{14.3}
\\
\Delta = {\bf n} \; ({\bf a} \times   {\bf b} ) =
\nonumber
\\
= [\; {\bf e} ({\bf n}' \times {\bf a}' ) \;] \; {  - 2 \;sh\;
\beta  \; ({\bf a}' {\bf n}') \over (ch\; \beta  - sh\; \beta \;
{\bf e}{\bf n}' ) \; (ch\; \beta  - sh\; \beta \; {\bf e}{\bf
a}')\;
 (ch\; \beta  - sh\; \beta \; {\bf e}{\bf b}') } \; ,
\label{14.4}
\end{eqnarray}

\noindent however, one  must take into account that the lengths  of the
 vectors involved ${\bf a},{\bf b},{\bf n}$ are different from 1.

\vspace{5mm}
Let us calculate the length of the light  vector in the  moving
reference frame. We should start with the  transformation law
\begin{eqnarray}
{\bf W} = { {\bf W}' + {\bf e} \;  [\; - sh\; \beta   + (ch\;
\beta -1)\;  ( {\bf e} \; {\bf W}' )\; ] \over ch\; \beta  - sh\;
\beta \; {\bf e}\;  {\bf W}'  } \; .
\label{14.5a}
\end{eqnarray}

\noindent
In the rest reference frame  $K'$ the speed of light is
 $k$ (it is convenient to divide  all velocities  by   $c$):
\begin{eqnarray}
{\bf W}' = k {\bf w}' \; , \qquad {\bf w}^{'2} = 1 \; .
\nonumber
\end{eqnarray}

\noindent Eq. (\ref{14.5a}) reads as
\begin{eqnarray}
{\bf W} = { k {\bf w}'  + {\bf e} \;  [\; -sh\; \beta   + k (ch\;
\beta -1)\;    {\bf e}  {\bf w}' \; ] \over ch\; \beta  -  k\;
sh\; \beta \; {\bf e}  {\bf w}'  } \; .
\label{14.5c}
\end{eqnarray}

\noindent The  modulus of ${\bf W}$ is
\begin{eqnarray}
{\bf W}^{2}  = \left [ { k {\bf w}'  + {\bf e} \;  [\;- sh\; \beta
+ k (ch\; \beta -1)\;    {\bf e}  {\bf w}' \; ] \over ch\; \beta
-  k\; sh\; \beta \; {\bf e}  {\bf w}'  } \right ] ^{2}=
\nonumber
\\
= { k^{2}   +2k ({\bf w}' {\bf e}) [\; -sh\; \beta   + k (ch\;
\beta -1)\;    {\bf e}  {\bf w}' \; ]  + [\; -sh\; \beta   + k
(ch\; \beta -1)\;    {\bf e}  {\bf w}' \; ] ^{2} \over [ch\; \beta
-  k\; sh\; \beta \; {\bf e}  {\bf w}' ]^{2} } \; .
\nonumber
\end{eqnarray}

\noindent With the  notation
\begin{eqnarray}
{\bf w}' {\bf e}=  \mu \; , \qquad \mu \in [-1,+1 ] \;
\label{14.6}
\end{eqnarray}

\noindent one  gets to
\begin{eqnarray}
{\bf W}^{2}  =
 { k^{2}   +2k \mu [\;- sh\; \beta   + k \mu
(ch\; \beta -1) ]  + [\; -sh\; \beta   + k\mu (ch\; \beta -1)\;
] ^{2} \over (ch\; \beta  -  k \mu \; sh\; \beta )^{2} } \; .
\nonumber
\end{eqnarray}

\noindent With simple  calculation
\begin{eqnarray}
 k^{2}   +2k \mu [\;- sh\; \beta   + k \mu
(ch\; \beta -1) ]  + [\; -sh\; \beta   + k\mu (ch\; \beta -1)\;
] ^{2} =
\nonumber
\\
= k^{2}   -2k \mu \; sh\; \beta   +2k ^{2} \mu^{2} (ch\; \beta -1)
+
\nonumber
\\
+ sh^{2}  \beta  -  2  k\mu \; sh\; \beta (ch\; \beta -1)+ k^{2}
\mu^{2} (ch^{2} \beta - 2 ch\; \beta + 1) =
\nonumber
\\
= k^{2} -  2  k\mu \; sh\; \beta \; ch\; \beta + sh^{2} \beta
-k^{2} \mu^{2} + k^{2} \mu^{2} ch^{2} \beta =
\nonumber
\\
= ( k^{2}-1) + ( ch^{2} \beta  -  2  k\mu \; sh\; \beta \; ch\;
\beta    +k^{2} \mu^{2} sh^{2} \beta )=
\nonumber
\\
= ( k^{2}-1) + ( ch \beta  -    k\mu \; sh\; \beta \; )^{2} \;
\nonumber
\end{eqnarray}

\noindent  we arrive at
\begin{eqnarray}
{\bf W}^{2}  =1 - {(1 - k^{2})  \over (ch\; \beta  -  k \mu \;
sh\; \beta )^{2} } \; < \;1 \; .
\label{14.7a}
\end{eqnarray}

\begin{quotation}

{\em Thus, the speed of light is not invariant quantity under the ordinary
Lorentz transformations; it depends on mutual orientation of
 ${\bf w}'$ and   ${\bf e}$ -- remembering  $ \mu =  {\bf e} {\bf w}' $.
 In any  moving reference frame $K$,
  the  light velocity $W$ in the  media is less than vacuum velocity.
 }
\end{quotation}

Eq.  (\ref{14.7a})  can be written as
\begin{eqnarray}
{\bf W}^{2}   = 1  -
 (1 - k^{2})    \; {  (1 -V^{2})    \over
(1  -  \mu kV   )^{2} } \; .
\label{14.7b}
\end{eqnarray}

There are two simple cases of  (anti) parallel motion of the  light and  reference frame:
\begin{eqnarray}
\mu   =+1: \qquad W  = {V+k    \over 1  -  kV      } \; ,
\nonumber
\\
\mu    =-1: \qquad
 W   = {V-k    \over
1  +  kV      } \; ;
\label{14.8a}
\end{eqnarray}

\noindent in usual  units these look as
\begin{eqnarray}
\mu   =+1: \qquad W  = {V+W'    \over 1  -  W'V /c^{2}     } \; ,
\nonumber
\\
\mu    =-1: \qquad  W   = {V- W'    \over 1  +  W'V/c^{2}      } \; .
\label{14.8b}
\end{eqnarray}

From the formulas obtained  (\ref{14.7a}) and
(\ref{14.5a}),  one can readily derive relationship describing
aberration of the light in the moving
reference frame. Indeed,
\begin{eqnarray}
{\bf W} =
{ k {\bf w}'  + {\bf e} \;  [\; -sh\; \beta   + k\mu
(ch\; \beta -1)\; ] \over
ch\; \beta  -  k \mu \; sh\; \beta  } = W\; {\bf w} \; ,
\nonumber
\\
 W   =  \sqrt{1 -
{(1 - k^{2})  \over
(ch\; \beta  -  k \mu \; sh\; \beta )^{2} }} \;  , \qquad \mu = {\bf e}  {\bf w}' \; .
\label{14.9b}
\end{eqnarray}

\noindent
So that a unit vector of the light in the $K$ frame is given by
\begin{eqnarray}
{\bf w} = \left [ 1 -
{(1 - k^{2})  \over
(ch\; \beta  -  k \mu \; sh\; \beta )^{2} }  \right ]^{-1/2}\;
{ k {\bf w}'  + {\bf e} \;  [\; -sh\; \beta   + k\mu
(ch\; \beta -1)\; ] \over
ch\; \beta  -  k \mu \; sh\; \beta  } =
\nonumber
\\
= { k {\bf w}'  + {\bf e} \;  [\; -sh\; \beta   + k\mu
(ch\; \beta -1)\; ] \over \sqrt{(ch\; \beta  -  k \mu \; sh\; \beta )^{2} - 1+k^{2} }}\; .
 \hspace{30mm}
\label{14.9c}
\end{eqnarray}

\vspace{5mm}
\noindent
This relation is much simplified when
the vectors  ${\bf w}'$  and  ${\bf e}$ are  perpendicular to each other:
\begin{eqnarray}
 \mu = 0, \qquad
{\bf w} = { k {\bf w}'  -  sh\; \beta\; {\bf e}     \;  \over
\sqrt{sh^{2} \beta   +k^{2}}   } \; ,
\label{14.10a}
\end{eqnarray}

\noindent
therefore the angle of aberration is as follows:
\begin{eqnarray}
\tan \alpha = {sh \; \beta \over k} \; ,
\label{14.10b}
\end{eqnarray}

\noindent
or in ordinary units
\begin{eqnarray}
\tan \alpha =  { V\over kc }\; {1 \over  \sqrt{1-V^{2}/c^{2}}} \; .
\nonumber
\end{eqnarray}

\begin{quotation}

{\em It should be especially   noted one other aspect of the problem: eq.
(\ref{14.9b}) means that the light velocity in the  reference frame  $K$ is a function of
direction of the propagation of the light. This fact is of most significance because it change basically
the general structure of special relativity in presence of a media. In such circumstances
there appears an absolute reference frame related to the rest media, the reference frame
 $K'$. In the   reference frame $K'$,
 the light velocity is an isotropic quantity that preserves its
 value in all  space directions. In any  other  reference frame, moving $K$,
 the light velocity is anisotropic --  it is a function of directions. }

\end{quotation}

 {\em  It is convenient to write eq. (\ref{14.7a}) in the form}

\begin{eqnarray}
1- {\bf W}^{2}  = (1 - W^{'2} ) \;\;
{  (1-V^{2})  \over
( 1 - {\bf W}' {\bf V})^{2} } \; .
\label{14.11a}
\end{eqnarray}

{\em Here, by symmetry reasons,  we can restrict ourselves  to  one angle variable so that
}

\begin{eqnarray}
 1- W^{2}(\alpha)      = (1 - W^{'2}) \; \;
{   (1-V^{2})  \over
( 1 - VW'  \; \cos \alpha )^{2} } \; .
\label{14.11b}
\end{eqnarray}

\noindent We may reach the formal simplicity with the use of
trigonometrical parametrization:
\begin{eqnarray}
V = th\; \beta \; , \qquad W' = th\; B'\; , \qquad W(\alpha) = th\; B(\alpha) \; ,
\nonumber
\end{eqnarray}

\noindent them eq.  (\ref{14.11b}) gives
\begin{eqnarray}
ch \; B (\alpha) = ch \; B' \; ch \; \beta - sh \; B' \; sh \; \beta \; \cos \alpha \; .
\label{14.11c}
\end{eqnarray}

\vspace{5mm}

\section{ The light and the  tensor formalism of 4-velocities
 $u^{a}$}

\hspace{5mm}
In general, the 4-velocity vector is defined as [30]
\begin{eqnarray}
u^{a}= (u^{0}, {\bf u} ) = (\;  {1 \over \sqrt{1-W^{2} }}\;
, \; {{\bf W} \over  \sqrt{1-W^{2} }}  \; ) \; , \qquad {\bf W} = W \; {\bf w} \; .
\label{15.1a}
\end{eqnarray}

To have the  case of the light one should set   $W=1$ , then eq. (\ref{15.1a}) gives
\begin{eqnarray}
u^{a}= (u^{0}, {\bf u} ) = \infty \; (\;  1 , \; {\bf w}  \; ) \; .
\label{15.1b}
\end{eqnarray}

\noindent
To add some details at this limiting procedure it is convenient
to employ the following  parametrization
\begin{eqnarray}
{\bf W} =  th \; B \; {\bf w} \; , \qquad
u^{a}=  (\;  ch \; B , \; sh\; B\;  {\bf w}  \; ) \; , \;\;
B \in [0, + \infty ) \; ,
\nonumber
\end{eqnarray}

\noindent to the  light limit  there corresponds  $B \longrightarrow +\infty$:
\begin{eqnarray}
\lim _{B \rightarrow \infty } \; ch\; B =
\lim _{B \rightarrow \infty } \; {e^{B} + e^{-B} \over 2} = \;+ \infty ,
\nonumber
\\
\lim _{B \rightarrow \infty } \; sh\; B =
\lim _{B \rightarrow \infty } \; {e^{B} - e^{-B} \over 2} = \;+ \infty .
\nonumber
\\
\lim _{B \rightarrow \infty } \; th\; B = 1 \;, \qquad
u^{a}= \infty \; (\;  1 , \; {\bf w}  \; ) \;
\nonumber
\end{eqnarray}

\noindent and
\begin{eqnarray}
{1 \over \sqrt{1-W^{2} }} =  ch \; B = { e^{B} +e^{-B}\over 2}, \qquad ch^{2} B
\approx    { e^{2B} +2 \over 4}  \sim  {\infty^{2} \over 4}  +{1\over 2} \; ,
\nonumber
\\
{W \over \sqrt{1-W^{2} }} =  sh \; B = { e^{B} -e^{-B}\over 2} , \qquad
sh^{2} B \approx  { e^{2B} -2 \over 4} \sim  {\infty^{2}\over 4} - {1\over 2}
\nonumber
\\
ch^{2} B - sh^{2} B \approx 1 \; ,
\nonumber
\\
W^{2} = {e^{2B} -2 \over e^{2B} + 2} =
1 - {4 \over e^{2B} + 2}
\; \sim \; {\infty^{2} -2 \over \infty^{2}  +2 } = 1 - { 4 \over \infty^{2}  +2 }\;  ,
\label{15.3b}
\\
ch\; B = { \sqrt{e^{2B} +2}  \over 2} = {e^{B} \over 2} \;\sqrt{1 + 2e^{-2B}} =
{e^{B} \over 2} \; (1 + e^{-2B}) \; \sim \;  {\infty \over 2} \; (1 + \infty^{-2}) \;  ,
\nonumber
\\
sh\; B = { \sqrt{e^{2B} -2}  \over 2} = {e^{B} \over 2} \;\sqrt{1 - 2e^{-2B}} =
{e^{B} \over 2} \; (1 - e^{-2B}) \; \sim \;  {\infty \over 2} \; (1 - \infty^{-2})\; .
\nonumber
\end{eqnarray}

Now let us  return to eq. (\ref{15.1a}) at very big $B$-s:
\begin{eqnarray}
u^{a}=    ( \;  {e^{B} \over 2} \; (1 + e^{-2B}),  \;
 {e^{B} \over 2} \; (1 - e^{-2B}) \; {\bf w}  \;   ) \sim \infty \; (1, {\bf w}) \; ;
\label{15.4}
\end{eqnarray}

\noindent  and examine how will act the Lorentz transform on such a  limiting 4-vector
First, consider simple 1-dimensional case  ${\bf W}=(W,0,0)$:
\begin{eqnarray}
u^{a}=   ( \;  {e^{B} \over 2} \; (1 + e^{-2B}), \; \;
 {e^{B} \over 2} \; (1 - e^{-2B}) , 0, 0  \;   ) \sim \infty\; (1,1,0,0)
\; .
\nonumber
\end{eqnarray}

\noindent
The Lorentz transformation
\begin{eqnarray}
t' = ch\; \beta \; t +   sh\; \beta \;  x \; , \qquad
 x'=    \; sh\; \beta  \; t + ch\; \beta \; x
\nonumber
\end{eqnarray}

\noindent act as follows
\begin{eqnarray}
u^{'0} =  ch\; \beta \;  ch B +   sh\; \beta \; sh \;B = ch\; (\beta +B)  \; ,
\nonumber
\\
 u^{'1}=  sh\; \beta  \; ch\; B  +
ch\; \beta \; sh \; B = sh \; (\beta + B) \;     \; ,
\nonumber
\end{eqnarray}

\noindent that is
\begin{eqnarray}
u^{'a} =  ( e^{\beta +B} \; {1 + e^{-2(\beta+B)}\over 2} \; ,
 e^{\beta +B} \; {1 - e^{-2(\beta+B)}\over 2}  ,\; 0,\; 0 )  \sim \infty'  (1,1,0,0)
\label{15.5b}
\end{eqnarray}

\noindent where  the notation is used
\begin{eqnarray}
\infty ' = {e^{\beta + B} \over 2} \; .
\nonumber
\end{eqnarray}

\noindent Here we should see the parameter $\beta$ as a finite one whereas
the $B$  must be seen as infinity.

Generalization of the above analysis to the case of arbitrary Lorentz transformation
can be done quite easily. Here we have
\begin{eqnarray}
u^{'a} = {1 \over \sqrt{1 -W^{'2}}} \; (1 ,\;
{\bf W}'  \; ) \; ,
\nonumber
\end{eqnarray}

\noindent where
\begin{eqnarray}
 W '  =  \sqrt{1 -
{(1 - W^{2})  \over
(ch\; \beta  +  W ({\bf e}  {\bf w} ) \; sh\; \beta )^{2} }} \;  ,
\qquad {\bf W}' = W' \; {\bf w}' \; ,
\nonumber
\\
{\bf w}' =
 {  {\bf W}  + {\bf e} \;  [\; sh\; \beta   + W ({\bf e}  {\bf w} )
(ch\; \beta -1)\; ] \over \sqrt{(ch\; \beta  +  W ({\bf e}  {\bf w} )
\; sh\; \beta )^{2} - (1-W^{2}) }}\; ,
\nonumber
\end{eqnarray}

\noindent  so
\begin{eqnarray}
{\bf W}' = { ch\; \beta  +  W ({\bf e}{\bf w})  \; sh\; \beta \over \sqrt{1 -W^{2}}}\;
\; ( \; 1 \; , \;
{  {\bf W}  + {\bf e} \;  [\; sh\; \beta   + W ({\bf e} {\bf w})
(ch\; \beta -1)\; ] \over  ch\; \beta  +  W ({\bf e}{\bf w}) \; sh\; \beta   } \;  )\; .
\label{15.6}
\end{eqnarray}

\noindent
From this, for  the light   we will produce
\begin{eqnarray}
W \longrightarrow \; 1, \qquad {1 \over \sqrt{1 -W^{2}}}  \longrightarrow
{e^{B}\over 2} \; \sqrt{1 + 2 e^{-2B}} \;\;  \longrightarrow   \;\; \infty \; ,
\qquad
{\bf W}' \longrightarrow {\bf w}' ,
\nonumber
\end{eqnarray}

\noindent
and previous relation takes the form
\begin{eqnarray}
u^{'a}  = \infty \; \varphi'   \;
 ( \; 1 \; , \; {\bf w}' )
\label{15.7a}
\end{eqnarray}

\noindent  where the function  $\varphi'$ at the  $\infty$   symbol
is
\begin{eqnarray}
\varphi ' (\beta, {\bf e}, {\bf w}) \; \equiv \;
 [ ch\; \beta  +   ({\bf e}{\bf w})  \; sh\; \beta ]  =
{  1 +   {\bf w}{\bf V} \over \sqrt{1 - V^{2}}}  \;,
\label{15.7b}
\end{eqnarray}

\noindent
and the new light vector  ${\bf w}'$   is given by
\begin{eqnarray}
{\bf w}' =
{  {\bf w}  + {\bf e} \;  [\; sh\; \beta   +  ({\bf e} {\bf w})
(ch\; \beta -1)\; ] \over  ch\; \beta  +   ({\bf e}{\bf w}) \; sh\; \beta   } \; .
\label{15.7с}
\end{eqnarray}

{\em
Thus,  action of the Lorentz transform on light 4-velocity vector
can be summarized
 in symbolical form as follows
}
\begin{eqnarray}
L\; [ \; u^{a} = \infty \; (1 ,{\bf w} ) \; ] =
\varphi ' \; \infty \; (1 ,{\bf w}') \;  ,
\label{15.8}
\\
\varphi ' =  ch\; \beta  +   ({\bf e}{\bf w})  \; sh\; \beta
=
{  1 +   {\bf w}{\bf V} \over \sqrt{1 - V^{2}}} \; .
\nonumber
\\
{\bf w}' =
{  {\bf w}  + {\bf e} \;  [\; sh\; \beta   +  ({\bf e} {\bf w})
(ch\; \beta -1)\; ] \over  ch\; \beta  +   ({\bf e}{\bf w}) \; sh\; \beta   } \; ,
\nonumber
\end{eqnarray}

{\em and for usual (non light like) velocities }
\begin{eqnarray}
L\; [\;  u^{a} \; ] = u^{'a} \; , \hspace{40mm}
\nonumber
\\
u^{a} =   {1 \over \sqrt{1-  W^{2} }} \;(\; 1, {\bf W} \;) \; ,
\qquad  \qquad u^{'a} =   {1 \over \sqrt{1-  W^{'2} }} \;(\; 1, {\bf W}' \;) =
\label{15.9}
\\
=
 {1 \over \sqrt{1-  W^{2} }}
 \;\; { (1  +   {\bf W}{\bf  V}   )  \over \sqrt{1 -V^{2}}  }\;
\; ( \; 1 \; , \;
{  {\bf W}  + {\bf e} \;  [\; sh\; \beta   +
(ch\; \beta -1) {\bf e} {\bf W}\; ] \over  ch\; \beta  +  W ({\bf e}{\bf w})
 \; sh\; \beta   } \;  )\; .
\nonumber
\end{eqnarray}

\vspace{5mm}

{\em Evidently, eq.  (\ref{15.9}) will coincide with  (\ref{15.8}) at the limit  $W \rightarrow 1$.}

\vspace{5mm}

\section{Relativistic velocities and Lobachewski geometry }

\hspace{5mm}
The problem considered of describing  the light 4-velocities in contrast
to usual 4-velocities has an interesting geometrical interpretation\footnote{
The composition law for velocities is intimately related with hyperbolic
geometry (i.e. geometry on spaces with constant negative curvature), as was first
pointed out by A. Sommerfeld [87], V. Vari\v{c}ak , Alfred A. Robb [88-90], and \'{E}mile Borel
 [91].   More
recently the subject was elaborated on by  Abraham A. Ungar (Ungar [92];
many others aspects see in [87]. } .
Indeed,   every 4-vector  $u^{a} = (u^{0}, {\bf u})$ obeys the condition
\begin{eqnarray}
(u^{0})^{2} - {\bf u}^{2} = 1 \;
\label{16.1}
\end{eqnarray}

\noindent which gives the known realization of the 3-dimensional Lobachewski space
 $H_{3}$ of constant negative curvature as a surface in a pseudo Euclidean 4-space. There exist  one-to-one correspondence
 between 4-velocities and points of the geometrical space of negative constant curvature $H_{3}$.

 One can see that all points at spatial infinity   making the bounding domain
 $\bar{H}_{3}$ are described by
\begin{eqnarray}
\underline{\bar{H}_{3} :}
 \qquad  u^{0} =    {e^{B}\over 2} (1 + e^{-2B})\;,
{\bf u} = {e^{B}\over 2}(1 - e^{-2B}){\bf w} \;,  {\bf w}^{2} = 1  \; ,
\;\; B \longrightarrow \infty
\label{16.2}
\end{eqnarray}

\noindent  and just such (infinite bound)  points should be associated with all light 4-velocities.

In the parametrization $(B,\theta,\phi)$
\begin{eqnarray}
u^{0} = ch \; B \; , \qquad {\bf u } = sh\; B \; {\bf w} ,
\nonumber
\\
 {\bf w} =(
\sin \theta \cos \phi, \sin \theta \sin \phi, \cos \theta ) \; ,
\label{16.3}
\\
B \in [0, + \infty )\; , \qquad
\theta \in [0, \pi ], \qquad \phi \in [0, 2 \pi ] \;
\nonumber
\end{eqnarray}

\noindent the infinite bound   $\bar{H}_{3}$ is defined by
\begin{eqnarray}
\underline{\bar{H}_{3}:}
 \qquad B =  \infty, \qquad
\theta \in [0, \pi ], \qquad \phi \in [0, 2 \pi ]  \; .
\label{16.4}
\end{eqnarray}

\noindent
With the help of new variables $(W=th\; B ,\theta,\phi)$
\begin{eqnarray}
{\bf W} = {{\bf u} \over u_{0} }  = (th \;  B)  \; {\bf W}  = W \; {\bf w} \; , \qquad
W = th\; B \; \in [0, +1) \; .
\nonumber
\end{eqnarray}

\noindent
this bound set  $\bar{H}_{3}$  can be described in terms of finite quantities:
\begin{eqnarray}
\underline{\bar{H}_{3}:}
 \qquad W = 1, \qquad
\theta \in [0, \pi ], \qquad \phi \in [0, 2 \pi ]  \; .
\label{16.6}
\end{eqnarray}

\vspace{5mm}

Geometrical considerations can be helpful additional instrument in studying  the problem.
For instance, let us pose the question: to what extent will differ ordinary velocity
and light one. As an appropriate  characteristic let us try the geometrical distance between
respective points in the space $H_{3}$. It is known the formula for metric of  $H_{3}$
in coordinates   $(B ,\theta, \phi ) \;$:
\begin{eqnarray}
dl^{2} = dB^{2} + sh\; ^{2} B (d\theta^{2} + \sin^{2} \theta d\phi^{2}) \; ,
\qquad (  W =th\; B) \; .
\label{16.7}
\end{eqnarray}

The distance between two points,
$(W_{1},\theta_{0},\phi_{0})$  and  $(W_{2}=1,\theta_{0},\phi_{0})$,
along the fixed direction  $(\theta_{0},\phi_{0})$ is given by
\begin{eqnarray}
\left. l  = \int _{B_{1}}^{B_{2}} \; \sqrt{dB^{2} +
sh\; ^{2} B (d\theta^{2} + \sin^{2} \theta d\phi^{2})} \right |_{\theta_{0},\phi_{0}}=
\nonumber
\\
=
\int _{B_{1}}^{B_{2}} \; dB = B_{2} - B_{1} = \infty -  \;arcth\; W_{1} \; .
\label{16.8}
\end{eqnarray}

\begin{quotation}

{\em
Thus, in geometrical terms, any ordinary velocity $W_{1}$  being just slightly different  from
the light one $W_{2}=1$    is located the infinite distance from the light velocity:
 $l = \infty - B_{1}$. It may be emphasized geometrical sense of $W$: it
 provides us with minimal  geometrical distance in  the corresponding Lobachewsky space:
}
\end{quotation}
\begin{eqnarray}
l=    arcth\; W \; , \qquad W  = th\;l \; ;
\label{16.9a}
\end{eqnarray}

\noindent
in  ordinary units it  looks as
\begin{eqnarray}
{W \over c} = th\; {l\over R} \; ,
\label{16.9b}
\end{eqnarray}

\noindent  where  $R$ stands for the curvature radius of the  space $H_{3}$.
 (all such spaces, $\{ \; R, H_{3}\; \}$ are similar to  each other.
).

One other distance characteristic  may be defined in the case of differently
oriented velocities (consider specially simple example):
\begin{eqnarray}
 {\bf W}_{1} = (W \cos \phi_{1}, W \sin \phi_{1},0) \;,
\nonumber
\\
{\bf W}_{1} = (W \cos \phi_{2}, W\sin \phi_{2},0) ,
\label{16.10a}
\\
L =  \{ \; (W, \theta=\pi/2, \phi \in [\phi_{1} , \phi_{2}  \; )\; \} \; ;
\nonumber
\end{eqnarray}

\noindent that is
\begin{eqnarray}
\left. L= \int_{\phi_{1}} ^{\phi_{2}}
\sqrt{dB^{2} + sh\; ^{2} B (d\theta^{2} + \sin^{2} \theta d\phi^{2})} \right | _{W,\theta=\pi/2}
=
\nonumber
\\
= sh \; B \; (\phi_{2} - \phi_{1})=
{W \over \sqrt{1-W^{2}} } \;  (\phi_{2} - \phi_{1})\;.
\label{16.10b}
\end{eqnarray}

{\em In geometrical terms, the Lorentz transforms of the velocity vectors}
\begin{eqnarray}
{\bf W} '=
{ {\bf W}  + {\bf e}  \; [\; sh\; \beta   +
(ch\; \beta -1)\;  {\bf e} \; {\bf W} \;] \over
1  +  {\bf V}\;  {\bf W}  } \; \sqrt{1 -V^{2}} =
\label{16.11}
\\
 =  \;  {  {\bf V}  +    {\bf e} ( {\bf e}{\bf W} )  \over
1 \;  +  \;  {\bf V}\;  {\bf W}  }
+ { {\bf W}  -  {\bf e} ({\bf e} \; {\bf W} ) \over 1  +  {\bf V}\;  {\bf W}  } \; \sqrt{1 -V^{2}}
 \; ,
\nonumber
\\
{1 \over \sqrt{1-  W^{'2} }}=
 \;\; { 1  \over  \sqrt{1-  W^{2} } } \; { (1  +   {\bf W}{\bf  V}   ) \over \sqrt{1 -V^{2}}  } \;
\nonumber
\end{eqnarray}

\noindent
\begin{quotation}

\noindent
{\em can be interpreted as  special  group actions  on the points of  Lobachewsky space $H_{3}$.
To the  case of the light in vacuum there corresponds such group actions upon the  points of
 the bound $\bar{H}_{3}$ of the space  (located in the infinity);
 at this the 3-dimensional problem (16.11)
effectively reduces to 2-dimensional one because the normalization condition holds
\begin{eqnarray}
{\bf W}^{'2} = {\bf W}^{2} = 1 \; .
\nonumber
\end{eqnarray}

\noindent
However, for the light in the media both vectors ${\bf W}'$ and  ${\bf W}$ do not belong this
bound $\bar{H}_{3}$, they both are ordinary finite points in the space  $H_{3}$.}

\end{quotation}

\section{On relativistic transformation of the geometrical form
of \\ the surface to a moving reference frame in  presence of a media
}

\vspace{5mm}

\hspace{5mm}
(IN VACUUM)

\vspace{5mm}

Let us recall  the above method  to determine  the form of any rigid surface in a moving reference frame
Let the surface $S'$ be given  in the rest reference frame
\begin{eqnarray}
 S': \qquad  \{ \; \varphi({\bf x}) =0  \; \} \; .
\label{17.1}
\end{eqnarray}

\noindent
To this geometrical structure one can pose in correspondence
a special set of  events in space-time:
\begin{eqnarray}
 \{\;  (t',{\bf x}'):  \qquad  {\bf x}' = {\bf W}' t' \;,
 \; \;  \varphi({\bf x}') =0  \; \}
\nonumber
\end{eqnarray}

\noindent which with respect to Lorentz formulas
\begin{eqnarray}
t' = ch\; \beta \; t +   sh\; \beta \; {\bf e}\;   {\bf x} \; ,
\nonumber
\\
{\bf x}'= {\bf e}  \; sh\; \beta  \; t + {\bf x}  +
(ch\; \beta -1)\; {\bf e} \; ({\bf e}  {\bf x})
\nonumber
\end{eqnarray}

\noindent
will take other form
\begin{eqnarray}
 \{ \; (t,{\bf x}): \qquad  {\bf x} = {\bf W} t \; , \;
 \varphi \; [ \;  {\bf x}  + {\bf e}  \;(\;  sh\; \beta  \; t +
(ch\; \beta -1)\;  {\bf e}  {\bf x} \; )\; ] =0  \;
   \} \; .
\nonumber
\end{eqnarray}

\noindent
In order to obtain an equation for the  transformed geometrical surface $S$,
one should exclude the time  variable $t$:
\begin{eqnarray}
{\bf x} = {\bf W} t  , \; t= \sqrt{ {\bf x}^{2}} \; , \qquad ( {\bf W}^{2} = 1 \; ) \; ;
\label{17.5}
\end{eqnarray}

\noindent  so  one  produces
\begin{eqnarray}
S: \qquad \varphi \;[  \;  {\bf x}  + {\bf e}  \;(\;  sh\; \beta  \; \sqrt{ {\bf x}^{2}} +
(ch\; \beta -1)\;  {\bf e}  {\bf x} \; )\;  ] = 0 \; .
\label{17.6}
\end{eqnarray}

\hspace{5mm} (IN THE MEDIA)

\vspace{5mm}

In the same line one can act for the case of a uniform media when
the light velocity ${\bf W}^{'2}$ changes  in accordance with special law:
\begin{eqnarray}
 {\bf W}^{2}  =1 - (1 - W^{'2}) \; { (1 - V^{2})   \over
(1 -  VW'\cos \alpha )^{2} } \;  .
\label{17.7}
\end{eqnarray}

\noindent
Again, one starts with certain $S'$ surface in the rest reference frame
\begin{eqnarray}
 S': \qquad  \{ \; \varphi({\bf x}) =0 \}
\nonumber
\end{eqnarray}

\noindent With the  help of light signals, one  to the $S'$ structure can be referred
a set of events in space-time:
\begin{eqnarray}
\{\;  (t',{\bf x}'):  \qquad  {\bf x}' = {\bf W}' t' \;,   \;
\sqrt{ {\bf x}^{'2} }  =   kt' \; , \;  \varphi({\bf x}') =0  \; \} .
\nonumber
\end{eqnarray}

\noindent Under Lorentz  formulas
\begin{eqnarray}
t' = ch\; \beta \; t +   sh\; \beta \; {\bf e}\;   {\bf x} \; ,
\nonumber
\\
{\bf x}'=  {\bf x}  +  {\bf e}  \;(\;  sh\; \beta  \; t   +
(ch\; \beta -1)\; {\bf e}  {\bf x} \; )
\; .
\nonumber
\end{eqnarray}

\noindent
these events change into
\begin{eqnarray}
 \{ \; (t,{\bf x}): \qquad  {\bf x} = {\bf W} t \; , \;
 \varphi \; [  {\bf x}  + {\bf e}  \;(\;  sh\; \beta  \; t +
(ch\; \beta -1)\;  {\bf e}  {\bf x}\; )\; ] =0  \;
  \} \; .
\nonumber
\end{eqnarray}

\noindent Now, it is the point to exclude the time variable
$t$ ,  however one   must take into account
the rule  (\ref{17.7}):
\begin{eqnarray}
{\bf x} = {\bf W} t  , \qquad  t= {  \sqrt{ {\bf x}^{2} }  \over \sqrt{{\bf W}^{2}} }
\; ,
\nonumber
\end{eqnarray}

\noindent  so we arrive at
\begin{eqnarray}
S: \qquad \varphi \left [  {\bf x}  + {\bf e}  \;(\;  sh\; \beta  \;
{  \sqrt{ {\bf x}^{2} }  \over \sqrt{{\bf W}^{2}} }
\;+
(ch\; \beta -1)\;  ({\bf e}  {\bf x})\;  \right  ] = 0 \; ,
\label{17.13a}
\end{eqnarray}

\noindent where
\begin{eqnarray}
 {\bf W}^{2}  =1 - (1 - W^{'2}) \;
{ (1 - V^{2})   \over
(1 -  VW'\cos \alpha )^{2} } \;  .
\nonumber
\end{eqnarray}

\begin{quotation}

{\em The relation (\ref{17.13a}) describing the  change in the form of
a rigid surface  in presence of   media gives a  general  solution to the  problem
under consideration. However, that solution in not explicit and in any particular case we
need some
additional analysis to have in hand the new transformed form of the surface in fact.
}

\end{quotation}

\section{Modified Lorenz transforms in a uniform media}

\hspace{5mm}
Formal scheme of Special Relativity in a uniform media can be constructed
on the base of the light velocity in the media $kc$ ([94-96]; see the recent paper [97]):
\begin{eqnarray}
c \qquad  \Longrightarrow \qquad k\; c , \qquad k<1 \; .
\nonumber
\end{eqnarray}

Modified Lorentz formulas are   defined so that the new interval preserves its form:
\begin{eqnarray}
k^{2}c^{2} \;  t^{2} -{\bf x}^{2}=
k^{2}c^{2} \; t^{'2} -{\bf x}^{'2}  \; .
\label{18.2}
\end{eqnarray}

In the simplest 1-dimensional case
\begin{eqnarray}
k^{2}c^{2} \;  t^{2} -  x^{2}=
k^{2}c^{2} \; t^{'2} -  x^{'2}  \;
\label{18.3a}
\nonumber
\end{eqnarray}

\noindent new Lorentz  formulas look as
\begin{eqnarray}
x' = { x -Vt \over \sqrt{1 - V^{2} / k^{2} c^{2}  }}, \qquad
t' = {t - xV /k^{2}c^{2} \over \sqrt{1 - V^{2} /k^{2}c^{2} }} \; ;
\nonumber
\end{eqnarray}

\noindent indeed, it is easily verified identity
\begin{eqnarray}
k^{2} c^{2} \;  t^{'2} - {\bf x}^ {'2}= {1 \over 1 - V^{2} /k^{2}c^{2} }\; \left [
k^{2} c^{2} (t - {xV \over k^{2}c^{2} } )^{2} - (x-Vt)^{2} \right ] =
k^{2}c^{2} \;  t^{2} -{\bf x}^{2} \; .
\nonumber
\end{eqnarray}

\noindent
From  (\ref{18.3a}) it follows  the invariance of the light velocity  $W_{light}=kc$ in the media:
\begin{eqnarray}
k^{2}c^{2} \;  t^{2} -  x^{2}= 0 , \qquad
k^{2}c^{2} \; t^{'2} -  x^{'2}  =0  \; : \qquad \Longrightarrow \qquad
{x^{2} \over t^{2}} = {x^{'2} \over t^{'2}} = k^{2} c^{2} =  \mbox{inv} \; .
\nonumber
\end{eqnarray}

The modified Lorentz formulas   can be written
\begin{eqnarray}
x' = { x -{V\over kc} kc t \over \sqrt{1 - V^{2} / k^{2} c^{2}  }}, \qquad
kc t' = {kc t - x \; {V \over kc} \over \sqrt{1 - V^{2} /k^{2}c^{2} }} \; ;
\label{18.4}
\end{eqnarray}

\noindent that is in the new variables
\begin{eqnarray}
kc t \Longrightarrow t , \qquad x \Longrightarrow x , \qquad {V\over kc} \Longrightarrow V
\nonumber
\end{eqnarray}

\noindent they  look
\begin{eqnarray}
x' = { x - V  t \over \sqrt{1 - V^{2} }}, \qquad
t' = {t - x \; V  \over \sqrt{1 - V^{2}  }} \; .
\nonumber
\end{eqnarray}

Extension to Lorentz transforms with arbitrary velocity vector ${\bf V}$ is given
evidently as follows
\begin{eqnarray}
t' = ch\; \beta \; t +   sh\; \beta \; {\bf e}\;   {\bf x} \; ,
\nonumber
\\
{\bf x}'= {\bf e}  \; sh\; \beta  \; t + {\bf x}  +
(ch\; \beta -1)\; {\bf e} \; ({\bf e}  {\bf x})=
\nonumber
\\
=
[\;  {\bf x}  -  {\bf e} \; ({\bf e}  {\bf x}) \;] + {\bf e}  \;
 [  sh\; \beta  \; t + ch\; \beta \; ( {\bf e}  {\bf x} ) \; ]
\label{18.5a}
\end{eqnarray}

\noindent  where $(kc t)$   stands for  $t$ and
\begin{eqnarray}
{{\bf V} \over kc} = {\bf e} \; th\; \beta \;  , \qquad {\bf e} ^{2} = 1 \; ,
\nonumber
\\
{1 \over \sqrt{1 - V^{2}/k^{2}c^{2} }} = ch\; \beta, \qquad
{V /kc \over \sqrt{1 - V^{2}/k^{2} c^{2} }} = sh\; \beta \; .
\label{18.5b}
\end{eqnarray}

\noindent
In ordinary variables, eq.  (\ref{18.5a})  has the form
\begin{eqnarray}
t' =  {   t +    {\bf V} {\bf x} /  k^{2}c^{2}
 \over \sqrt{1 -V^{2} / k^{2} c^{2} } }
 \; ,
\nonumber
\\
{\bf x}'=
[\;  {\bf x}  -  {\bf e} \; ({\bf e}  {\bf x}) \;] +   \;
 {   {\bf e}({\bf e}  {\bf x} )  + {\bf V}  t  \over \sqrt{1 -V^{2} / k^{2} c^{2} } } \; .
\label{18.5c}
\end{eqnarray}

\noindent
From (\ref{18.5c}) one  can readily produce the  modified light velocity addition rule
\begin{eqnarray}
{\bf W} ' = { [\;   {\bf W}  -  {\bf e} \; ({\bf e}  {\bf W}) \;] + {\bf e}  \;
 [  sh\; \beta   + ch\; \beta \; ( {\bf e}  {\bf W} ) \; ] \over
 ch\; \beta  +   sh\; \beta \; {\bf e}\;   {\bf W} } =
\nonumber
\\
= { {\bf W}  -  {\bf e} \; ({\bf e}  {\bf W}) \over
 1   +   th\; \beta \; {\bf e}\;   {\bf W} } \;  \;  ch^{-1} \; \beta
 \;\; + \;\;
 { {\bf e}  \;
 th\; \beta   +  {\bf e}  ( {\bf e}  {\bf W} )  \over
 1  +   th\; \beta \; {\bf e}\;   {\bf W} } \;
 \label{18.6a}
 \end{eqnarray}

\noindent or in ordinary units
\begin{eqnarray}
{\bf W} ' =  { {\bf W}  -  {\bf e} \; ({\bf e}  {\bf W}) \over
 1   +    {\bf V}\;   {\bf W} /k^{2}c^{2} } \;  \;  \sqrt{1 - V^{2}/k^{2} c^{2} }\;\;  +
\;\;  { {\bf V}  \;
 +  {\bf e}  ( {\bf e}  {\bf W} )  \over
 1  +    {\bf V}\;   {\bf W} /k^{2}c^{2}  } \; .
 \nonumber
 \end{eqnarray}

It may be verified straightforwardly that the  value of light velocity in the media
is invariant under modified Lorentz formula (\ref{18.6a}):
\begin{eqnarray}
{\bf W}^{2} = k^{2} c^{2} ; \qquad \Longrightarrow \qquad  {\bf W}^{'2} = k^{2} c^{2}
\label{18.7}
\end{eqnarray}

\noindent
Indeed, starting from
\begin{eqnarray}
{\bf W}' /  kc  = \left [\;
{  {\bf W} \over  kc }  + {\bf e}  \;  [ \;
{ V   \over kc \; \sqrt{1 -V^{2} / k^{2}c^{2} }  }  +
( {1  \over  \sqrt{1 -V^{2} /k^{2}c^{2} } }  - 1 ) \;
 { {\bf e} {\bf W} \over kc }  \; )\; \right ] \times
\nonumber
\\
 \times \left [
{1 \over  \sqrt{1 -V^{2} / k^{2} c^{2} } }      +  { V\;  ({\bf e}{\bf W})
 \over  k^{2}c^{2}\;  \sqrt{1 -V^{2} / k^{2} c^{2} }  } \right ] ^{-1} \;
 \nonumber
 \end{eqnarray}

\noindent for the first factor we have
\begin{eqnarray}
 \left [\;
{  {\bf W} \over  kc }  + {\bf e}  \;  [ \;
{ V   \over kc \; \sqrt{1 -V^{2} / k^{2}c^{2} }  }  +
( {1  \over  \sqrt{1 -V^{2} /k^{2}c^{2} } }  - 1 ) \;
 { {\bf e} {\bf W} \over kc }  \; )\; \right ] ^{2}  =
\nonumber
\\
= { {\bf W}^{2} \over  k^{2}c^{2} } + 2 \; { ( {\bf e} {\bf W}) \over k^{2} c^{2} } \;
\left [\;
{ V \over \sqrt{1 -V^{2} / k^{2} c^{2} } }  +  ( \;
{1 \over \sqrt{1 -V^{2} / k^{2} c^{2} }} -1) \;
({\bf e} {\bf W})\; \right  ] +
\nonumber
\\
+ {V^{2} \over k^{2}c^{2} (1 -V^{2}/k^{2}c^{2}) } + 2\; {V \; ({\bf e} {\bf W}) \over
k^{2}c^{2} \sqrt{1 -V^{2} / k^{2} c^{2} }}  \;
 ( {1 \over \sqrt{1 -V^{2} / k^{2} c^{2} }} -1 ) +
\nonumber
\\
 +
{({\bf e}{\bf W})^{2} \over k^{2} c^{2}}  \; \left ( {1 \over 1 -V^{2} /k^{2}c^{2} } -
{2 \over \sqrt{1 -V^{2} / k^{2} c^{2} }} +1  \; \right ) =
\nonumber
\\
=
( { {\bf W}^{2} \over  k^{2}c^{2} }  - 1 ) \; + \;
  1  +   { V^{2} /k^{2}c^{2} \over 1 - V^{2} /k^{2}c^{2}} +
{2 V \; ({\bf e}{\bf W}) \over kc \; (1 -V^{2}/k^{2}c^{2})} +
{ ({\bf e}{\bf W})^{2} \over k^{2}c^{2} }  \; [ {1 \over 1 -V^{2}/k^{2}c^{2} } -1 ] =
\nonumber
\\
= (  {{\bf W}^{2} \over  k^{2} c^{2} } -1    ) +
\left [ {1 \over  \sqrt{1 -V^{2} / k^{2} c^{2} } }    +  { V\;  ({\bf e}{\bf W})
 \over  k^{2}c^{2}   \sqrt{1 -V^{2} / k^{2} c^{2} }  }  \; \right ] ^{2} \; .
\nonumber
\end{eqnarray}

\noindent
Therefore, we  have arrived at the rule
\begin{eqnarray}
 {\bf W}^{'2}  = k^{2} c^{2}  + (  {\bf W}^{2}  -k^{2}c^{2}    ) \;
\left [ {1 \over  \sqrt{1 -V^{2} / k^{2} c^{2} } }    +  { V\;  ({\bf e}{\bf W})
 \over  k^{2}c^{2}   \sqrt{1 -V^{2} / k^{2} c^{2} }  }  \; \right ] ^{-2} \; ,
\label{18.8a}
\end{eqnarray}

\noindent it can be  rewritten as
\begin{eqnarray}
 {\bf W}^{'2}  = k^{2} c^{2}  + {  {\bf W}^{2}  -k^{2}c^{2}   \over
[ ch \; \beta + sh\; \beta \; {\bf e}{\bf W} /kc ]^{2}}
\label{18.8b}
\end{eqnarray}

\noindent
For the case of the  light, eq.  (\ref{18.8b}) reduces to eq.   (\ref{18.7})

\section{On transforming the form of rigid surface when going to
a moving reference frame in presence of a uniform media, with modified Lorentz formulas
  }

\vspace{5mm}

\hspace{5mm}
(IN MEDIA )

\vspace{5mm}

General method remains the same.  However, now the  modified Lorentz formulas
should  be  used. Let in the rest reference frame $K'$ certain surface be given
\begin{eqnarray}
S': \qquad  \{ \; \varphi({\bf x}) =0  \; \} \; .
\nonumber
\end{eqnarray}

\noindent
With  the surface  $S'$ one may associate (with the help of light signal in the  media)
quite definite  set of space-time events:
\begin{eqnarray}
 \{\;  (t',{\bf x}'):  \qquad  {\bf x}' = {\bf W}' t' \;,
 \; \;  \varphi({\bf x}') =0  \;, \; W = kc \; \} .
\nonumber
\end{eqnarray}

With the  help of \underline{modified} Lorentz formulas
\begin{eqnarray}
(kc t') =   (kc t)\; ch\; \beta  +   sh\; \beta \; {\bf e}\;   {\bf x} \; ,
\nonumber
\\
{\bf x}'= {\bf e}  \; sh\; \beta  \; ( kc t) + {\bf x}  +
(ch\; \beta -1)\; {\bf e} \; ({\bf e}  {\bf x})
\label{19.3}
\end{eqnarray}

\noindent
where
\begin{eqnarray}
 ch\;  \beta = {1 \over \sqrt{1 - V^{2}/k^{2}c^{2}}},
\qquad  sh \;\beta = {V \over kc\; \sqrt{1 - V^{2}/k^{2}c^{2}}}
\nonumber
\end{eqnarray}

\noindent
the set of events changes  to
\begin{eqnarray}
 \{ \; (t,{\bf x}): \qquad  {\bf x} = {\bf W} t \; , \;
 \varphi \; [ \;  {\bf x}  + {\bf e}  \;( \; kct \;  sh\; \beta   +
(ch\; \beta -1)\;  {\bf e}  {\bf x} \; )\; ] =0  \;
   \} \; .
\label{19.4}
\end{eqnarray}

In order to have an equation for the surface $S$ in the  reference frame  $K$,  one
 should exclude the time  variable  $t$:
\begin{eqnarray}
{\bf x} = {\bf W} t  , \qquad  kct = \sqrt{ {\bf x}^{2}} \;  ;
\nonumber
\end{eqnarray}

\noindent  so that
\begin{eqnarray}
S: \qquad \varphi \;[  \;  {\bf x}  + {\bf e}  \;(\;  sh\; \beta  \; \sqrt{ {\bf x}^{2}} +
(ch\; \beta -1)\;  {\bf e}  {\bf x} \; )\;  ] = 0 \; .
\label{19.6}
\end{eqnarray}

\vspace{5mm}
\noindent
In eq. (\ref{19.6}), the  media's presence  enters through the modified
hyperbolic functions:
\begin{eqnarray}
 ch\;  \beta = {1 \over \sqrt{1 - V^{2}/k^{2}c^{2}}} \; ,
\qquad  sh \;\beta = {V \over kc\; \sqrt{1 - V^{2}/k^{2}c^{2}}} \; .
\nonumber
\end{eqnarray}

For instance, eq. (\ref{19.6})  describing the  chance of $R$-circle into
ellipse when going from rest reference frame to the  moving one
\underline{(IN VACUUM)}
\begin{eqnarray}
{(x- 2R\; sh\; \beta )^{2}\over R^{2}\; ch^{2} \beta}
 +  { y^{2}  \over R^{2}} =1 \; ,
\label{19.9a}
\\
 ch\;  \beta = {1 \over \sqrt{1 - V^{2}/c^{2}}},
\qquad  sh \;\beta = {V \over c\; \sqrt{1 - V^{2}/c^{2}}} \;
\nonumber
\end{eqnarray}

\noindent
in case of the \underline{MEDIA} is modified by  presence of the  parameter $k$:
\begin{eqnarray}
{(x- 2R\; sh\; \beta )^{2}\over R^{2}\; ch^{2} \beta}
 +  { y^{2}  \over R^{2}} =1 \; ,
\label{19.9b}
\\
 ch\;  \beta = {1 \over \sqrt{1 - V^{2}/k^{2}c^{2}}},
\qquad  sh \;\beta = {V \over kc\; \sqrt{1 - V^{2}/k^{2}c^{2}}} \; .
\nonumber
\end{eqnarray}

\subsection*{20. Conclusions }

Let us summarize some results.

\begin{quotation}

I. {\em The influence of the relativistic motion of the reference frame
on the light reflection law has been  investigated in detail.}

The method  used is based on applying the relativistic
aberration affect for three light signals: incident, normal and reflected rays.
Two choices for a normal light signal in the rest reference frame are used:
one going to and  another going from the reflecting surface.
 The form of the reflection law in the moving reference frame is substantially modified
 and includes an additional  parameter which is the  velocity vector of the reference frame.
 The relationship produced proves   invariance under  Lorentz  group
 transformations.
  It is shown that the reflected ray, as measured by a moving observer, would not
  in general be in the same plane as the incident and normal rays. So a plane geometric figure,
  determined by the incident, normal and reflected rays, may be observed as having three
  dimensions.

II. {\em A general method to transform the form of any rigid surface in 3-dimensional space
with respect to the arbitrary directed relative motion of the reference frame has been
detailed.}

This method is based on the light signals processes and the invariance of the light
velocity under Lorentz transformations. It is shown that a moving observer will measure
a plane surface as a hyperboloid. That observer will also measure a spherical surface
as an ellipsoid. A right line in the plane is seen by a moving observer as a  hyperbola.

III. {\em Extending of the above analysis to the case of uniform media has been given.}

In the case of a uniform media the light velocity is not invariant under Lorentz
 transformation and is  anisotropic.
 An alternative scheme of a modified Lorentz symmetry based on the invariance of
 the light velocity in the media is presented. This  significantly modifies the
 formulation of the reflection law in the moving reference frame and also the form of any surface in the moving reference frame compared with the vacuum case.

IV.
{\em Some geometrical aspects of the relativistic velocity concept  in terms of  the
 Lobachewsky 3-geometry are briefly discussed.}

\end{quotation}

\end{document}